\def\TIPITOT{
\font\dodicirm=cmr12
\font\dodicii=cmmi12
\font\dodicisy=cmsy10 scaled\magstep1
\font\dodiciex=cmex10 scaled\magstep1
\font\dodiciit=cmti12
\font\dodicitt=cmtt12
\font\dodicibf=cmbx12 scaled\magstep1
\font\dodicisl=cmsl12
\font\ninerm=cmr9
\font\ninesy=cmsy9
\font\eightrm=cmr8
\font\eighti=cmmi8
\font\eightsy=cmsy8
\font\eightbf=cmbx8
\font\eighttt=cmtt8
\font\eightsl=cmsl8
\font\eightit=cmti8
\font\seirm=cmr6
\font\seibf=cmbx6
\font\seii=cmmi6
\font\seisy=cmsy6
%%%%%%%%%%%%%%%%%%%%%%%%%%%%%%%%%%%%%%%
\font\dodicitruecmr=cmr10 scaled\magstep1
\font\dodicitruecmsy=cmsy10 scaled\magstep1
\font\tentruecmr=cmr10
\font\tentruecmsy=cmsy10
\font\eighttruecmr=cmr8
\font\eighttruecmsy=cmsy8
\font\seventruecmr=cmr7
\font\seventruecmsy=cmsy7
\font\seitruecmr=cmr6
\font\seitruecmsy=cmsy6
\font\fivetruecmr=cmr5
\font\fivetruecmsy=cmsy5
%%%% definizioni per 10pt %%%%%%%%
\textfont\truecmr=\tentruecmr
\scriptfont\truecmr=\seventruecmr
\scriptscriptfont\truecmr=\fivetruecmr
\textfont\truecmsy=\tentruecmsy
\scriptfont\truecmsy=\seventruecmsy
\scriptscriptfont\truecmr=\fivetruecmr
\scriptscriptfont\truecmsy=\fivetruecmsy
%%%%% cambio grandezza %%%%%%
\def \ottopunti{\def\rm{\fam0\eightrm}% switch to 8-point type
\textfont0=\eightrm \scriptfont0=\seirm \scriptscriptfont0=\fiverm
\textfont1=\eighti  \scriptfont1=\seii  \scriptscriptfont1=\fivei
\textfont2=\eightsy \scriptfont2=\seisy \scriptscriptfont2=\fivesy
\textfont3=\tenex   \scriptfont3=\tenex \scriptscriptfont3=\tenex
\textfont\itfam=\eightit  \def\it{\fam\itfam\eightit}%
\textfont\slfam=\eightsl  \def\sl{\fam\slfam\eightsl}%
\textfont\ttfam=\eighttt  \def\tt{\fam\ttfam\eighttt}%
\textfont\bffam=\eightbf  \scriptfont\bffam=\seibf
\scriptscriptfont\bffam=\fivebf  \def\bf{\fam\bffam\eightbf}%
\tt \ttglue=.5em plus.25em minus.15em
\setbox\strutbox=\hbox{\vrule height7pt depth2pt width0pt}%
\normalbaselineskip=24pt
\let\sc=\seirm  \let\big=\eightbig  \normalbaselines\rm
\textfont\truecmr=\eighttruecmr
\scriptfont\truecmr=\seitruecmr
\scriptscriptfont\truecmr=\fivetruecmr
\textfont\truecmsy=\eighttruecmsy
\scriptfont\truecmsy=\seitruecmsy
}\let\nota=\ottopunti}
\newfam\msbfam   %per uso in \TIPITOT
\newfam\truecmr  %per uso in \TIPITOT
\newfam\truecmsy %per uso in \TIPITOT

\overfullrule0pt

%%%%%%%%%%%%%%%%% EQUAZIONI CON NOMI SIMBOLICI
%%%
%%% Per assegnare un nome simbolico ad una equazione basta
%%% scrivere \Eq(...) o, in \eqalignno, \eq(...) o,
%%% nelle appendici, \Eqa(...) o \eqa(...);
%%% dentro le parentesi e al posto di ... si puo' scrivere qualsiasi commento;
%%% per avere i nomi simbolici segnati a sinistra delle formule si deve
%%% dichiarare il documento come bozza, iniziando il testo con
%%% \BOZZA. Sinonimi: \Eq,\EQ,\EQS; \eq,\eqs; \Eqa,\Eqas;\eqa,\eqas.
%%% All' inizio di ogni paragrafo si devono definire il
%%% numero del paragrafo e della prima formula dichiarando
%%% \numsec=... \numfor=...  (brevetto Eckmannn).
%%% Si possono citare formule seguenti; le corrispondenze fra nomi
%%% simbolici e numeri effettivi sono memorizzate nel file \jobname.aux, che
%%% viene letto all'inizio, se gia' presente. E' possibile citare anche
%%% formule che appaiono in altri file, purche' sia presente il
%%% corrispondente file .aux; basta includere all'inizio l'istruzione
%%% \include{nomefile}
%%%
%%%%%%%%%%%%%%%%%%%%%%%%%%%%%%%%%%%%%%%%%%%%%%%%%%%%%%%%%%%%%%%

\global\newcount\numsec
\global\newcount\numfor
\global\newcount\numtheo
\global\advance\numtheo by 1

\def\senondefinito#1{\expandafter\ifx\csname#1\endcsname\relax}

\def\SIA #1,#2,#3 {\senondefinito{#1#2}%
\expandafter\xdef\csname #1#2\endcsname{#3}\else
\write16{???? ma #1,#2 e' gia' stato definito !!!!} \fi}

\def\etichetta(#1){(\veroparagrafo.\veraformula)%
\SIA e,#1,(\veroparagrafo.\veraformula) %
\global\advance\numfor by 1%
\write15{\string\FU (#1){\equ(#1)}}%
\write16{ EQ #1 ==> \equ(#1) }}

\def\letichetta(#1){\veroparagrafo.\verotheo
\SIA e,#1,{\veroparagrafo.\verotheo}
\global\advance\numtheo by 1
\write15{\string\FU (#1){\equ(#1)}}
\write16{ Sta \equ(#1) == #1 }}

\def\tetichetta(#1){\veroparagrafo.\veraformula %%%%copy four lines
\SIA e,#1,{(\veroparagrafo.\veraformula)}
\global\advance\numfor by 1
\write15{\string\FU (#1){\equ(#1)}}
\write16{ tag #1 ==> \equ(#1)}}

\def\FU(#1)#2{\SIA fu,#1,#2 }

\def\etichettaa(#1){(A\veroparagrafo.\veraformula)%
\SIA e,#1,(A\veroparagrafo.\veraformula) %
\global\advance\numfor by 1%
\write15{\string\FU (#1){\equ(#1)}}%
\write16{ EQ #1 ==> \equ(#1) }}

\def\BOZZA{
\def\alato(##1){%
 {\rlap{\kern-\hsize\kern-1.4truecm{$\scriptstyle##1$}}}}%
\def\aolado(##1){%
 {%\vtop to \profonditastruttura
{%\baselineskip
 %\profonditastruttura\vss
 \rlap{\kern-1.4truecm{$\scriptstyle##1$}}}}}
}

\def\alato(#1){}
\def\aolado(#1){}

\def\veroparagrafo{\number\numsec}
\def\veraformula{\number\numfor}
\def\verotheo{\number\numtheo}

\def\Eq(#1){\eqno{\etichetta(#1)\alato(#1)}}
\def\eq(#1){\etichetta(#1)\alato(#1)}
\def\leq(#1){\leqno{\aolado(#1)\etichetta(#1)}}%%%%%this line for \leqno
\def\teq(#1){\tag{\aolado(#1)\tetichetta(#1)\alato(#1)}}%%%%%this line for\tag
\def\Eqa(#1){\eqno{\etichettaa(#1)\alato(#1)}}
\def\eqa(#1){\etichettaa(#1)\alato(#1)}
\def\eqv(#1){\senondefinito{fu#1}$\clubsuit$#1
\write16{#1 non e' (ancora) definito}%
\else\csname fu#1\endcsname\fi}
\def\equ(#1){\senondefinito{e#1}\eqv(#1)\else\csname e#1\endcsname\fi}

%%%% next six lines by paf (no responsibilities taken)
\def\Lemma(#1){\aolado(#1)Lemma \letichetta(#1)}%
\def\Theorem(#1){{\aolado(#1)Theorem \letichetta(#1)}}%
\def\Statement(#1){\aolado(#1){Statement \letichetta(#1)}}%
\def\Proposition(#1){\aolado(#1){Proposition \letichetta(#1)}}%
\def\Corollary(#1){{\aolado(#1)Corollary \letichetta(#1)}}%
\def\Remark(#1){{\noindent\aolado(#1){\bf Remark \letichetta(#1).}}}%
\def\Definition(#1){{\noindent\aolado(#1){\bf Definition
\letichetta(#1)$\!\!$\hskip-1.6truemm}}} 
\def\Example(#1){\aolado(#1) Example \letichetta(#1)$\!\!$\hskip-1.6truemm}

\def\include#1{
\openin13=#1.aux \ifeof13 \relax \else
\input #1.aux \closein13 \fi}

\openin14=\jobname.aux \ifeof14 \relax \else
\input \jobname.aux \closein14 \fi
\openout15=\jobname.aux

%%%%%%%%%%%%%%%%%%% end of Gallavotti's macros %%%%%%%%%%%%%%%%%%%%%%%%%

%%%%%%%%%%%%%%%%%%%% amssym.def included here %%%%%%%%%%%%%%%%%%%%%%%%%%
%%%%%%%%%%%%%%%%%%%%%%%%%%%%%%%%%%%%%%%%%%%%%%%%%%%%%%%%%%%%%%%%%%%%%%%%
\expandafter\ifx\csname amssym.def\endcsname\relax \else\endinput\fi
%
%  Store the catcode of the @ in the csname so that it can be restored later.
\expandafter\edef\csname amssym.def\endcsname{%
       \catcode`\noexpand\@=\the\catcode`\@\space}
%  Set the catcode to 11 for use in private control sequence names.
\catcode`\@=11
%
%  Include all definitions related to the fonts msam, msbm and eufm, so that
%  when this file is used by itself, the results with respect to those fonts
%  are equivalent to what they would have been using AMS-TeX.
%  Most symbols in fonts msam and msbm are defined using \newsymbol;
%  however, a few symbols that replace composites defined in plain must be
%  defined with \mathchardef.

\def\undefine#1{\let#1\undefined}
\def\newsymbol#1#2#3#4#5{\let\next@\relax
 \ifnum#2=\@ne\let\next@\msafam@\else
 \ifnum#2=\tw@\let\next@\msbfam@\fi\fi
 \mathchardef#1="#3\next@#4#5}
\def\mathhexbox@#1#2#3{\relax
 \ifmmode\mathpalette{}{\m@th\mathchar"#1#2#3}%
 \else\leavevmode\hbox{$\m@th\mathchar"#1#2#3$}\fi}
\def\hexnumber@#1{\ifcase#1 0\or 1\or 2\or 3\or 4\or 5\or 6\or 7\or 8\or
 9\or A\or B\or C\or D\or E\or F\fi}

\font\tenmsa=msam10
\font\sevenmsa=msam7
\font\fivemsa=msam5
\newfam\msafam
\textfont\msafam=\tenmsa
\scriptfont\msafam=\sevenmsa
\scriptscriptfont\msafam=\fivemsa
\edef\msafam@{\hexnumber@\msafam}
\mathchardef\dabar@"0\msafam@39
\def\dashrightarrow{\mathrel{\dabar@\dabar@\mathchar"0\msafam@4B}}
\def\dashleftarrow{\mathrel{\mathchar"0\msafam@4C\dabar@\dabar@}}

\def\ulcorner{\delimiter"4\msafam@70\msafam@70 }
\def\urcorner{\delimiter"5\msafam@71\msafam@71 }
\def\llcorner{\delimiter"4\msafam@78\msafam@78 }
\def\lrcorner{\delimiter"5\msafam@79\msafam@79 }
\def\yen{{\mathhexbox@\msafam@55 }}
\def\checkmark{{\mathhexbox@\msafam@58 }}
\def\circledR{{\mathhexbox@\msafam@72 }}
\def\maltese{{\mathhexbox@\msafam@7A }}

\font\tenmsb=msbm10
\font\sevenmsb=msbm7
\font\fivemsb=msbm5
\newfam\msbfam
\textfont\msbfam=\tenmsb
\scriptfont\msbfam=\sevenmsb
\scriptscriptfont\msbfam=\fivemsb
\edef\msbfam@{\hexnumber@\msbfam}
\def\Bbb#1{{\fam\msbfam\relax#1}}
\def\widehat#1{\setbox\z@\hbox{$\m@th#1$}%
 \ifdim\wd\z@>\tw@ em\mathaccent"0\msbfam@5B{#1}%
 \else\mathaccent"0362{#1}\fi}
\def\widetilde#1{\setbox\z@\hbox{$\m@th#1$}%
 \ifdim\wd\z@>\tw@ em\mathaccent"0\msbfam@5D{#1}%
 \else\mathaccent"0365{#1}\fi}
\font\teneufm=eufm10
\font\seveneufm=eufm7
\font\fiveeufm=eufm5
\newfam\eufmfam
\textfont\eufmfam=\teneufm
\scriptfont\eufmfam=\seveneufm
\scriptscriptfont\eufmfam=\fiveeufm

%  Restore the catcode value for @ that was previously saved.
\csname amssym.def\endcsname

%%%%%%%%%%%%%%%%%%%%% end of amssym.def %%%%%%%%%%%%%%%%%%%%%%%%%

%%%%%%%%%%%% local macros %%%%%%%%%%%%%%%%%%%%%%%%%
\def\g{\gamma}
\def\text{\hbox}
\def\a{\alpha}
\def\b{\beta}
\def\si{\sigma}
\def\L{\Lambda}
\def\l{\lambda}
\def\d{\delta}
\def\G{\Gamma}
\def\p{\partial}
\def\z{\zeta}
\def\D{\Delta}
\def\rho{\varrho}
\def\Z{{\Bbb Z}}
\def\t{\theta}
\def\th{\theta}
\def\T{\Theta}
\def\Th{\Theta}
\def\R{{\Bbb R}}
\def\e{\varepsilon}
\def\pg{[\th_i^{\e_i}]}

\newsymbol\square 1003
\newsymbol\blacksquare 1004
\def\qed{$\blacksquare$}

\def\ind#1{ {\bf 1 \hskip-.26em I}_{\lower1ex\hbox{$\scriptstyle #1$}} } 

%Macro to indicate correction

% Once corrections are done, change next two lines to 
%\def\corr#1{}

%%%%%%%%%%%%%%%%%%%% text starts here %%%%%%%%%%%%%%%%%%%%%%%%%%%%%%%%%%

%%%%%%%%%%%%%%%%%%%%%% inizialization %%%%%%%%%%%%%%%%%%%%%%%%%%%%%%%%%%
%%%%%% put % in front of \BOZZA to remove labels on the left %%%%%%%%%%%
\BOZZA
\TIPITOT
\magnification=\magstep1
\baselineskip12pt

\centerline {\dodicibf Liquid-Vapor Phase Transitions}
\centerline {\dodicibf for Systems with Finite Range Interactions}

\bigskip
\bigskip
\centerline{ 
J.L. Lebowitz $^{1,2}$,
\footnote{ }{\eightrm $^1$ Departments of Mathematics and Physics,
Rutgers University, 
Piscataway, NJ  08854-8019, e-mail: lebowitz@math.rutgers.edu} 
\footnote{ }{\eightrm $^2$ Work supported in part by NSF
grant DMR 95--23266 and AFOSR grant 4-26435, and by Nato Grant CRG.9604948} 
\hskip.3cm A. Mazel $^{1,2,3}$,
\footnote{ }{\eightrm $^3$ International Institute of Earthquake
prediction Theory and Theoretical Geophysics, 113556 Mos\-cow, Russia} 
\hskip.3cm E. Presutti$^4$ 
\footnote{ }{\eightrm $^4$ Dipartimento di Matematica, 
Universit\`a di Roma Tor Vergata, Via della Ricerca Scientifica, 
00133 Roma, Italy, e-mail:  presutti@axp.mat.utovrm.it} }

\bigskip\noindent
{\bf Abstract.} {\eightrm We consider particles in $\R^d, d \geq 2$ interacting
via  attractive pair and  repulsive four-body potentials  of the Kac
type.  Perturbing about mean field theory, valid when the interaction range
becomes infinite, we prove rigorously the existence of a liquid-gas phase
transition, when the interaction range is finite but long compared to the
interparticle spacing for a range of temperature.}


\bigskip
\bigskip 
\centerline{{\bf 1. Introduction }}
\bigskip
\numsec= 1
\numfor= 1
\nobreak
An outstanding problem in equilibrium statistical mechanics is to derive
rigorously the existence of a liquid-vapor phase transition in a continuum
particle system.  While such transitions are observed in all types of
macroscopic systems there is at present no proof from first principles of
their existence in particles interacting with any kind of reasonable
potential, say Lennard-Jones or hard core plus attractive square well. Pair
potentials of this type obtained by comparison of experiment with low
density expansions, accurately describe the behavior of both gases and 
liquids in the ranges of temperatures and pressures where boiling and
condensation takes place. In fact computer simulations using classical
statistical mechanics of such systems, containing several hundred to
several hundred thousand particles, convincingly show, via extrapolations
which take into account finite size effects, that systems with these type
of interactions will have true liquid-vapor phase transitions in the
thermodynamic (infinite volume) limit. In this paper we go some way toward
a proof of such a transition, i.e.\ we prove for the first time the
existence of a liquid vapor transition in a continuum particle system with
finite range interactions and no special symmetry.

Historically the first proof of liquid-vapor type phase transitions was
given for lattice systems (which are isomorphic to Ising spins). These
systems can be thought of as idealizations of real fluids in which however
the natural continuous spatial translation invariance symmetry is replaced
by that of the lattice $\Z^d, d \geq 2$. It was Peierls [P] who first gave
a convincing argument (later made fully rigorous by Dobrushin [D1] and
Griffiths [Gr]) of the coexistence of different phases in such systems.
The power of Gibbsian statistical mechanics to produce such rigorous
results was brought home to the general science community by the dramatic
work of Onsager [O] explicitly solving the two
dimensional Ising model (or lattice gas), with nearest neighbor
interactions. Since that time there have been found many other exactly
solvable two dimensional lattice systems [Bax]. At the same time the
development of various
types of inequalities as well as the powerful
Pirogov-Sinai formalism [PS] have resulted in a comprehensive rigorous
theory of phase transitions in lattice systems, at sufficiently low
temperatures.

For continuum systems there is at present proof of a phase transition for
the two component Widom-Rowlinson model [WR] in which the dominant interaction
is a strong repulsion between particles of different species.  The physical
arguments of WR were made rigorous by Ruelle [R1] who was able to
generalize the Peierls argument to prove phase coexistence in this
system. Ruelle's proof strongly exploits the symmetry between the two
components present in the WR model. The same is true, at least to some
extent, of various extensions of this model [LL], [BKL], [Ge]. For
continuum systems without some special symmetry the only proofs of phase
transitions so far are for systems with interactions which decay very
slowly or not at all. Such one dimensional models with many particle
interactions were analyzed and proven to exhibit phase transitions by
Fisher and Felderhof [FF].  More recently Johansson [J] has considered
interactions in one dimension which decay as $r^{-\alpha}$, $\alpha\in
(1,2)$, proving that at low temperatures there is phase transition in the
sense that the pressure is not differentiable.

The mean field or van der Waals type of phase transition was also first derived
rigorously by Kac, Uhlenbeck and Hemmer [KUH] for a class of one
dimensional models, hard rods of radius one with an added pair potential
   $$
\phi_{\g}(q_i,q_j)=- \a {1 \over 2} \g
\exp [-\g |q_i -q_j|],\quad \g, \a >0
   \Eq(1.1)
   $$
 in the limit $\g \to 0$, see also Baker, [Bak].
This was later generalized by Lebowitz and Penrose [LP] to
$d$-dimensional systems with suitable short range interactions and Kac
potentials of the form
   $$
\phi_{\g}(q_i,q_j)= -\a \g^d J(\g |q_i - q_j|)
   \Eq(1.2)
   $$
with 
   $$
\int_{\R^d} J(r) dr =1, \quad J(r) >0
   \Eq(1.3)
   $$
In the thermodynamic limit followed by the limit $\g \to 0$ the Helmholtz
free energy $a$ takes the form, for a fixed temperature $\b^{-1}$,
   $$
\lim_{\g \to 0} a(\rho, \g)=C E\{ a_0(\rho)- {1 \over 2} \a \rho^2
\}
   \Eq(1.4)
   $$
Here $\rho$ is the particle density, $a_0$ is the free energy density of
the reference system, i.e. the system with $\alpha=0$ in \equ(1.2).  $a_0$
is convex in $\rho$ (by general theorems) and $CE\{ f(x)\}$ is the largest
convex lower bound of $f$. For $\a$ large enough the term in the curly
brackets in \equ(1.4) has a double well shape and the $CE$ corresponds to
the Gibbs double tangent construction. This is equivalent to Maxwell's
equal area rule applied to a van der Waals' type equation of state where
it gives the coexistence of liquid and vapor phases [LP]; see also [vK].

In this paper we prove the coexistence of liquid and gas phases for systems
with finite range interactions as small perturbations at finite $\g >0$
from the mean field behavior at $\g =0$.  The same approach was recently
applied to the lattice case for Ising models, [CP], [BZ], [BP], where the
Peierls argument applies directly, because of the spin flip symmetry of the
model.  The absence of symmetries in our case requires instead the whole
machinery of the Pirogov-Sinai theory.  This theory was developed for and
is very powerful in proving phase transitions of lattice systems at low
temperatures, when the pure phases are close to the ground states.  To
apply the theory to our system we will perform a coarse graining in which
we divide the space into large cubes of side $\ell$ and introduce variables
$\rho_x$ which are the particle densities in the cubes labeled by $x$.
After integrating out all the other variables, we will be left with a new
system described by the variables $\rho_x$.  We will then show that the
distribution of the $\rho_x$ is still Gibbsian with a new Hamiltonian and
temperature.  The essential point is that the ratio between the new and old
temperatures scales as $\ell^{-d}$.  By suitably choosing the side of the
cubes, we will then enter into the low temperature regime where the Peierls
and the Pirogov-Sinai (hereafter denoted by P-S) methods apply.  In this
new perturbative scheme, the unperturbed state is described by mean field
(formally $\gamma=0$) and the small parameter of the expansion is the
inverse interaction range $\gamma$, instead of the temperature in the
traditional approach. By choosing a suitable range of values of chemical
potential and temperature we will then be able to put ourselves at the
vapor-liquid phase coexistence.

 To insure stabilization against collapse, which would be induced by a
purely attractive pair potential, the natural choice is to replace the
point particles by hard spheres. Our approach however does not work in such
a case, as we need a cluster expansion for the unperturbed reference system
(i.e. without the Kac interaction) at values of the chemical potential or
density for which it is not proved to hold.  We avoid the problem by
considering point particles and insuring stability by introducing a
positive four body potential of the same range as the negative two body
potential. In this way we avoid having to control strong short range
interactions, something beyond our present day abilities for dense
continuum systems.

\bigskip
\bigskip 
\centerline{{\bf 2. Definitions and Results. }}
\bigskip
\numsec= 2
\numfor= 1
\nobreak
For ease of reference we give below the main definitions followed by the
main result.

\bigskip
\goodbreak
\centerline{{\it Particle Configurations and Phase Space}}

\medskip
\nobreak

We consider a system of identical  particles in $\Bbb R^d$, $d\ge 2$.
 Particles are points  in $\Bbb R^d$,
  {\it particle configurations} are countable,  locally finite collections
of particles. The set of all particle configurations in $\L$, $\L\subset
\Bbb R^d$, is the phase space ${\cal Q}^{(\Lambda)}$ and we simply write
${\cal Q}$ when $\L= \Bbb R^d$. The particle configurations are denoted by
$q$ or by $q^{(\L)}$ when we want to underline that they are in
${\cal Q}^{(\L)}$. We write $q=(q_1,..,q_n)$ when the configuration
consists of $|q|=n$ particles positioned at points $q_1,..,q_n \in \Bbb
R^d$.

\bigskip
\goodbreak
\centerline{{\it Free Measure}}
\medskip
\nobreak

For a bounded measurable subset $\L$ of $\Bbb R^d$ the free measure $dq$
on ${\cal Q}^{(\L)}$ (also called the Liouville measure) is
   $$
\int_{{\cal Q}^{(\L)}}dq\;  f(q ) =
\sum_{n=0}^{\infty}{1\over n!}
\int_{\L}dq_1\cdots \int_{\L}dq_n\,f(q_1,..,q_n)
   \Eq(2.1)
   $$
where  $f$ is any bounded measurable function on  
${\cal Q}^{(\L)}$.

\bigskip
\goodbreak
\centerline{{\it Hamiltonian}}
\medskip
\nobreak

The energy of the configuration $q=(q_i)$ is given for our system by
the formal Hamiltonian
   $$
\eqalign{
H_{\g,\l}(q) 
%&= -\l |q| 
%- \sum_{q_{i_1},q_{i_2} \in q}
%J_{\g}^{(2)}(q_{i_1},q_{i_2})
%+\sum_{q_{i_1},q_{i_2},q_{i_3},q_{i_4} \in q}
%J_{\g}^{(4)}(q_{i_1},q_{i_2},q_{i_3},q_{i_4}) \cr
&= -\l |q| 
- {1\over 2!}\sum_{i_1} \sum_{i_2 \not= i_1}
J_{\g}^{(2)}(q_{i_1},q_{i_2}) \cr
&+{1\over 4!} \sum_{i_1} \sum_{i_2 \not= i_1}
\quad \sum_{i_3 \not= i_1,i_2}
\quad \sum_{i_4 \not= i_1,i_2,i_3} 
J_{\g}^{(4)}(q_{i_1},q_{i_2},q_{i_3},q_{i_4}) \cr}
   \Eq(2.2)
   $$
Here $\l$ is the chemical potential while $J_{\g}^{(2)}(\cdot,\cdot)$
and $J_{\g}^{(4)}(\cdot,\cdot,\cdot,\cdot)$ are respectively two and
four-body potentials. For notational simplicity we choose $\g \in
\{2^{-n}, n \in \Bbb N\}$; $\g$ is a scaling factor in the definition
of $J_{\g}^{(2)}(\cdot,\cdot)$ and
$J_{\g}^{(4)}(\cdot,\cdot,\cdot,\cdot)$ which are chosen of the form
  $$
J_{\g}^{(2)}(q_{i_1},q_{i_2})=\g^{2d}\int dr\, \prod_{j=1}^2
\ind{|r-q_{i_j}|\le \g^{-1}R_d}
   \Eq(2.4)
   $$
   $$
J_{\g}^{(4)}(q_{i_1},q_{i_2},q_{i_3},q_{i_4})=
\g^{4d}\int dr\, \prod_{j=1}^4
\ind{|r-q_{i_j}|\le \g^{-1}R_d}
   \Eq(2.5)
   $$
where $\ind{A}$ is the characteristic function of the set $A$ and $R_d$
is the radius of the ball in $\Bbb R^d$ having a unit volume. Denoting
by $B_{\g}(r)$ a ball of radius $\g^{-1}R_d$ centered at $r \in \Bbb
R^d$ we have
   $$
J_{\g}^{(2)}(q_{i_1},q_{i_2})=\g^{2d} |B_{\g}(q_{i_1}) \cap B_{\g}(q_{i_2})|
   \Eq(2.5.1)
   $$
and
   $$
J_{\g}^{(4)}(q_{i_1},q_{i_2},q_{i_3},q_{i_4})=\g^{4d} |\cap_{j=1}^4
B_{\g}(q_{i_j})|
   \Eq(2.5.2)
   $$
Let
$J^{(2)}=J_1^{(2)}$ and $J^{(4)}=J_1^{(4)}$. Then the scaling properties
of $J_{\g}^{(2)}$ and $J_{\g}^{(4)}$ can be expressed by
   $$
J_{\g}^{(2)}(q_{i_1},q_{i_2})=\g^d J^{(2)}(\g q_{i_1},\g q_{i_2})
   \Eq(2.5.3)
   $$
and
   $$
J_{\g}^{(4)}(q_{i_1},q_{i_2},q_{i_3},q_{i_4})=
\g^{3d} J^{(4)}(\g q_{i_1},\cdots,\g q_{i_4}),
   \Eq(2.5.4)
   $$
which is similar to \equ(1.2). It is also clear that
   $$
\int J_{\g}^{(2)}(0,r)\; dr = \int J_{\g}^{(4)} (0,r_1, r_2, r_3)\; dr_1 dr_2
dr_3 = 1
   \Eq(2.5.5)
   $$
as in \equ(1.3).

As both $J_{\g}^{(2)}$ and $J_{\g}^{(4)}$ are positive we have in \equ(2.2) a
competition between an attractive pair and a repulsive four-body
potential. When the scaling parameter $\g$ is small (but finite), the
model has a large but finite interaction radius, $2R_d\g^{-1}$, and a
small interaction strength between any given two or four
particles. Nevertheless, the total strength of the interaction between a
given particle and all other particles in a configuration $q$ of bounded
nonvanishing density is of order 1. These are characteristic properties of
the Kac potentials which, as noted earlier, usually reproduce the van der
Waals theory [LP] in the scaling limit $\g \to 0$. The specific form
\equ(2.4)-\equ(2.5) of the interaction $J^{(2)}$ and $J^{(4)}$ makes the
analysis simpler, see the discussion in the beginning of Section~3.

\bigskip
\goodbreak
\centerline{{\it Gibbs Measures}}
\medskip
\nobreak

For any two configurations $q=(q_i)$ and $\bar q=(\bar q_j)$ denote by
$q \cup \bar q =(q_i, \bar q_j)$ the configuration containing all particles
from both $q$ and $\bar q$. The {\it conditional energy} of $q^{(\L)}$
given a configuration $\bar q \in {\cal Q}$ is 
   $$ 
H_{\g,\l}(q^{(\L)}|\bar q)=
 H_{\g,\l}(q^{(\L)} \cup \bar q) 
-H_{\g,\l}(\bar q)
   \Eq(2.3)
   $$
This conditional energy consists of two parts: the energy,
$H_{\g,\l}(q^{(\L)})$, of the configuration $q^{(\L)}$ itself and the
{\it interaction energy} 
   $$
U_{\g}(q^{(\L)}|\bar q)=H_{\g,\l}(q^{(\L)}|\bar q)-H_{\g,\l}(q^{(\L)})
   \Eq(2.3.1)
   $$
between configurations $q^{(\L)}$ and $\bar q$. 

Let $\L^c=\Bbb R^d
\setminus \L$ be the complement of $\L$.
Given an inverse temperature $\b$ the {\it Gibbs measure}
$\mu^{(\L)}_{\g,\b,\l}(dq^{(\L)}|\bar q^{(\L^c)})$ in a bounded
measurable set $\L$ with a boundary condition $\bar q^{(\L^c)}$ is the
following probability measure on ${\cal Q}^{(\L)}$ 
        $$
\mu^{(\L)}_{\g,\b,\l}(dq^{(\L)}|\bar q^{(\L^c)})=
{e^{-\b H_{\g,\l}(q^{(\L)}|\bar q^{(\L^c)})} \over 
\Xi_{\g,\b,\l}(\L|\bar q^{(\L^c)})} dq^{(\L)} 
   \Eq(2.11)
        $$
where $\Xi_{\g,\b,\l}(\L|\bar q^{(\L^c)})$ is the {\it partition
function}
        $$
\Xi_{\g,\b,\l}(\L|\bar q^{(\L^c)})= 
\int_{{\cal Q}^{(\L)}}dq^{(\L)} e^{-\b H_{\g,\l}(q^{(\L)}|\bar q^{(\L^c)})}
   \Eq(2.12)
        $$

{\it The infinite
volume Gibbs measures} $\mu_{\g,\b,\l}(dq)$ are probabilities on ${\cal
Q}$ such that for any bounded measurable set $\L$ and
$\mu_{\g,\b,\l}$-almost any boundary condition $\bar q^{(\L^c)} \in
{\cal Q}^{(\L^c)}$ the conditional measure
$\mu_{\g,\b,\l}(dq^{(\L)}|\bar q^{(\L^c)})$ is equal to
$\mu^{(\L)}_{\g,\b,\l}(dq^{(\L)}|\bar q^{(\L^c)})$ given by \equ(2.11). 

We say that a translation invariant Gibbs measure has a particle density
$\rho>0$ if for any bounded set $\L$ the expectation of $|q\cap \L|$ is
equal to $\rho|\L|$.

\bigskip
\goodbreak
\centerline{{\it The Main Result}}
\medskip
\nobreak

Our main purpose is to investigate the phase diagram of the model
\equ(2.2) and to prove that a phase transition of the liquid - vapor type
takes place for some values of the parameters. Let $\b_c=\left( {3 \over
2} \right)^{3 \over 2}$ and $\b_0 > \b_c$ be a number defined via
\equ(2.22) below.

\medskip\noindent{\bf \Theorem (s2.1)} 

\nobreak
{\sl For any $\b \in (\b_c,\b_0)$ there exist functions $\g_0(\b)$ and
$\l(\g, \b)$ such that for $0 <\g < \g_0(\b)$ the model \equ(2.2) has at
least two distinct Gibbs measures $\mu^{\pm}_{\g,\b,\l(\g,\b)}(dq)$.
These measures are translation invariant and ergodic, with an
exponential decay of correlations. They have particle densities
respectively equal to $\rho^\star_{\g,\b,-}>0$ and
$\rho^\star_{\g,\b,{+}} > \rho^\star_{\g,\b,{-}}$. The quantities
$\l(\g,\b)$ and $\rho^\star_{\g,\b,\pm}$  have
limits as $\g \to 0$, denoted by $\l(\b)$ and $\rho_{\b,\pm}$, and there
exist positive constants $c$ and $\a$ such that, $|\lambda(\gamma,\beta)
- \lambda(\beta)| + \sum_{\si=\pm} |\rho^\star_{\gamma,\beta,\si} -
\rho_{\beta,\si}| \le c \gamma^\a$. }

\medskip
\goodbreak
The limit quantities $\l(\b)$ and $\rho_{\b,\pm}$ are given exactly by
mean field type formulas [LP] and our proof of Theorem~\equ(s2.1) is a
perturbation theory constructed around the mean field picture carried out by
using the P-S theory [PS]. The restriction $\b \in (\b_c,\b_0)$ rather
than $\b >\b_c$ allows for a simplified proof; we hope to
present a proof for all $\b >\b_c$ in a future paper. While
technically different and applied to different settings our methods are
nevertheless conceptually close to those of [DZ].

\medskip
\goodbreak
\centerline{{\it Contents of the Remaining Sections}}
\medskip
\nobreak

In Section~3 we give an outline of the proof formulating a number of
statements which are proven in subsequent sections. In Section~4 we prove
Peierls estimates on contours, and in Section~5 we use cluster expansion to
investigate properties of the effective Hamiltonian obtained after the
coarse graining transformation. In Section~6 we study the restricted
ensembles proving the basic property of the P-S scheme, namely that it is
possible to adjust the value of the chemical potential in such a way that
the pressures in the two restricted ensembles are equal. We will conclude
from this the proof of Theorem 2.1. Two auxiliary statements used in the
proof are included in the text as Sections~7~and~8.

\bigskip
\bigskip
\goodbreak
\centerline{{\bf 3. Scheme of Proof. }}
\numsec= 3
\numfor= 1
\bigskip
\nobreak

The starting point of our approach, which as explained in the
Introduction is of crucial importance, consists of coarse
graining: we partition the space into cubes
$C_x^{(\ell_2)}$, $x$ the centers of the cubes, of side $\ell_2:=
\gamma^{-1+\alpha_2}$, $\alpha_2$ a small positive number, (the
subscript 2 foresees the use of other scales that will be introduced
later in the proof). Given a particle configuration $q$, we call
$\rho_x$, the particle densities in each cube and we integrate out all
other variables, i.e. we consider the marginal over the variables
$\rho_x$. The new measure is still Gibbsian and its new, effective
Hamiltonian (written as a function of the $\rho_x$) can be characterized
with a remarkable accuracy by cluster expansion techniques, see Section
5. As mentioned earlier, the main point of this procedure is that the
new effective inverse temperature becomes $\beta \ell_2^d$. We can thus
enter into the very low temperature regime by taking $\gamma$ small
enough, with $\ell_2$ correspondingly large (in Section 5 we will absorb the
temperature into the Hamiltonian). We are then in the right setup for
the P-S theory. An analysis \`a la [LP], which we omit here, would show
that the new effective Hamiltonian converges formally in the limit
$\gamma\to 0$ to the mean field free energy functional ${\cal
F}_{\b,\l}(\rho)$; see \equ(2.13) below. We will begin our analysis by
characterizing the ground states of ${\cal F}_{\b,\l}(\rho)$ and
thereafter use the P-S theory to investigate the perturbations at $\g>0$.

\bigskip
\goodbreak

\centerline{\it Mean Field Free Energy Functional and Ground
states}
\medskip

\nobreak

The mean field Gibbs free energy functional 
${\cal F}_{\b,\l}(\rho)$, $\rho=\rho(r)$ denoting  a
 non-negative bounded measurable function in $\Bbb R^d$, is
        $$
\eqalign{
{\cal F}_{\b,\l}(\rho)&= 
\int  dr  {\rho(r)\over \b}(\log\rho(r) -1) 
- \int  dr \l\rho(r) \cr
&-{1\over 2!} \int dr_1 dr_2 J^{(2)}(r_1,r_2) \rho(r_1) \rho(r_2) \cr
&+{1\over 4!} \int  dr_1 \cdots dr_4 J^{(4)}(r_1, \dots, r_4)
\rho(r_1) \cdots \rho(r_4) }
        \Eq(2.13)
        $$
The first integral is
 the  entropy contribution to the free energy (more precisely the
product of the temperature times 
the entropy of the ideal gas with the changed sign).  The other three terms
arise from the corresponding interaction terms in \equ(2.2). More
details concerning the relation between \equ(2.13) and \equ(2.2) can be
found in Section~4.

The {\it mean field ground states} are, by definition,
  the minimizers
of ${\cal F}_{\b,\l}(\rho)$. To find them we set
       $$
{\cal R}(r,\rho)= \int_{|r-r_1|\le  R_d} dr_1 \rho(r_1)
               \Eq(2.14)
        $$
and
        $$
{\cal I}(r,\rho)=  \int_{|r-r_1|\le  R_d} dr_1
{\rho(r_1)\over \b} \big( \log\rho(r_1) -1 \big)
               \Eq(2.14.1)
        $$
${\cal R}(r,\rho)$ and ${\cal I}(r,\rho)$ are
respectively the mean density and the mean entropy of $\rho$ in the ball
$B(r)$. With this notation we can rewrite \equ(2.13) 
as
        $$
{\cal F}_{\b,\l}(\rho)= \int  dr\left( {\cal I}(r,\rho)
- \l {\cal R}(r,\rho) 
- {1\over 2!}{\cal R}(r,\rho)^2
+ {1\over 4!}{\cal R}(r,\rho)^4 \right),
        \Eq(2.15)
        $$
which is true only because of the special form, \equ(2.4) and \equ(2.5),
of $J^{(2)}$ and $J^{(4)}$. In fact \equ(2.15) is the main reason for
choosing $J^{(2)}$ and $J^{(4)}$ of this form. By convexity
        $$
{\cal I}(r,\rho) \ge {{\cal R}(r,\rho) \over \b}
\Big( \log {\cal R}(r,\rho)-1 \Big) 
        \Eq(2.16)
        $$
and equality is achieved only if ${\cal R}(r,\rho)=\rho(r)$. 

It follows from \equ(2.15) that if $s^\star \ge 0$ is a minimizer of
the function  
   $$
F_{\b,\l}(s) = {s\over \b} \Big( \log s -1 \Big) - \l s
-{s^2\over 2!}+{s^4\over 4!}
   \Eq(2.17)
   $$
then $\rho(r)={\cal R}(r,\rho) \equiv s^\star$ is a minimizer of ${\cal
F}_{\b,\l}(\rho)$. The second derivative
   $$
F''_{\b,\l}(s)= {1 \over \b s} -1 + {s^2 \over 2} 
   \Eq(2.17.1)
   $$
of $F_{\b,\l}(s)$ is positive if $\b < \b_c=\left( {3 \over 2}
\right)^{3 \over 2}$. Hence for $\b < \b_c$ and any $\l$ the
function $F_{\b,\l}(s)$ is convex and has a unique
 minimizer which is the root of the equation
    $$
-s+{1\over 3!} s^3 +{1\over \b} \log s - \l = 0
   \Eq(2.20)
   $$
For $\b >\b_c$ this equation has three positive roots
with two of them, $s=\rho_{\b,\l,-}$ and $s=\rho_{\b,\l,+}$, being local
minimizers of $F_{\b,\l}(s)$. Furthermore, there exists a unique
$\l=\l(\b)$ for which both local minima are the global ones and the
function $F_{\b,\l}(s)$ has a ``double well'' shape with the same
``depth'' of the wells. We set $\rho_{\b,\pm} = \rho_{\b,\l(\b),\pm}$.
Clearly for $\b > \b_c$ and $\l={\l(\b)}$ the densities $\rho(r) \equiv
\rho_{\b,-}$ and $\rho(r) \equiv \rho_{\b,+}$ are distinct mean field
gound states. For later purposes we remark that
   $$
 -1 <\l(\b) <0
   \Eq(2.20.1)
   $$
which can be checked by direct calculation.

\bigskip
\goodbreak
\centerline{\it Mean Field Equations and Contraction Property}
\medskip
\nobreak

As the ground states are minimizers of ${\cal F}_{\b,\l}(\rho)$, they
satisfy the mean field equation $\delta {\cal F}_{\b,\l}(\rho)/\delta
\rho(r)=0$. By an explicit calculation we then find that they are fixed
points of the transformation
 $$
\rho(r) \to \Phi\big(\rho(\cdot),r\big)
   \Eq(2.20.2)
   $$
where
   $$
\Phi\big(\rho(\cdot),r\big):= \exp\Big\{
\lambda + \int dr_1 J^{(2)}(r,r_1)\rho(r_1)
- {1 \over 3!} \int dr_1..\int dr_3 J^{(4)}(r,..,r_3) \rho(r_1)
\rho(r_2) \rho(r_3) 
\Big\}
   \Eq(2.20.3)
   $$
Setting $\rho^\star=\rho_{\b,-}$ or $\rho^\star=\rho_{\b,+}$ the
derivative $\psi(r)= \delta \Phi\big(\rho(\cdot),r \big)/\delta \rho(r)$
computed at $\rho(r)\equiv \rho^\star$ is equal to 
    $$
\psi(r)= \b \rho^\star J^{(2)}(0,r)[1-\left({\rho^\star} \right)^2/2]
   \Eq(2.20.4)
    $$
We define $\b_0$ in Theorem \equ(s2.1)
so that for all $\b\in (\b_c,\b_0)$
    $$
\int dr |\psi(r)|=
\b \rho^\star \big |1-{\rho^\star}^2/2 \big | <1
   \Eq(2.22)
    $$
If \equ(2.22) holds, the transformation $\Phi$ is a contraction (in a
neighborhood 
of the ground states and using sup norms).  

The existence of $\b_0>\b_c$ follows by noting that since $\b
\rho^\star<1$ (by \equ(2.17.1) and because
$F''_{\b,\lambda}(\rho^\star)>0$) the only condition to check for
\equ(2.22) to hold is $\b \rho^\star[1-{\rho^\star}^2/2]>-1$. By
\equ(2.17.1) it can be rewritten as
   $$
F''_{\b,\lambda}(\rho^\star)< { 2 \over \b 
\rho^\star}
   \Eq(2.20.5)
   $$
and the result then follows from the fact that 
$ F''_{\b,\lambda}(\rho^\star)\to 0$ as $\b \to \b_c$ while both
$\rho_{\b,\pm}$ remain bounded.

We will see in Section~5 that for $\b\in (\b_c,\b_0)$ and $\g>0$ small
enough, the Dobrushin uniqueness condition is satisfied by the effective
Hamiltonian of the system restricted to the ground state ensemble
(defined later in this section). This uniqueness condition appears to be
the $\g>0$ analogue of the contraction property \equ(2.22). We will in
the sequel restrict to $\b\in (\b_c,\b_0)$ which guarantees the
following properties.

For $\b_c <\b <\b_0$ there exists a positive number $\z(\b)$
such that
    $$
\max_{\si_1=\pm 1, \si_2=\pm 1}  \b (\rho_{\b,\si_1}+2\z(\b)) 
\left| 1- {(\rho_{\b,\si_1} +\si_2 2\z(\b))^2 \over 2} \right| = a(\b) <1 
   \Eq(2.21)
   $$
Consequently there exist a positive number $\d(\b)$ such that for any
$\l \in (\l(\b)- \d(\b), \l(\b) + \d(\b))$ one has
    $$
\max_{\si_1=\pm 1, \si_2=\pm 1}  \b (\rho_{\b,\l,\si_1}+\z(\b)) 
\left| 1- {(\rho_{\b,\l,\si_1} +\si_2 2\z(\b))^2 \over 2} \right| \le a(\b) 
   \Eq(2.21.1)
   $$

\medskip
We expect that for $\g$ small enough the particle configurations will
have densities close to the mean field ground states. To investigate the
issue we need a preliminary spatial scale for computing particle 
densities and a notion of closeness between densities. We will then have
a picture of the particle configurations in terms of spatial regions
where there is agreement or disagreement with the ground states.

\bigskip
\goodbreak
\centerline{{\it The Partitions 
${\cal D}^{(\ell)}$, the Cubes $C^{(\ell)}_x$ and the Densities
$\rho^{(\ell)}_x$}}

\medskip
\nobreak
Let ${\cal D}^{(\ell)}$, $\ell \in \{2^n,\, n\in \Bbb N\}$, be
decreasing partitions of $\Bbb R^d$ into cubes $C^{(\ell)}$ of side
$\ell$, i.e. ${\cal D}^{(\ell)}$ is coarser than ${\cal D}^{(\ell')}$ if
$\ell > \ell'$. For any $r\in \Bbb R^d$ we denote by $C^{(\ell)}(r)$ the
cube of ${\cal D}^{(\ell)}$ that contains $r$. We suppose that the
centers of the cubes $C^{(1)}$ are in $\Bbb Z^d$. Consequently the
centers of cubes $C^{(\ell)}$ are in ${\Bbb Z}^d_{\ell}= \ell \Bbb Z^d +
\left( {\ell-1 \over 2}, \ldots , {\ell-1 \over 2} \right)$. For $x \in
\Bbb Z^d_{\ell}$ we denote $C^{(\ell)}_x$ the cube of ${\cal
D}^{(\ell)}$ centered at $x$.

Given a region $\Lambda$, $[\Lambda]^{(\ell)}$ is
 the maximal ${\cal
D}^{(\ell)}$ measurable subset of $\Lambda$, i.e. the union of all
$C^{(\ell)}$ contained in $\L$. We also identify $[\Lambda]^{(\ell)}$
with $\{ x \in {\Bbb Z}^d_{\ell}|\; x \in [\Lambda]^{(\ell)} \}$. The
${\Bbb R}^d$ volume of $[\Lambda]^{(\ell)}$ is denoted by
$|[\Lambda]^{(\ell)}|$ while its $\Bbb Z^d_{\ell}$ volume, i.e. the
number of lattice points, is denoted by $||[\Lambda]^{(\ell)}||$.

The density of a configuration $q$ in a box $C^{(\ell)}_x$ is
   $$
\rho^{(\ell)}_x(q)=\ell^{-d} \left| q \cap C^{(\ell)}_x \right|
   \Eq (3.2.1)
   $$
Accordingly $\rho^{(\ell)}(q)=(\rho^{(\ell)}_x(q))$ is called a (${\cal
D}^{(\ell)}$ measurable) {\it density configuration} corresponding to a
particle configuration $q$. We denote
$\rho^{(\ell)}(\L)=(\rho^{(\ell)}_x(\L))$ a density configuration in
$[\Lambda]^{(\ell)}$ not related to any particle configuration.

The notation $[q]^{(\ell)}$ is used for the particle configuration
obtained from $q=(q_i)$ by shifting all $q_i$ to the centers of the
corresponding boxes $C^{(\ell)}(q_i)$. For a configuration
$q^{(\Lambda)}$ we set
   $$
H_{\g,\l}(\rho^{(\ell)}(q^{(\Lambda)}))=H_{\g,\l}([q^{(\Lambda)}]^{(\ell)})
   \Eq(6.14)
   $$
and
   $$
H_{\g,\l}(\rho^{(\ell)}(q^{(\Lambda)})| \bar q^{(\Lambda^c)}
)=H_{\g,\l}([q^{(\Lambda)}]^{(\ell)}| \bar q^{(\Lambda^c)}),
   \Eq(6.15)
   $$
where $\Lambda$ is a region and $\bar q^{(\Lambda^c)}$ is a boundary
condition.

\bigskip
\goodbreak
\centerline{{\it Spatial Scales, $\eta$ Functions and Ground States
Configurations}}
\medskip
\nobreak

Several scales are of special interest for us
   $$
\ell_1=\ell_{1,\g} = \g^{-1+\a_1},\qquad
\ell_2=\ell_{2,\g} = \g^{-1+\a_2},\qquad
\ell_3=\ell_{3,\g} = \g^{-1-\a_3},\qquad
     \Eq (3.2)
    $$
where the numbers $0< \a_i <1$ are rational, $\a_1=1/2$ and $\a_2+\a_3 \ll
(2d)^{-1}$.

The scale $\ell_2$ is our coarse graining scale. Agreement or
disagreement with the ground state is indicated by $\eta$-functions,
$\eta= \eta(q)= (\eta_x(q))$, $x \in \Bbb Z^d_{\ell_2}$:
   $$
\eta_x(q)=\cases{
-1, & for $|\rho^{(\ell_2)}_x(q) - \rho_{\b,-}| \le \z$ \cr
+1, & for $|\rho^{(\ell_2)}_x(q) - \rho_{\b,+}| \le \z$ \cr
0,  & otherwise \cr}
   \Eq (3.2.2)
   $$
$\z=\z(\b)$ being taken from \equ(2.21). Particle configurations $q$ or
density configurations $\rho^{(\ell)}(q)$ are called compatible with
$\eta$ if $\eta=\eta(\rho^{(\ell)}(q))=\eta(q)$.

We  say that a particle configuration $q$ belongs to the {\it ground
state ensemble} of the {\it $\pm$ phase} or equivalently {\it
liquid/vapor phase} if $\eta(q) \equiv \pm 1$. For brevity we will
sometimes simply say that $q$ is a $\pm$ ground state configuration.

For a scale $\ell$ we define the corresponding {\it standard ground
state configurations} $q^{(\ell)}_-$ and $q^{(\ell)}_+$ as those which
have $\rho^{(\ell)}(q^{(\ell)}_-) \equiv \rho_{\b,-}$ and
$\rho^{(\ell)}(q^{(\ell)}_+) \equiv \rho_{\b,+}$ with all the particles
placed at the centers of the corresponding boxes $C^{(\ell)}$. When
$\ell=\ell_2$ we will drop the superscript $(\ell)$.

The other scales, $\ell_3$ and $\ell_1$, are used to construct contours
(see below) and to do some approximate calculations (see Section~4)
respectively.

\medskip
Our proof of the phase transition or coexistence of phases is based on
the following qualitative picture describing typical particle
configurations in the two pure phases. In the gas phase a typical
configuration $q$ coincides in most of $\R^d$ with some typical
configuration in the gas ground state ensemble. Inside this ``sea" of
the gas ground state there are rare ``islands" occupied by the liquid
ground state. These two types of ground states are separated from each
other by regions in $\R^d$ which are called Peierls contours. The
``excess'' free energy of $q$ with respect to the free energy of the gas
ground state occupying all of $\Bbb R^d$ is concentrated in the vicinity
of these contours and is proportional to their volume. A similar,
inverse, picture describes typical configurations of the liquid phase;
the liquid ground state now forms a ``sea'' with rare ``islands'' of the
gas ground state. Thus the typical configurations of the gas and liquid
phases are distinct with the density of particles being an order
parameter distinguishing them. A rigorous verification of the above
picture is fairly straightforward for simple lattice gases; Ising
systems, in which the gas and liquid ground states correspond
respectively to ``all sites empty'' and ``all sites occupied''. It
requires however considerable work for continuum systems and we start
with precise definitions.

\bigskip
\goodbreak
\centerline{{\it Correct and Incorrect Sets, Contours}}
\medskip
\nobreak

A family of cubes from the partition ${\cal D}^{(\ell)}$ is called {\it
$\star$-connected} if its closure is a connected subset of $\Bbb R^d$;
we consider from now on $d \ge 2$. We then say that $r$ is a {\it
($+$)-correct} point of a configuration $q$ if $\eta(q)=+1$ in
$[C^{(\ell_3)}(r)]^{(\ell_2)}$ and in each of the neighboring ${\cal
D}^{(\ell_3)}$ measurable cubes $\star$-connected to $C^{(\ell_3)}(r)$.
Similarly one defines {\it ($-$)-correct} points. Finally, if $r$ is not
a correct point of $q$ it is an {\it incorrect} one. The connected
components of the incorrect points of $q$ form the supports of the
contours. The pair consisting of the support of the contour and the
restriction of $\eta(q)$ to this support is called a contour of $q$ and
is denoted $\G(q)$. Observe that any $q$ uniquely defines its contours
$\G_i(q)$.

Axiomatically a {\it contour} $\G=({\rm Supp}(\G), \eta^{\G})$ is
defined as a pair which consists of a bounded $\star$-connected ${\cal
D}^{(\ell_3)}$ measurable set ${\rm Supp}(\G)$ called the spatial {\it
support} of $\G$ and a spin valued function $\eta^{\G}$ on ${\rm
Supp}(\G)$, with the condition that there exists at least one
configuration $q$ that gives rise to $\Gamma$.

To motivate the next definitions and to outline further constructions
let us consider first an oversimplified problem.

\vfil\goodbreak\medskip
\def\boxit#1#2#3{\vbox{\hrule height#1pt
\hbox{\vrule width#1pt\kern#2pt
\vbox{\kern#2pt\hbox{#3}\kern#2pt}\kern#2pt\vrule width#1pt}\hrule height#1pt}}
\vbox{
\centerline{\boxit{2}{10}{\boxit{1}{30}{\boxit{1}{10}{\boxit{2}{40}{\phantom
{test}}}}}}
\vskip-120pt\nobreak
\centerline{A \hskip33pt Supp($\Gamma$) \hskip50pt  D \hskip50pt
Supp($\Gamma$) \hskip33pt A}
\vskip15pt\nobreak
\centerline{ B \hskip31pt C \hskip103pt C \hskip31pt B }
\vskip100pt\nobreak
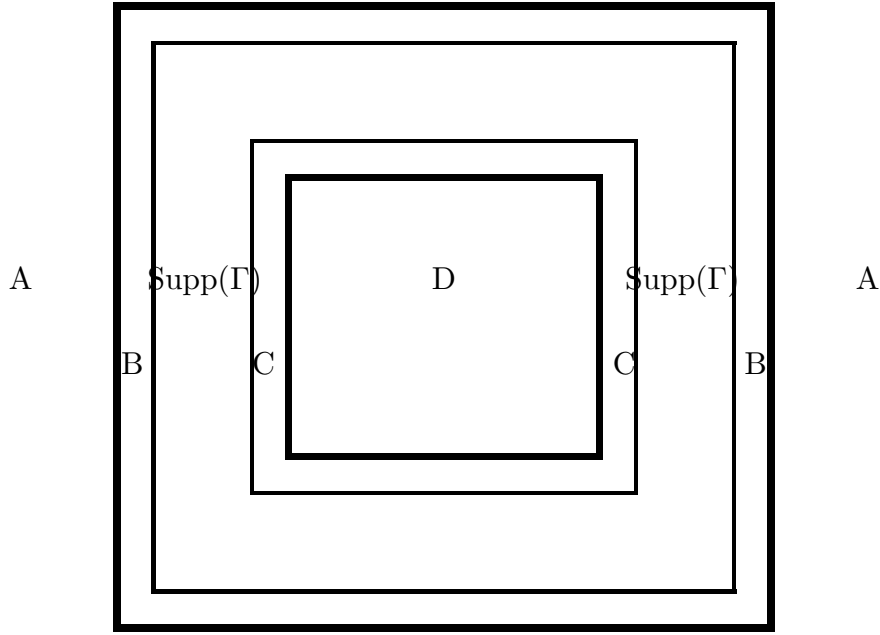
\centerline{Fig.1}}

\medskip\goodbreak
We choose boundary conditions $\bar q^{(\L^c)}$ in the $+$ phase and we
impose that in the region $A$ of Fig.1 the configuration is in the $+$
phase, in $D$ it is in the $-$ phase while in ${\rm Supp}(\Gamma)$ it
agrees with $\eta^{\Gamma}$.

The simplified contour ensemble gives rise to the simplified partition
function $Z$ (we are using here the notation that will be discontinued
in the sequel)
    $$
Z = \sum_{ \G}
\int_{{\cal Q}^{(\L)}} dq^{(\L)} \; \ind{\eta(q^{(A)})\equiv 1} \;\;
\ind{\eta(q^{(D)})\equiv -1} \;\;\ind{\eta(q^{({\rm Supp}(\Gamma))})\equiv 
\eta^\Gamma} \;\;
e^{-\b H_{\g,\l}(q^{(\L)}|\bar q^{(\L^c)})}
   \Eq(3.6e1)
   $$
Here the external sum is taken over all contours $\G$ of the type
described by Fig.1 and the goal now is to rewrite $Z$ as an integral
over the $+$ ground state ensemble, i.e. as the integral over $q^{(\L)}$
with $\{\eta(q^{(\L)})\equiv 1\}$. For $B$ and $C$ strips of
width $\g^{-1}$ in ${\rm Supp}(\Gamma)$, as in Fig. 1,  we write
    $$
Z = \sum_{ \G }
\int_{{\cal Q}^{(\L)}} dq^{(\L)} \; \ind{\eta(q^{(\L)})\equiv 1} \;\;
e^{-\b H_{\g,\l}(q^{(\L)}|\bar q^{(\L^c)})}W(\Gamma|q^{(B)})
   \Eq(3.6e2)
   $$
where  the statistical weight $W(\Gamma|q^{(B)})$ is
   $$
W\big(\G |  q^{(B)}   \big)  =
\phantom{1234567890123456789012345678901234567890123456789012345}
   \Eq(3.7e1)
   $$
\nobreak
   $$
 {{\displaystyle \int_{{\cal Q}^{({\rm Supp}(\G) \setminus
B)}} }}
dq\; \ind{\eta(q)=\eta^{\G}} \;
e^{-\b H_{\g,\l}(q| q^{(B)})}
{\displaystyle
\int_{{\cal Q}^{(D)}}dq^{(D)}
e^{-\b H_{\g,\l}(q^{(D)}| q^{(C)})}\ind{\eta(q^{(D)}) \equiv -1}}\over
{\displaystyle \int_{{\cal Q}^{({\rm Supp}(\G) \setminus B)}}}
dq\; \ind{\eta(q) \equiv 1} \;
e^{-\b H_{\g,\l}(q| q^{(B)})}
{\displaystyle
\int_{{\cal Q}^{(D)}}dq^{(D)}
e^{-\b H_{\g,\l}(q^{(D)}| q^{(C)})}\ind{\eta(q^{(D)}) \equiv 1}}
   $$ 
In \equ(3.6e2) the configuration $q^{(\L)}$ is not related to
$\eta^\Gamma$ and the contours $\G$ appear only through their
statistical weights as extra variables in the partition functions. This
is the contour model in the present over-simplified context. The
advantage of the contour model is to work in a ground state ensemble,
the price is the extra variables $\G$. The game, in the general case
with many contours, is to get good estimates on their statistical
weights \equ(3.7e1) in order to control their proliferation.

Let us examine this issue still in the context of the oversimplified
example. We should think of the integral in \equ(3.7e1) over the region
${\rm Supp}(\Gamma) \setminus B$ as a ``surface term'' and of the
integral over $D$ as a ``volume term''. The second one looks therefore
more likely to take large values and we consider it first, as it appears in 
the numerator in \equ(3.7e1) denoted by $N$. The
integral is over all configurations in the region $D$ that are in the
ground state ensemble of the $-$ phase. The boundary conditions for this
partition function are fixed in the region $C$ and they are also
configurations of the $-$ phase. In fact the $-$ phase extends over the
whole $\ell_3$ cubes $\star$-connected to $D$, recall the definition of
contours.

The intuition (which is made rigorous in Section 6) is that since $C$ is
well inside the $-$ phase, then the configurations in a neighborhood of
$C$ are actually not only in the $-$ ground state ensemble but they are 
very close to the standard ground state configuration. We thus write
(recall that $N$ denotes the numerator in \equ(3.7e1))
   $$
N= T_1T_2T_3
   \Eq(3.8e1)
   $$
where $T_1$, $T_2$ and $T_3$ are defined as follows.

Splitting $1/2$-$1/2$ the interaction between ${\rm Supp}(\G)$ and $D$
and recalling the intuition that the configurations involved in such
interactions are very close to the $-$ ground state configuration, we
write
   $$
T_1 =
 \int_{{\cal Q}^{({\rm Supp}(\G) \setminus
B)}}dq\; \ind{\eta(q)=\eta^{\G}} \;
e^{-\b H_{\g,\l}(q| q^{(B)},q_-^{(D)})+\b U_{\g}(q_-^{(C)}|
q_-^{(D)})/2}
   \Eq(3.8e2)
   $$

Denoting by $f_{\l,-}$ the thermodynamic limit of the pressure associated with
the partition function restricted to the $-$ ground state ensemble we
set
   $$
T_2 = e^{\b f_-|D|}
   \Eq(3.8e3)
   $$

The third term $T_3$ is implicitly defined so that \equ(3.8e1) holds. $T_3$
takes into account the errors made by replacing in \equ(3.8e2) the
interaction 
   $$
{1\over 2} U_{\g}(q^{(C)}|q^{(D)}) \to
U_{\g}(q^{(C)}|q_-^{(D)}) - {1\over 2} U_{\g}(q_-^{(C)}|q_-^{(D)})
   \Eq(3.8e1.1)
   $$
and the error which comes from $T_2$ where the partition function is
replaced by its thermodynamic limit in the sense of \equ(3.8e3).

  The error term $T_3$ appears to be bounded by
   $$
|\log T_3| \le c \g^{1/4} \;|{\rm surface \; of }\; D|
   \Eq(3.8e4)
   $$
where $c$ is a constant. The above inequality is proven in Section 6 in
the general case.

After an analogous decomposition in the denominator, we see that the
dangerous volume terms simplify between numerator and denominator if we
are able to choose the chemical potential in such a way that
$f_{\l,-}=f_{\l,+}$. Once again, the possibility to solve this equation
relies on a rather explicit representation of the pressures,
$f_{\l,\pm}$, which one is able to derive.

The ratio of the terms $T_1$ (from numerator and denumerator) is bounded
by
   $$
\exp \left( -c 
\ell_2^d \ell_3^{-d} \z^2 |{\rm Supp}(\G)| \right)
   \Eq(3.8e5)
   $$
This is the famous Peierls estimate which is proven in the general case
in Section 4. Its proof is relatively simpler than the proof of
\equ(3.8e4) and relies directly on properties of the mean free field
energy functional. This step is similar to the analogous one in Ising
models [BP].

Combining \equ(3.8e4) and \equ(3.8e5) and recalling that $d \a_1 + d
\a_2$ is assumed to be much less than $1/4$ we conclude that the whole
statistical weight $W(\G | q^{(B)})$ is bounded as in \equ(3.8e5).

Let us now go from the oversimplified example to the general case.

\bigskip
\centerline{{\it  More About Contours}}
\medskip
\nobreak

Given a (large enough) region $V$ and $\ell>0$ we call 
   $$
\p^{(\ell)} V = \{ r \in V^c: 
\text {dist}(r,V) \le \ell\};\quad
\d^{(\ell)} V = \{ r \in V : 
\text {dist}(r,V^c) \le \ell\}
     \Eq (3.3)
    $$
and we set $ \p V = \p^{(\g^{-1})} V$ and $ \d V = \d^{(\g^{-1})} V$.

Denoting by ${\rm Ext}(\G)$ and by ${\rm Int}_m(\G)$ respectively the
unbounded and the bounded connected components of $({\rm Supp}(\G))^c$
we observe that $\eta^{\G}$ takes the same values $\si(\G)$ and
$\si_m(\G)$ on each of the sets $\p^{(\ell_3)} {\rm Ext}(\G)$ and
$\p^{(\ell_3)} {\rm Int}_m(\G)$ correspondingly. The regions ${\rm
Ext}(\G)$ and ${\rm Int}(\G)=\cup_m {\rm Int}_m(\G)$ are respectively
called the {\it exterior} and the {\it interior} of $\G$, in the
previous example they are $A$ and $D$.

For a contour $\G$
denote
   $$
\d_m(\G)=\p {\rm Int}_m(\G),
   \Eq(3.5.0)
   $$
(the set $C$ in the example of Fig. 1)
   $$
\d^=(\G) =\p {\rm Ext}(\G)\bigcup \left(
\bigcup_{m:\; \si_m(\G)= \si(\G)} \d_m(\G) \right),
   \Eq(3.5)
   $$
(the set $B$ in Fig. 1, the last term being absent in
Fig. 1)
   $$
\d^{\neq}(\G) =
\bigcup_{m:\; \si_m(\G)\not= \si(\G)} \d_m(\G),
   \Eq(3.5.1)
   $$
(this is the set $C$ in Fig. 1) and
   $$
\d(\G)=\d^=(\G) \cup \d^{\neq}(\G)
=\d {\rm Supp}(\G)
   \Eq(3.5.2)
   $$
(this is $B \cup C$ in Fig. 1)

Two contours are called {\it disjoint} if their supports are not
$\star$-connected. We say that a contour {\it belongs} to a region if
its support is a subset of that region. In any collection of pairwise
disjoint contours there is a uniquely defined subcollection of {\it
external contours}, i.e. contours which do not belong to the interior of
any other contour in the collection. 

A contour $\G$ {\it surrounds a point} $r$ if $r \in {\rm Supp}(\G) \cup
{\rm Int}(\G)$. A contour $\G_3(q)$ {\it separates} contours $\G_1(q)$
and $\G_2(q)$ if ${\rm Supp}(\G_1(q)) \in {\rm Int}(\G_3(q))$ while
${\rm Supp}(\G_2(q)) \in {\rm Ext}(\G_3(q))$.

Consider an arbitrary configuration $q$ which differs from a ground
state configuration of a given phase $\si$ only inside a bounded region
$\L$. Then all contours of $q$ belong to $\L$ and for any external
contour $\G^{\rm ext}(q)$ one has $\si(\G^{\rm ext})=\si$. Note that
non-external contours $\G(q)$ may have $\si(\G(q))=-\si$. Moreover, the
contours of $q$ satisfy the following {\it matching condition}. If ${\rm
Supp}(\G_1(q)) \in {\rm Int}_m(\G_2(q))$ and there is no $\G_3(q)$
separating $\G_1(q)$ and $\G_2(q)$ then $\si(\G_1(q))=\si_m(\G_2(q))$.
This matching condition is highly non local and represents the main
difficulty in understanding contour statistics.

The idea of the P-S theory is to remove matching condition and to
construct an equivalent contour model with no matching rules present.
Generally this can be done by modification of the statistical weights of
contours and shifting the difficulty to the estimate of these modified
statistical weights. 

\medskip
Generally the study of the distribution of contours is important because
to prove the existence of distinct translation invariant $+$ and $-$
phases it is enough to show that the probability of the event that an
external contour surrounds the origin is sufficiently small. In a
translation invariant measure this implies that only finitely many
contours surround any given point $r$. Then one of the ground states
occupies the infinite region $\cap_i {\rm Ext}(\G^{\rm ext}_i(q))$ which
is exactly the ``sea'' discussed earlier. Furthermore, the probability
of the event that $\eta_0(q)=1$ (here $0$ denotes the origin) can be
taken as the order parameter distinguishing two phases. This probability
is less than $1/2$ for the $-$ phase and it is larger than$1/2$ for the
$+$ phase as follows from a bound on the probability of a contour of the
form \equ(3.8e5) (which will be shown to be true for all contours). As
we will see even in the general case when there is more than just one
single contour in the contour model the bound \equ(3.8e5) implies that
the contours are very rare and one can control them with an analysis not
too distant from the oversimplified example. 

Let us recall now that the bound \equ(3.8e5) on the statistical weight
of a contour was obtained after adjusting the chemical potential so that
the pressures in the two ground states ensembles are equal. In the
general case this is not a simple task because there 
are contours inside contours, i.e. inside the regions of type $D$ of the
example in Fig. 1 there are other contours and so on. One needs to
treat contours recursively as in the one-contour example to obtain at
the end of the recursion the contour model with the properly modified
statistical weights of contours. Only after that may one try to equalize
pressures in these contour models. The problem is that even if the bound
\equ(3.8e5) holds the pressures in the contour models will
depend on the contours and we are then in a sort of loop: to have the
good bound \equ(3.8e5) on the contours we need to make the
pressures equal, but to control the pressures we need a good bound on the
contours.

As explained by the P-S theory it is possible to survive such an
impasse, we will do it by following the Zahradnik approach [Z]. We will
define a new statistical weight with a cut off which by its definition cannot
exceed the value \equ(3.8e5) (with a suitable constant $c$). In this
context it will be easy to iterate the procedure of the example with one
contour to derive a contour model representation of the true partition
function. Such {\it auxiliary} partition functions (one for each phase)
give rise to the corresponding pressures and we will adjust the chemical
potential so that these two pressures (one for each phase) are equal. It
will then turn out (see [Z]) that for this particular value of the
chemical potential the contours satisfy the bound \equ(3.8e5) without
the need of the cutoff so that for this special value of the chemical
potential the auxiliary and true partition functions are equal.

The next subsection contains exact definitions and statements which are
necessary for application of general P-S theorem (see [PS], [Z]).

\bigskip
\goodbreak
\centerline{{\it Auxiliary Partition Functions}}
\medskip
\nobreak
 
Given a phase $\si$ and boundary condition $\bar q^{(\L^c)}$ belonging
to the ground state ensemble of this phase we define the {\it auxiliary
partition functions} and the {\it truncated statistical weight} as
   $$
Z^A_{\g,\b,\l}(\L|\bar q^{(\L^c)}) =
\int_{{\cal Q}^{(\L)}} dq \; \ind{\eta(q)\equiv \si} \;\;
e^{-\b H_{\g,\l}(q^{(\L)}|\bar q^{(\L^c)})}
\sum_{ \{\G_i\}^{\si} \in \L} 
\prod_i W^T({\G_i}| q^{(\d^=(\G_i))})
   \Eq(3.8)
   $$
and
   $$
W^T({\G}| q^{(\d^=(\G))})=\min \left(
W^A( {\G} | q^{(\d^=(\G))}), 
e^{-{c
\over 3} \ell_2^d \ell_3^{-d}
\z^2|{\rm Supp}(\G)|} \right)
    \Eq(3.9)
   $$
where $c = c_{\equ(3.8e5)}$
   $$
W^A(\G| \bar q^{(\d^=(\G))})
\phantom{1234567890123456789012345678901234567890123456789012345}
   \Eq(3.10)
   $$
\nobreak
   $$
={{\displaystyle \int_{{\cal Q}^{({\rm Supp}(\G) \setminus
\d^=(\G))}} }
dq\; \ind{\eta(q)=\eta^{\G}} \;
e^{-\b H_{\g,\l}(q|\bar q^{(\d^= (\G))})}
{\displaystyle \prod_{m:\; \si_m(\G)\ne \si(\G)}}
Z^A_{\g,\b,\l} \big({\rm Int}_m(\G)| 
q^{(\p {\rm Int}_m(\G))}\big)
\over
{\displaystyle \int_{{\cal Q}^{({\rm Supp}(\G) \setminus
\d^=(\G))}}}
dq\; \ind{\eta(q)=\si(\G)} \;
e^{-\b H_{\g,\l}(q|\bar q^{(\d^= (\G))})}
{\displaystyle \prod_{m:\; \si_m(\G)\ne \si(\G)}}
Z^A_{\g,\b,\l} \big({\rm Int}_m(\G)| 
q^{(\p {\rm Int}_m(\G))}\big) }
   $$ 
To show that the definition is well posed, we first say that a statement
related to a ${\cal D}^{(\ell)}$ measurable region is proved by {\it
induction in volume} if this statement is true for a ${\cal D}^{(\ell)}$
measurable region $\L$ as soon as it is true for all ${\cal D}^{(\ell)}$
measurable subsets of $\L$. We then observe that \equ(3.8)-\equ(3.10)
constitute a single inductive definition. At the initial step of this
induction in volume one considers only contours with empty interior and
uses \equ(3.9)-\equ(3.10) to define their truncated statistical weights.
Then \equ(3.8) allows one to calculate the auxiliary partition functions
for sufficiently small regions admitting inside them only contours
without interior. Afterwards \equ(3.9)-\equ(3.10) are used again to define the
truncated statistical weights of contours having sufficiently small but
non empty interiors and so on.

We hope that a notational similarity of the numerator and the
denominator of \equ(3.10) does not hide from the reader the fact that
these expressions are very different. For example, in the numerator of
\equ(3.10) configurations $q^{(\p {\rm Int}_m(\G))}$ belong to the
ground state ensemble of the phase $\si=\si(\G)$ while in the
denominator of \equ(3.10) these configurations belong to the ground
state ensemble of the opposite phase, $-\si$.

\medskip\goodbreak\noindent
{\bf \Statement (s3.2)}

\nobreak
{\sl If the truncated statistical weight $W^T({\G}| q^{(\d^=(\G))})$ of
any contour is smaller than the second term in the argument of the
$\min$ in \equ(3.9) then the auxiliary partition function coincides with
the true one. }

\medskip
The proof of the statement is standard in the P-S theory and we omit it
(see [Z]).

\medskip
Another standard observation is that for any region $\L$ with the
boundary condition $\bar q^{(\p \L)}$ belonging to the ground state
ensemble of the phase $\si$ the statistics of external contours in the
contour model coincides with that of the true particle model. Note that
the statistics of non-external contours are different in the contour and
particle models. For example, the contours of the opposite phase,
$-\si$, never appear in the contour model. As we explained before the
absence of the matching rules makes the analysis of the contour model
much easier than the analysis of the initial particle model. The price
paid for this simplification is a rather involved expression for $W(\G |
\bar q^{(\d^=(\G))})$.

\medskip\goodbreak\noindent
{\bf \Statement (s3.3)}

\nobreak
{\sl For all $\g$ small enough and all chemical potentials $\lambda \in
\big(\l(\b)-\g^{\a}, \l(\b)+\g^{\a} \big)$, $\a \ge 1/2$ there are
$f^A_{+,\lambda,\g}$, resp. $f^A_{-,\lambda,\g}$, such that for any
sequence of cubes $\L\to \Bbb R^d$ and any sequence of boundary
conditions $\bar q^{(\L^c)}$ belonging to the $+$ (resp. $-$) ground
state ensemble
   $$
\lim_{\L \to \Bbb R^d} {1\over \b|\L|}\log
Z^A_{\g,\b,\l}(\L|\bar q^{(\L^c)}) = f^A_{\pm,\lambda,\g}
\Eq(3.10e1)
   $$
Moreover there exists $\lambda(\g,\b)$ such that
   $$
\lim_{\g \to 0} \lambda(\g,\b)
= \lambda(\b)
\Eq(3.10e2)
   $$
and}
   $$ f^A_{+,\lambda(\g,\b),\g}= f^A_{-,\lambda(\g,\b),\g}
\Eq(3.10e3)
   $$

\medskip
The statement is proved in Section 6 using cluster expansion techniques,
together with a rather explicit representation of the auxiliary
pressures $f^A_{\pm,\lambda,\g}$.

\medskip
In analogy to  the first term $T_1$ in the decomposition  \equ(3.8e1) 
of the statistical weight we introduce the function
   $$
w(\G  | \bar q^{(\d^=(\G))})=
\phantom{1234567890123456789012345678901234567890123456789012345}
   \Eq(3.7.2)
   $$
\nobreak
   $$
{{\displaystyle \int_{{\cal Q}^{({\rm Supp}(\G) \setminus
\d(\G))}} }
dq\; \ind{\eta(q)=\eta^{\G}} \;
e^{-\b H_{\g,\l}(q|\bar q^{(\d^= (\G))}, q_{-\si(\G)}^{(\d^{\ne} (\G))}  )}
{\displaystyle \prod_{m:\; \si_m(\G)\ne \si(\G)}} e^{ {\beta \over 2} U_{\g}(
 q_{-\si(\G)}^{(\G))}
| q_{-\si(\G)}^{(\d^{\ne} (\G))})}   }
\over
{\displaystyle \int_{{\cal Q}^{({\rm Supp}(\G) \setminus
\d(\G))}}}
dq\; \ind{\eta(q)=\si(\G)} \;
e^{-\b H_{\g,\l}(q|\bar q^{(\d^= (\G))},q_{\si(\G)}^{(\d^{\ne} (\G))}  )}
{\displaystyle \prod_{m:\; \si_m(\G)\ne \si(\G)}}
e^{ {\beta \over 2} U_{\g}(
 q_{\si(\G)}^{(\G))}
| q_{\si(\G)}^{(\d^{\ne} (\G))})  }
   $$ 
In Section 4 we will prove the Peierls bound (which is the analogue of
\equ(3.8e5)):

\medskip\goodbreak\noindent
{\bf \Statement (s3.4)}  

\nobreak
{\sl There is $c>0$
such that for all $\g$ and $\lambda \in \big(\l(\b)-\g^{\a},
\l(\b)+\g^{\a} \big)$, $\a \ge 1/2$ }
   $$
w(\G  | \bar q^{(\d^=(\G))}) \le
\exp \left( -c 
\ell_2^d \ell_3^{-d} \z^2 |{\rm Supp}(\G)| \right)
   \Eq(3.7.3)
   $$

\medskip
Finally in Section 6 we will prove 

\medskip\goodbreak\noindent
{\bf \Statement (s3.5)} 

\nobreak
{\sl There is $c>0$ such that for all sufficiently small $\g$ and
$\lambda \in \big(\l(\b)-\g^{\a}, \l(\b)+\g^{\a} \big)$, $\a \ge 1/2$ }
   $$
\eqalign{
\Big|
\log W^A(\G| \bar q^{(\d^=(\G))}) &-  \log  w(\G  | \bar q^{(\d^=(\G))}) \cr
&- \sum_{m:\; \si_m(\G)\ne \si(\G)  }
\Big( \b | {\rm Int}_m(\G)|
[f^A_{-\si(\G),\lambda,\g} - f^A_{\si(\G),\lambda,\g}] \Big)
\Big| \cr
&\le c |{\rm Supp}(\G)| \g^{1/4} \cr}
   \Eq(3.10e6)
   $$

\medskip\noindent
Note that the factor $\g^{1/4} $ is not optimal but it is enough for
our purposes. 

\medskip
By combining the above statements and choosing $\l = \l(\b,\g)$ we then
obtain that the auxiliary and the true partition functions are equal and
consequently that the probability of a contour in the original system is
bounded as in \equ(3.7.3). This proves the existence of two different
phases and together with the exponential decay established in Sections~5
and~6 concludes the proof of Theorem~\equ(s2.1).

\bigskip
\bigskip
\goodbreak 
\centerline{{\bf 4. Peierls Estimate }}
\bigskip
\numsec= 4
\numfor= 1
\numtheo=1
In this section we prove Statement~\equ(s3.4), i.e. we compare the partition
functions in the numerator and denominator of \equ(3.7.2).

The partition function in the denominator of \equ(3.7.2) is taken over
ground state configurations of the phase $\si=\si(\G)$ placed in the
region ${\rm Supp}(\G) \setminus \d(\G)$ with the boundary condition
$\bar q^{(\d(\G))}$ specified on $\d(\G)$. This boundary condition,
$\bar q^{(\d(\G))}$, is different in $\d^{=}(\G)$ and in $\d^{\neq}(\G)$:
it is a configuration of the ground state ensemble of the phase
$\si$ in the region $\d^{=}(\G)$ and it is the standard configuration of
the phase $\si$ in $\d^{\neq}(\G)$.

The partition function from the numerator of \equ(3.7.2) is taken over
contour configurations, i.e. ones compatible with $\eta^{\G}$, in the
same region ${\rm Supp}(\G) \setminus \d(\G)$ with the same boundary
condition $\bar q^{(\d^{=}(\G))}$ in $\d^{=}(\G)$ but with the different
boundary condition given on $\d^{\neq}(\G)$. Contrary to the denominator
this boundary condition is the standard configuration of the opposite
phase, $-\si$. 

\bigskip\nobreak
\centerline{{\it Scheme of Estimate}}
\medskip
\nobreak

The estimate of the ratio in \equ(3.7.2) is performed in several steps.
At each step we achieve a certain simplification at the price of a non
essential error which is a factor not exceeding $\exp \Big( \g^{\a}
\ell_2^d \ell_3^{-d} \z^2 |{\rm Supp}(\G)| \Big)$ with $\a>0$. For $\g$
small enough the product of the finite number of such factors is, of
course, dominated by $\exp \left( -c \ell_2^d \ell_3^{-d} \z^2 |{\rm
Supp}(\G)| \right)$.

At the first step of the estimate we replace every configuration
$q=(q_i)$ contributing to the numerator or denominator of \equ(3.7.2) by
$[q]^{(\ell_1)}$. We show that the error caused by this transformation
is not important. The calculation is straightforward for the denominator
of \equ(3.7.2) since in the contributing configurations the density of
particles is bounded. One needs to be more careful with the numerator of
\equ(3.7.2). For some of the contributing configurations it may happen
that in some boxes $C^{(\ell_1)} \in {\rm Supp}(\G) \setminus \d(\G)$
the number of particles is too large. In this case we replace the
configuration inside this box by a configuration with a bounded density
and we use superstability to check that the error is small. As soon as
all consideration are reduced to the case of bounded density it becomes
clear that one needs to check \equ(3.7.3) for $\l=\l(\b)$ only.
Indeed, varying $\l$ in the interval $\lambda \in \big(\l(\b)-\g^{\a},
\l(\b)+\g^{\a} \big)$ with $\a \ge 1/2$ produces an error in
$H_{\g,\l}(q|\bar q^{(\d^= (\G))}, q_{\pm \si(\G)}^{(\d^{\ne} (\G))} )$ which
in absolute value does not exceed $\g^{1/2}|{\rm Supp}(\G)|$. Hence the
total error in \equ(3.7.2) is negligible with respect to the right hand
side of \equ(3.7.3)

After shifting particles into the centers of the corresponding boxes
integrals over the particle configurations $q$ become sums over density
configurations $\rho^{(\ell_1)}(q)= (\rho^{(\ell_1)}_x(q))$. Accordingly
the energy of the configuration $q$ is replaced, again up to non essential
error, by the discrete version of the mean field free energy functional
\equ(2.13) where the function $\rho(r)$ of a continuous argument $r \in
{\Bbb R}^d$ is replaced by its lattice version $\rho^{(\ell_1)}_x$, $x \in
{\Bbb Z}^d_{\ell_1}$.

The next observation is that the entropy factor coming from the summation
over different density configurations $\rho^{(\ell_1)}$ can be neglected
and one can consider only a contribution of two density configurations
providing the minimum of the free energy functional in the numerator and
denominator of \equ(3.7.2) respectively. This reduces everything to a
variational problem for the discrete version of the functional
\equ(2.13).

Utilizing \equ(2.21.1) one can see that for the minimizing density
configuration of this variational problem the influence of the boundary
conditions decays exponentially and at the distance of order
$\g^{-1-\a_3}$ from $\L^c$ it practically disappears. Thus, modulo non
essential errors, we can replace the initial variational problem by a
similar one in a smaller region with {\it standard} boundary conditions
only.

At the final step we consider for the denominator of \equ(3.7.2) a
suitably reduced region ${\rm Supp}(\G) \setminus \d(\G)$ with the
standard boundary condition $\rho_{\b,\si}$. The minimizing density is
clearly the same $\rho_{\b,\si}$ continued inside the region.
Investigating then the minimizing density configuration for the
numerator of \equ(3.7.2) one takes into account the fact that there are
sufficiently many boxes $C^{(\ell_2)}$ inside ${\rm Supp}(\G) \setminus
\d(\G)$ where the indicator function $\ind{\eta(q) =\eta^{\G}}$ forces
the minimizing configuration to be different from $\rho_{\b,\pm}$. This
leads to an excess of energy, with respect to the energy of
$\rho_{\b,\pm}$, which is enough to produce the factor $\exp \left( -c
\ell_2^d \ell_3^{-d} \z^2 |{\rm Supp}(\G)| \right)$.

We now turn to the details.

\bigskip
\goodbreak
\centerline{{\it Stability Estimates}}
\medskip
\nobreak
First we check that the accumulation of an infinite number of particles in
a bounded region is impossible. This is a consequence of

\medskip\noindent
{\bf \Lemma (s6.1)} 

\nobreak
{\sl 
For any particle $q_0$ and any configuration $q=(q_1, q_2, \ldots)$
the interaction energy
   $$
U_{\g}(q_0| q)=- \sum_{q_{i_1} \in q}
J_{\g}^{(2)}(q_{0},q_{i_1})
+\sum_{q_{i_1},q_{i_2},q_{i_3} \in q}
J_{\g}^{(4)}(q_0,q_{i_1},q_{i_2},q_{i_3})
   \Eq(6.1)
   $$
is bounded from below
   $$
U_{\g}(q_0| q) \ge H_{\min} 
   \Eq(6.2)
   $$
by an absolute constant 
   $$
H_{\min}=\min_{{\cal N} >0,\; 1>\g>0} \quad
{1 \over 3!} \left( {\cal N}^3 -3 \g^d {\cal N}^2 
+ 2  \g^{2d} {\cal N} \right)  -{\cal N}
   \Eq(6.7)
   $$
}

\medskip\noindent 
{\bf Proof.} Given $q$ let
   $$
{\cal N}(r,q) = \g^d \sum_{q_{i} \in q}
\ind{|r - q_i| \le \g^{-1}R_d}
   \Eq(6.3)
   $$
be a mean number of particles of $q$ situated inside $B_{\g}(r)$, $r \in
{\Bbb R}^d$. Then
   $$
\eqalign{
- \sum_{q_{i_1} \in q} J_{\g}^{(2)}(q_{0},q_{i_1})
&=-\g^{2d} \sum_{q_{i} \in q} \int dr \;\;
\ind{|r - q_0| \le \g^{-1}R_d}
\ind{|r - q_i| \le \g^{-1}R_d} \cr
&=-\g^{d} \int dr \;\;
\ind{|r - q_0| \le \g^{-1}R_d} {\cal N}(r,q)\cr}
   \Eq(6.4)
   $$
Similarly
   $$
\eqalign{
\sum_{q_{i_1},q_{i_2},q_{i_3} \in q}
J_{\g}^{(4)}(q_0,q_{i_1},q_{i_2},q_{i_3})
&=\g^{d} \int dr \;\; 
\ind{|r - q_0| \le \g^{-1}R_d} \cr
&\times {1 \over 3!} \left( {\cal N}^3(r,q) -3 \g^d {\cal N}^2(r,q) 
+ 2  \g^{2d} {\cal N}(r,q) \right) \cr}
   \Eq(6.5)
   $$
Hence 
   $$
U_{\g}(q_0| q) \ge \g^{d} \int dr \;\;
\ind{|r - q_0| \le \g^{-1}R_d} H_{\min}= H_{\min}, 
   \Eq(6.6)
   $$
which proves lemma. \qed

\medskip
Lemma~\equ(s6.1) provides the lower bound 
   $$
H_{\g,\l}(q^{(\Lambda)} | \bar q^{(\Lambda^c)}) \ge H_{\min} |q^{(\Lambda)}|=
H_{\min} |\Lambda| \rho,
   \Eq(6.7.1)
   $$
where $\rho= {|q^{(\Lambda)}|/ |\Lambda|}$ is the corresponding density.
To obtain an upper bound take some positive $\rho_{\max} > \rho_{\b,+}$.

\medskip\noindent
{\bf \Lemma (s6.1.1)} 

\nobreak
{\sl 
Consider configurations $q^{(\Lambda)}$ and $\bar q^{(\Lambda^c)}$ such
that $q^{(\Lambda)} \cup \bar q^{(\Lambda^c)}$ has at most $\rho_{\max}
(2\g^{-1})^d$ particles in any intersecting $\Lambda$ cube with the side
$2\g^{-1}$. Then
   $$
H_{\g,\l}(q^{(\Lambda)} | \bar q^{(\Lambda^c)}) \le  H_{\max}(\rho_{\max})
|\Lambda|, 
   \Eq(6.7.2)
   $$
where
   $$
H_{\max}(\rho_{\max})=|\l(\b)|\rho_{\max} + 2^{3d}\rho_{\max}^4
   \Eq(6.7.3)
   $$
}

\medskip \noindent {\bf Proof.} It is clear that $|q^{(\Lambda)}| \le
|\Lambda| \rho_{\max}$. The strength of four-body interaction between any
four particles is less than $\g^{3d}$. The number of interacting
quadruples of particles such that one of the elements of the quadruple is
a given particle is less than $(\rho_{\max} 2^d \g^{-d})^3$ Hence the
total four-body interaction contributing to $H_{\g,\l}(q^{(\Lambda)} | \bar
q^{(\Lambda^c)})$ is less than $|\Lambda| \rho_{\max}^4 2^{3d}$. The
two-body interaction is negative and does not contribute to the
estimate. \qed

\bigskip
\goodbreak
\centerline{{\it Bad Boxes}}
\medskip
\nobreak

In this subsection we treat boxes containing too many particles.

\medskip\noindent
{\bf \Lemma (s6.2)} 

\nobreak 
{\sl 
Consider a box $C^{(\ell)}$, $\ell < {1 \over 2} \g^{-1}$, a configuration
$\bar q \in {\cal Q}$ and an integer 
   $$
N \ge |C^{(\ell)}| e^{c\b}= \ell^d e^{c\b}
   \Eq(6.8)
   $$
Then
   $$
\int_{{\cal Q}^{(C^{(\ell)})}} dq\; \ind{|q^{(C^{(\ell)})}|=N}\; 
e^{-\b H_{\g,\l}(q^{(C^{(\ell)})}| \bar q)} \le e^{-N}
   \Eq(6.9)
   $$
and
   $$
\int_{{\cal Q}^{(C^{(\ell)})}} dq\; \ind{|q^{(C^{(\ell)})}| \ge |C^{(\ell)}|
e^{c\b}}\; e^{-\b H_{\g,\l}(q^{(C^{(\ell)})}| \bar q)} \le 2e^{-|C^{(\ell)}|
e^{c\b}}
   \Eq(6.9.1)
   $$
}

\medskip \noindent
{\bf Proof.} We check only \equ(6.9) as it easily implies \equ(6.9.1).
According to Lemma~\equ(s6.1) the interaction
$U_{\g}(q^{(C^{(\ell)})}|\bar q)$ between particles of
$q^{(C^{(\ell)})}$ and $\bar q$ satisfies the estimate
   $$
U_{\g}(q^{(C^{(\ell)})}|\bar q) \ge  H_{\min} N
   \Eq(6.10)
   $$
It is clear that
   $$
H_{\g,\l}(q^{(C^{(\ell)})}) \ge  -\g^{d} {N(N-1) \over 2!}
+ {1 \over 2^d} \g^{3d} {N(N-1)(N-2)(N-3)\over 4!} - |\l| N
   \Eq(6.11)
   $$
as the maximal volume of the intersection of two balls of radius
$\g^{-1}R_d$ is $\g^{-d}$ and the minimal volume of the intersection of
four such balls with the centers in $C^{(\ell)}$ is larger than ${1 \over
2^d}\g^{-d}$. Thus the logarithm of the left hand side of \equ(6.9) does
not exceed
   $$
\eqalign{
&-N \log N + N + N \log |C^{(\ell)}| -\b H_{\min} N \cr
&+\b |\l| N +\b \g^{d} {N(N-1) \over 2!} 
-\b{1 \over 2^d}\g^{3d} {N(N-1)(N-2)(N-3)\over 4!} \cr}
   \Eq(6.12)
   $$
It is not hard to check that for sufficiently small $\g$ and
sufficiently large absolute constant $c_{\equ(6.8)}$ the last expression
is smaller than $-N$ as soon as $N \ge |C^{(\ell)}|
e^{c_{\equ(6.8)}\b}$. Note that the most dangerous term in \equ(6.12) is
$\b \g^{d} {N(N-1) \over 2!}$. It is dominated by $\b{1 \over
2^d}\g^{3d} {N(N-1)(N-2)(N-3)\over 4!}$ for $N > c \g^{-d}$ and by $N
\log N- N \log |C^{(\ell)}|$ for $|C^{(\ell)}| e^{c_{\equ(6.8)}\b} \le N
\le c \g^{-d}$. This implies \equ(6.9) and hence \equ(6.9.1). \qed 

\medskip
Set $\rho_{\max}=2e^{c_{\equ(6.8)}\b}$. An easy consequence of
Lemma~\equ(s6.2) is 

\medskip\noindent{\bf \Lemma (s6.3)} 

\nobreak 
{\sl For any contour $\G$ and any $\ell \le \ell_2$ the
partition function in the numerator of \equ(3.7.2) does not exceed
   $$
\eqalign{
e^{\ell^{-d}|{\rm Supp}(\G)|}
{\displaystyle \int_{{\cal Q}^{({\rm Supp}(\G) \setminus
\d(\G))}}} dq\;\;
\ind{\eta(q)=\eta^{\G},\; \rho^{(\ell)}(q) \le \rho_{\max}} \;
&e^{-\b H_{\g,\l}(q|\bar q^{(\d^= (\G))}, q_{-\si}^{(\d^{\neq} (\G))})} \cr
&\times {\displaystyle \prod_{m:\; \si_m(\G)\ne \si(\G)}} e^{ {\beta \over 2}
U_{\g}( q_{-\si(\G)}^{(\G))} | q_{-\si(\G)}^{(\d^{\ne} (\G))})} \cr}
   \Eq(6.13)
   $$
}

\medskip\noindent
{\bf Proof.} To obtain \equ(6.13) we perform a partial integration over
the configurations in the bad boxes. Suppose that $q$ is fixed outside
the cube $C^{(\ell)} \in {\rm Supp}(\G)$ where $q$ has more than
$\rho_{\max}|C^{(\ell)}|$ particles. Now applying Lemma~\equ(s6.2) we
integrate over all such $q$-s leaving exactly ${1 \over 2} \rho_{\max}
|C^{(\ell)}|$ particles in $C^{(\ell)}$. That means that we integrate
over the particles in excess to ${1 \over 2} \rho_{\max} |C^{(\ell)}|$
of them assuming that the total number of particles in $C^{(\ell)}$ is
larger than $\rho_{\max}|C^{(\ell)}|$. This produces an extra factor $1+
2e^{-|C^{(\ell)}| e^{c\b}} \le e$. We continue this procedure box by box
and the total number of boxes in ${\rm Supp}(\G)$ is $\ell^{-d}|{\rm
Supp}(\G)|$. \qed

\bigskip
\goodbreak
\centerline{{\it Reduction to Variational Problem}}
\medskip
\nobreak
It was shown in the previous subsection that we may restrict our
considerations to the configurations with the bounded density. Now we
estimate an error which is produced by shifting particles in such a
configuration into the centers of the corresponding boxes $C^{(\ell)}$.

\medskip\noindent{\bf \Lemma (s6.4)} 

\nobreak 
{\sl Take 
$\ell < \g^{-1}$ and configurations $q^{(\Lambda)}$ and $\bar
q^{(\Lambda^c)}$ with $\rho^{(\ell)}(q^{(\Lambda)} \cup \bar q^{(\p
\Lambda)})) \le \rho_{\max}$. Then
   $$
| H_{\g,\l}(q^{(\Lambda)} | \bar q^{(\Lambda^c)} )
-H_{\g,\l}(\rho^{(\ell)}(q^{(\Lambda)})| \bar q^{(\Lambda^c)} )| \le
\ell \g |\Lambda| (2^d \rho_{\max}^2+2^{3d} \rho_{\max}^4)
   \Eq(6.16)
   $$
}

\medskip\noindent
{\bf Proof.} It is clear that $|q^{(\Lambda)}| \le |\Lambda|
\rho_{\max}$. Given two interacting particles the absolute value of the
error in their interaction due to shifting these particles at the centers
of the corresponding boxes is less than $\g^{d} \ell \g$. Given four
interacting particles the absolute value of the error in their interaction
due to shifting these particles at the centers of the corresponding boxes
is less than $\g^{3d} \ell \g$. As in Lemma~\equ(s6.1.1) the number of
interacting quadruples of particles such that one of the elements of the
quadruple is a given particle is less than $(\rho_{\max} 2^d \g^{-d})^3$.
Similarly the number of pairs of interacting particles such that one of
the elements of the pair is a given particle is less than $(\rho_{\max}
2^d \g^{-d})$. Hence the total error is less than $\ell \g |\Lambda| (2^d
\rho_{\max}^2+2^{3d} \rho_{\max}^4)$. \qed

\medskip
This lemma allows us to replace the integrals over $dq$ in the numerator
and denominator of \equ(3.7.2) by the sums over density configurations
$\rho^{(\ell_1)}$. Namely, consider the partition function on the right
hand side of \equ(6.13). The integral over $q \in {\cal Q}^{({\rm
Supp}(\G) \setminus \d(\G))}$ with $\eta(q)=\eta^{\G}$ and
$\rho^{(\ell_1)}(q) \le \rho_{\max}$ can be calculated in two steps. 

First one can fix a density configuration $\rho^{(\ell_1)}
=(\rho^{(\ell_1)}_x)$, $x \in [{\rm Supp}(\G) \setminus
\d(\G)]^{(\ell_1)}$ and integrate over configurations $q^{({\rm
Supp}(\G) \setminus \d(\G))}$ with
$\rho^{(\ell_1)}(q)=\rho^{(\ell_1)}$.

Afterwards one can sum over all $\rho^{(\ell_1)} \le \rho_{\max}$
compatible with $\eta^{\G}$. The obvious upper estimate for this sum is
the total number of density configurations $\rho^{(\ell_1)}_x \le
\rho_{\max}$, $x \in [{\rm Supp}(\G) \setminus \d(\G)]^{(\ell_1)}$ times
the maximal contribution given by a single density configuration.

The number of density configurations $\rho^{(\ell_1)} \le \rho_{\max}$
compatible with $\eta^{\G}$ is less than
   $$
(\rho_{\max} |C^{(\ell_1)}|)^{ |{\rm Supp}(\G) \setminus
\d(\G)| \over |C^{(\ell_1)}|}
   \Eq(6.19)
   $$ 

Given $\rho^{(\ell_1)}$ the integral over configurations $q$ with
$\rho^{(\ell_1)}(q)=\rho^{(\ell_1)}$ does not exceed
   $$
\eqalign{
\exp \Big( &-\b H_{\g,\l}(\rho^{(\ell_1)}|\bar q^{(\d^= (\G))},
q_{-\si}^{(\d^{\neq} (\G))}) +  {\beta \over 2} \sum_{m:\;
\si_m(\G)\ne \si(\G)} 
U_{\g}( q_{-\si(\G)}^{(\G))}| q_{-\si(\G)}^{(\d^{\ne} (\G))}) \cr
&+ \log |C^{(\ell_1)}| \sum_{x \in [{\rm Supp}(\G) \setminus
\d(\G)]^{(\ell_1)}} \rho^{(\ell_1)}_x |C^{(\ell_1)}| \cr
&- \sum_{x \in [{\rm Supp}(\G) \setminus \d(\G)]^{(\ell_1)}}
\log \big((\rho^{(\ell_1)}_x |C^{(\ell_1)}|)!\big) \cr
&+ \b \ell_1 \g |{\rm Supp}(\G) \setminus \d(\G)| (2^d
\rho_{\max}^2+2^{3d} \rho_{\max}^4) \Big) \cr}
   \Eq(6.17)
   $$
as follows from Lemma~\equ(s6.4).
 
Suppose that the density configuration $\bar \rho^{(\ell_1)}$ gives the
minima of
   $$
\eqalign{
\widetilde {\cal F}_{\g,\b,\l}(\rho^{(\ell_1)}|\bar q^{(\d^=
(\G))}, q_{-\si}^{(\d^{\neq} (\G))}, \star) &= \b
H_{\g,\l}(\rho^{(\ell_1)}|\bar q^{(\d^= (\G))},
q_{-\si}^{(\d^{\neq} (\G))}) \cr
&-  {\beta \over 2} \sum_{m:\; \si_m(\G)\ne \si(\G)}
U_{\g}( q_{-\si(\G)}^{(\G))}| q_{-\si(\G)}^{(\d^{\ne} (\G))}) \cr
&- |C^{(\ell_1)}| \log |C^{(\ell_1)}| 
\sum_{x \in [{\rm Supp}(\G) \setminus \d(\G)]^{(\ell_1)}}
\rho^{(\ell_1)}_x \cr
&+ \sum_{x \in [{\rm Supp}(\G) \setminus \d(\G)]^{(\ell_1)}}
\log \big( (\rho^{(\ell_1)}_x |C^{(\ell_1)}|)! \big) \cr }
   \Eq(6.18)
   $$ 
among all $\rho^{(\ell_1)} \le \rho_{\max}$ compatible with
$\eta^{\G}$. 
Then
   $$
\eqalign{
-\widetilde {\cal F}_{\g,\b,\l(\b)}(\bar \rho^{(\ell_1)}|\bar q^{(\d^=
(\G))}, q_{-\si}^{(\d^{\neq} (\G))}, \star)
&+\Big| {\rm Supp}(\G) \setminus \d(\G) \Big| 
\Big( \g^{1/2} + \ell_1^{-d}  \cr
&+ \ell_1^{-d} \log (\rho_{\max}|C^{(\ell_1)}|)
+ \b \ell_1 \g (2^d \rho_{\max}^2+2^{3d} \rho_{\max}^4)
\Big) \cr}
   \Eq(6.20)
   $$ 
is the upper bound for the $\log$ of the numerator of \equ(3.7.2). Note
that the term $\g^{1/2}$ in \equ(6.20) is the price paid for taking
$\l=\l(\b)$.

Similarly if $\hat \rho^{(\ell_1)}$ gives the minima of $\widetilde
{\cal F}_{\g,\b,\l(\b)}(\rho^{(\ell_1)}|\bar q^{(\d^= (\G))},
q_{\si}^{(\d^{\neq} (\G))}, \star)$ among all density configurations
$\rho^{(\ell_1)}$ from the ground state ensemble of the phase $\si$ then
   $$
-\widetilde {\cal F}_{\g,\b,\l(\b)}(\hat \rho^{(\ell_1)}|\bar q^{(\d^=
(\G))}, q_{\si}^{(\d^{\neq} (\G))}, \star)
-\Big| {\rm Supp}(\G) \setminus \d(\G) \Big| 
\Big( \g^{1/2} + \b \ell_1 \g (2^d \rho_{\max}^2+2^{3d} \rho_{\max}^4)  \Big)
   \Eq(6.21)
   $$ 
is the lower bound for the $\log$ of the denominator of \equ(3.7.2).

The notations with $\widetilde {~~}$ and $\star$ foresee forthcoming
simplifications and variations. In particular $\widetilde {\cal
F}_{\g,\b,\l(\b)}(\rho^{(\ell_1)}|\bar q^{(\d^= (\G))},
q_{\si}^{(\d^{\neq} (\G))})$ is defined as $\widetilde {\cal
F}_{\g,\b,\l(\b)}(\rho^{(\ell_1)}|\bar q^{(\d^= (\G))},
q_{\si}^{(\d^{\neq} (\G))}, \star)$ without the term ${\beta \over 2}
\sum_{m:\; \si_m(\G)\ne \si(\G)} U_{\g}( q_{-\si(\G)}^{(\G))}|
q_{-\si(\G)}^{(\d^{\ne} (\G))})$ in \equ(6.17).

Now denoting $b_{\g,\ell}=||[B_{\g}(\cdot)]^{(\ell)}||$ let 
   $$
I_{\g}^{(2)}(x_1,x_2)=
%\left({\ell^d \over b_{\g,\ell}} \right)^2
b_{\g,\ell}^{-2}\; ||[B_{\g}(x_1)]^{(\ell)} \cap
[B_{\g}(x_2)]^{(\ell)} ||
   \Eq(6.21.1)
   $$ 
and
   $$
I_{\g}^{(4)}(x_1, \dots, x_4)=
%\left({\ell^d \over b_{\g,\ell}} \right)^4 
b_{\g,\ell}^{-4}\; ||\cap_{j-1}^4 [B_{\g}(x_j)]^{(\ell)} ||
   \Eq(6.21.2)
   $$ 
be discrete versions of $J_{\g}^{(2)}(x_1,x_2)$ and $J_{\g}^{(4)}(x_1,
\dots, x_4)$. For any region $[\L]^{(\ell)}$ (we do not exclude the case
$[\L]^{(\ell)}={\Bbb Z}^d_{\ell}$) and any density configuration
$\rho^{(\ell)}(\L)=(\rho^{(\ell)}_x)$, $x \in [\L]^{(\ell)}$ define a
functional
   $$
\eqalign{
{\cal F}_{\g,\b,\l}(\rho^{(\ell)}(\L))&= \ell^d \left( \sum_{x \in
[\L]^{(\ell)}} {\rho^{(\ell)}_x \over \b}(\log\rho^{(\ell)}_x -1) -
\sum_{x \in [\L]^{(\ell)}} \l\rho^{(\ell)}_x \right. \cr
&-{1\over 2!} \sum_{x_2, x_1 \in [\L]^{(\ell)}} 
I_{\g}^{(2)}(x_1,x_2) \rho^{(\ell)}_{x_1} \rho^{(\ell)}_{x_2} \cr
&+\left. {1\over 4!} \sum_{x_1,x_2,x_3,x_4 \in [\L]^{(\ell)}}
I_{\g}^{(4)}(x_1, \dots, x_4)
\rho^{(\ell)}_{x_1} \cdots \rho^{(\ell)}_{x_4} \right) \cr }
   \Eq(6.22)
   $$
which is a discrete analogue of the mean field free energy functional
\equ(2.13). We also define a conditional functional
   $$
{\cal F}_{\g,\b,\l}(\rho^{(\ell)}(\L) | \bar \rho^{(\ell)}(\L^c))=
{\cal F}_{\g,\b,\l}(\rho^{(\ell)}(\L) + \bar \rho^{(\ell)}(\L^c))-
{\cal F}_{\g,\b,\l}( \bar \rho^{(\ell)}(\L^c))
   \Eq(6.23)
   $$
with the boundary condition $\bar \rho^{(\ell)}=(\bar \rho^{(\ell)}_x)$,
$x \in [\L^c]^{(\ell)}$. Here the similarity with \equ(2.3.1) is obvious
and the meaning of $\rho^{(\ell)} + \bar \rho^{(\ell)}$ becomes
straightforward if we set $\rho^{(\ell)} \equiv 0$, $x \in
[\L^c]^{(\ell)}$ and $\bar \rho^{(\ell)} \equiv 0$, $x \in [\L]^{(\ell)}$.
Setting
   $$
U_{\g,\b}(\rho^{(\ell)}(\L) | \bar \rho^{(\ell)}(\L^c))=
{\cal F}_{\g,\b,\l}(\rho^{(\ell)}(\L) | \bar \rho^{(\ell)}(\L^c))-
{\cal F}_{\g,\b,\l}( \rho^{(\ell)}(\L))
   \Eq(6.23.1)
   $$
we introduce
   $$
{\cal F}_{\g,\b,\l}(\rho^{(\ell)}(\L) | \bar \rho^{(\ell)}(\L^c), \star)
= {\cal F}_{\g,\b,\l}(\rho^{(\ell)}(\L) | \bar \rho^{(\ell)}(\L^c)) - {1
\over 2} U_{\g,\b}(\rho^{(\ell)}_{\b,\si}(\L) | 
\rho^{(\ell)}_{\b,\si}(\L^c)),
   \Eq(6.23.2)
   $$
where $\rho^{(\ell)}_{\b,\si} \equiv \rho_{\b,\si}$ and $\si=+$ or
$\si=-$. With some abuse of notation we use for ${\cal
F}_{\g,\b,\l}(\rho^{(\ell)}(\L)| \rho^{(\ell)}(\bar q^{(\p \L)}))$ an
alternative notations ${\cal F}_{\g,\b,\l}(\rho^{(\ell)}(\L)| \bar q^{(\p
\L)})$ or \hfil \goodbreak \noindent ${\cal
F}_{\g,\b,\l}(\rho^{(\ell)}(\L)|\bar
\rho^{(\ell)}(q^{(\p \L)}))$. The role which is played by functional
\equ(6.23) is clarified in 

\medskip\noindent{\bf \Lemma (s6.5)} 

\nobreak 
{\sl Consider a ${\cal D}^{(\ell)}$ measurable region $\L$ with the
boundary condition $\bar q^{(\p \L)}$ which is a ground state
configuration on every connected component of $\p \L$. Then for $\ell \le
\g^{-1}$ and any $\rho^{(\ell)}=(\rho^{(\ell)}_x)$, $x \in
[\L]^{(\ell)}$ such that $\rho^{(\ell)}_x \le \rho_{\max}$ one has
   $$
|\widetilde {\cal F}_{\g,\b,\l}(\rho^{(\ell)}(\L)|\bar q^{(\p \L)}) -
{\cal F}_{\g,\b,\l}(\rho^{(\ell)}(\L)| \bar \rho^{(\ell)}(q^{(\p \L)}))| \le
\ell \g 5(2^d \rho_{\max}^2+2^{3d} \rho_{\max}^4 + \rho_{\max}^4) |\L|
   \Eq(6.24)
   $$
   $$
|\widetilde {\cal F}_{\g,\b,\l}(\rho^{(\ell)}(\L)|\bar q^{(\p \L)}, \star) -
{\cal F}_{\g,\b,\l}(\rho^{(\ell)}(\L)| \bar \rho^{(\ell)}(q^{(\p
\L)}), \star)| \le 
\ell \g 5(2^d \rho_{\max}^2+2^{3d} \rho_{\max}^4 + \rho_{\max}^4) |\L|
   \Eq(6.24.1)
   $$
}

\medskip\noindent
{\bf Proof.} Estimate \equ(6.24.1) is an obvious consequence of
\equ(6.24) and we concentrate on \equ(6.24). The difference between
$\widetilde {\cal F}_{\g,\b,\l}$ and ${\cal F}_{\g,\b,\l}$ has two
sources. The first one is due to replacement of balls $B_{\g}(\cdot)$ in
the definition of $J^{(\cdot)}_{\g}$ by their lattice versions
$[B_{\g}(\cdot)]^{(\ell)}$ in the definition of $I^{(\cdot)}_{\g}$.
Clearly the difference between $|B_{\g}(\cdot)|=\g^{-d}$ and
$|[B_{\g}(\cdot)]^{(\ell)}|$ is less than $\g^{-d} \ell \g$. Hence the
error produced by the discretization of $B_{\g}(\cdot)$ can be estimated
exactly as in Lemma~\equ(s6.4). 

The second source is due to not properly counted contribution of pairs
$x_1, x_2$ with $x_1=x_2$ and quadruples $x_1,x_2,x_3,x_4$ with not all
$x_{i_j}$ being different. To estimate from above the absolute value of
this error let us consider the following five contributions to the
energy of $\rho^{(\ell)}$.
\item{(i)} Self-interaction of $C^{(\ell)}_x$ due to the pair interaction of
particles in $C^{(\ell)}_x$.
\item{(ii)} Self-interaction of $C^{(\ell)}_x$ due to the four-body
interaction of particles in $C^{(\ell)}_x$.
\item{(iii)} Interaction between $C^{(\ell)}_{x_1}$ and $C^{(\ell)}_{x_2}$
due to the four-body interaction of two particles in $C^{(\ell)}_{x_1}$
with two particles in $C^{(\ell)}_{x_2}$.
\item{(iv)} Interaction between $C^{(\ell)}_{x_1}$ and $C^{(\ell)}_{x_2}$
due to the four-body interaction of three particles in $C^{(\ell)}_{x_1}$
with one particle in $C^{(\ell)}_{x_2}$.
\item{(v)} Interaction between $C^{(\ell)}_{x_1}$, $C^{(\ell)}_{x_2}$ and
$C^{(\ell)}_{x_3}$ due to the four-body interaction of two particles in
$C^{(\ell)}_{x_1}$ with one particle in $C^{(\ell)}_{x_2}$ and one
particle in $C^{(\ell)}_{x_3}$.

\medskip\noindent
All five contributions can be estimated by similar arguments. For that
reason we present these arguments only for cases (i) and (v).

The strength of self-interaction of $C^{(\ell)}_x$ due to the pair
interaction of particles in $C^{(\ell)}_x$ is less than $\g^d (\rho_{\max}
\ell^d)^2$. The number of boxes $C^{(\ell)}_x$ in the region $\L$ is
$\ell^{-d} |\L|$. Hence the total contribution is less than $(\ell \g)^d
\rho_{\max}^2 |\L|$.

The strength of interaction between $C^{(\ell)}_{x_1}$, $C^{(\ell)}_{x_2}$
and $C^{(\ell)}_{x_3}$ due to the four-body interaction of two particles
in $C^{(\ell)}_{x_1}$ with one particle in $C^{(\ell)}_{x_2}$ and one
particle in $C^{(\ell)}_{x_3}$ is less than $\g^{3d} (\rho_{\max}
\ell^d)^4$. The number of boxes $C^{(\ell)}_{x_2}$ and $C^{(\ell)}_{x_3}$
interacting with given box $C^{(\ell)}_{x_1}$ is less than $(\g
\ell)^{-2d}$. The number of boxes $C^{(\ell)}_{x_1}$ is $\ell^{-d}
|\L|$. Hence the total contribution is less than $(\ell \g)^d
\rho_{\max}^4 |\L|$. \qed

\medskip
Looking for the minima of ${\cal F}_{\g,\b,\l} (\rho^{(\ell)}| \bar
\rho^{(\ell)}(q^{(\p \L)}))$ it is simpler to understand
$\rho^{(\ell)}_x$ as continuous variables. Therefore if the minima of
${\cal F}_{\g,\b,\l} (\rho^{(\ell)}| \bar \rho^{(\ell)}(q^{(\p \L)}))$ is
achieved on density configuration $\hat \rho^{(\ell)}$ then it may happen
that at least for some $x \in [\L]^{(\ell)}$ the number $ \ell^d \hat
\rho^{(\ell)}_x$ is not an integer. The solution to this problem is given
by

\medskip\noindent{\bf \Lemma (s6.6)} 

\nobreak 
{\sl For the density configuration $\tilde \rho^{(\ell)}_x= \ell^{-d} [
\ell^d \hat \rho^{(\ell)}_x]$ one has
   $$
|{\cal F}_{\g,\b,\l}(\hat \rho^{(\ell)}(\L)| \bar \rho^{(\ell)}(q^{(\p \L)}))-
{\cal F}_{\g,\b,\l}(\tilde \rho^{(\ell)}(\L)| \bar \rho^{(\ell)}(q^{(\p
\L)}))| \le c \ell^{-d} \rho_{\max} |\L|
   \Eq(6.25)
   $$
(Here $[\;\cdot\;]$ denotes the integer part of a number.)
}

\medskip\noindent
{\bf Proof.} We proceed as in the proof of Lemma~\equ(s6.4). The total
number of points $x \in [\L]^{(\ell)}$ is $\ell^{-d} |\L|$. Given point
$x$ the absolute value of the difference in the corresponding
self-interaction is less than $\b^{-1} \log \rho_{\max} + |\l|
\rho_{\max}$. The difference in two-point interaction between points
$x_1, x_2 \in [\L]^{(\ell)}$ in absolute value does not exceed $2 \g^d
\rho_{\max} \ell^d$. The number of points $x_2$ interacting with given
$x_1$ is less than $(\g \ell)^{-d}$. The difference in four-point
interaction between points $x_1, x_2,x_4, x_4 \in [\L]^{(\ell)}$ in
absolute value does not exceed $4 \g^{3d} (\rho_{\max} \ell^d)^3$. The
number of points $x_2, x_3, x_4$ interacting with given $x_1$ is less
than $(\g \ell)^{-3d}$. Combining this estimates one obtains the lemma.
\qed

\bigskip
\goodbreak
\centerline{{\it Variational Problem (Dependence on Boundary Condition)}}
\medskip
\nobreak

The results of the previous subsection reduce the Peierls estimate to
the following variational problem:
\item{(i)} Find the minimum of ${\cal
F}_{\g,\b,\l(\b)}(\rho^{(\ell_1)}({\rm Supp}(\G) \setminus \d(\G))| \bar
q^{(\d^= (\G))}, q_{\si}^{(\d^{\neq} (\G))}, \star)$ over the density
configurations $\rho^{(\ell_1)}$ $=(\rho^{(\ell_1)}_x)$, $x \in [{\rm
Supp}(\G) \setminus \d(\G)]^{(\ell_1)}$ such that $\rho^{(\ell_1)}$
belongs to the ground state ensemble of the phase $\si$.
\item{(ii)} Find the minimum of ${\cal
F}_{\g,\b,\l(\b)}(\rho^{(\ell_1)}({\rm Supp}(\G) \setminus \d(\G))| \bar
q^{(\d^= (\G))}, q_{-\si}^{(\d^{\neq} (\G))}, \star)$ over the
configurations $\rho^{(\ell_1)}=(\rho^{(\ell_1)}_x)$, $x \in [{\rm
Supp}(\G) \setminus \d(\G)]^{(\ell_1)}$ such that $\rho^{(\ell_1)}$ is
compatible with $\eta^{\G}$.
\item{(iii)} Estimate from below the difference between the minimal
value of \hfil\goodbreak ${\cal F}_{\g,\b,\l(\b)}(\rho^{(\ell_1)}({\rm
Supp}(\G) \setminus \d(\G))| \bar q^{(\d^= (\G))}$,
$q_{-\si}^{(\d^{\neq} (\G))}, \star)$ and the minimal value of
\hfil\goodbreak ${\cal F}_{\g,\b,\l(\b)}(\rho^{(\ell_1)}({\rm Supp}(\G)
\setminus \d(\G))| \bar q^{(\d^= (\G))}, q_{\si}^{(\d^{\neq} (\G))},
\star)$.

\medskip\noindent
The existence of the minima above is obvious as $0 < \rho^{(\ell_1)}_x
< \rho_{\max}$. Following the approach of Section~3 define
       $$
{\cal R}_{\g,\ell}(x,\rho^{(\ell)})= b_{\g,\ell}^{-1} \sum_{x_1 \in
[B_{\g}(x)]^{(\ell)}} \rho^{(\ell)}_{x_1}
        \Eq(6.26)
        $$
and
        $$
{\cal I}_{\g,\ell}(x,\rho^{(\ell)})= b_{\g,\ell}^{-1} \sum_{x_1 \in
[B_{\g}(x)]^{(\ell)}} {\rho^{(\ell)}_{x_1}\over \b} \big(
\log\rho^{(\ell)}_{x_1} -1 \big),
        \Eq(6.27)
        $$
where we deliberately use a general scale $(\ell)$ instead of $(\ell_1)$
as the constructions below are of general origin. In complete similarity
with \equ(2.15)
        $$
{\cal F}_{\g,\b,\l}(\rho^{(\ell)})= 
\ell^d \sum_{x \in {\Bbb Z}^d_{\ell}} 
\left( {\cal I}_{\g,\ell} (x,\rho^{(\ell)})
- \l {\cal R}_{\g,\ell}(x,\rho^{(\ell)}) 
- {1\over 2!}{\cal R}_{\g,\ell}(x,\rho^{(\ell)})^2
+ {1\over 4!}{\cal R}_{\g,\ell}(x,\rho^{(\ell)})^4 \right) 
        \Eq(6.28)
        $$
implying that $\rho^{(\ell)}({\Bbb Z}^d_{\ell}) \equiv \rho_{\b,\l,-}$
and $\rho^{(\ell)}({\Bbb Z}^d_{\ell}) \equiv \rho_{\b,\l,+}$ are the
global minimizers for ${\cal F}_{\g,\b,\l}(\rho^{(\ell)})$. The local
minimizers of ${\cal F}_{\g,\b,\l}(\cdot)$, i.e. the minima of ${\cal
F}_{\g,\b,\l} (\cdot| \bar \rho^{(\ell)}(\L^c))$ or ${\cal F}_{\g,\b,\l}
(\cdot| \bar \rho^{(\ell)}(\L^c), \star)$, are studied in the lemma
below. Note that in this lemma we consider not only $\l=\l(\b)$ but all
$\l \in (\l(\b)-\d(\b), \l(\b) +\d(\b))$. Though the lemma discusses
${\cal F}_{\g,\b,\l} (\cdot| \bar \rho^{(\ell)}(\L^c))$ the same
argument covers the case of ${\cal F}_{\g,\b,\l} (\cdot | \bar
\rho^{(\ell)}(\L^c), \star)$. 

\medskip\noindent
{\bf \Lemma (s6.7)} 

\nobreak
{\sl Consider a ${\cal D}^{(\ell)}$ measurable region $\L$ and take a
boundary condition $\bar \rho^{(\ell)}(\L^c)$ with $\displaystyle
\max_{x \in [\p \L]^{(\ell)}} |\bar \rho^{(\ell)}_x- \rho_{\b,\l,\si}|
\le \z$, where $\si$ is one of the phases, $+$ or $-$. Then the unique
minimum of ${\cal F}_{\g,\b,\l} (\rho^{(\ell)}(\L)| \bar
\rho^{(\ell)}(\L^c))$ among density configurations $\rho^{(\ell)}(\L)$
with $\displaystyle \max_{x \in [\L]^{(\ell)}} |\rho^{(\ell)}_x-
\rho_{\b,\l,\si}| \le \z$ is achieved on the density configuration $\hat
\rho^{(\ell)}=(\hat \rho^{(\ell)}_x)$, $x \in [\L]^{(\ell)}$ such that
        $$
|\hat \rho^{(\ell)}_x - \rho_{\b,\l,\si}| \le {\z a(\b)^{[\g {\rm
dist}\;(x, \L^c)]} \over 1-a(\b)},
        \Eq(6.29)
        $$
where $[ \cdot ]$ denotes the integer part.
}

\medskip\noindent
{\bf Proof.} Calculating ${\p \over \p \rho^{(\ell)}_x} {\cal
F}_{\g,\b,\l} (\rho^{(\ell)}(\L)| \bar \rho^{(\ell)}(\L^c))$, $x \in
[\L]^{(\ell)}$ one obtains the necessary condition for $\hat
\rho^{(\ell)}$
   $$
\eqalign{ 0&={1 \over \b} \log \hat \rho^{(\ell)}_x - \l - \sum_{x_1}
I_{\g}^{(2)}(x,x_1) \hat \rho^{(\ell)}_{x_1} \cr &+{1\over 3!}
\sum_{x_1,x_2,x_3} I_{\g}^{(4)}(x, x_1,x_2,x_3) \hat \rho^{(\ell)}_{x_1}
\hat \rho^{(\ell)}_{x_2} \hat \rho^{(\ell)}_{x_3}, \cr }
   \Eq(6.30)
   $$

It is clear that for $\bar \rho^{(\ell)}_x \equiv \rho_{\b,\l,\si}$, $x
\in [\L^c]^{(\ell)}$ one has $\hat \rho^{(\ell)}_x \equiv
\rho_{\b,\l,\si}$, $x \in [\L]^{(\ell)}$. Introduce an auxiliary
parameter $t \in [0,1]$ and an {\it interpolated boundary condition}
   $$
\bar \rho^{(\ell)}_x(t)=(1-t)\rho_{\b,\l,\si}+ t \bar \rho^{(\ell)}_x
\;, x \in [\L^c]^{(\ell)}
   \Eq(6.31)
   $$
Let $\hat \rho^{(\ell)}(t)$ be the solution of \equ(6.30) with the
boundary condition $\bar \rho^{(\ell)}(t)$. Then
   $$
\eqalign{
\hat \rho^{(\ell)}_x(t)&=\exp \Bigg( \b\l +\b \sum_{x_1 \in
[B_{\g}(x)]^{(\ell)}} I_{\g}^{(2)}(x,x_1) \hat \rho^{(\ell)}_{x_1}(t) \cr
&-{\b \over 3!} \sum_{x_1,x_2,x_3 \in [B_{\g}(x)]^{(\ell)}} I_{\g}^{(4)}(x,
x_1,x_2,x_3) \hat \rho^{(\ell)}_{x_1}(t) \hat \rho^{(\ell)}_{x_2}(t) \hat
\rho^{(\ell)}_{x_3}(t) \Bigg) \cr }
   \Eq(6.32)
   $$
for all $x \in [\L]^{(\ell)}$. Taking the derivative with respect to $t$
one obtains
   $$
\eqalign{
{d \over dt} \hat \rho^{(\ell)}_x(t)&=\b \hat \rho^{(\ell)}_x(t) \Bigg( 
\sum_{x_1 \in [B_{\g}(x)]^{(\ell)}} I_{\g}^{(2)}(x,x_1) {d \over dt} \hat
\rho^{(\ell)}_{x_1}(t) \cr
&-{1 \over 2!} \sum_{x_1,x_2,x_3 \in [B_{\g}(x)]^{(\ell)}}
I_{\g}^{(4)}(x, x_1,x_2,x_3) {d \over dt} \hat \rho^{(\ell)}_{x_1}(t) \hat
\rho^{(\ell)}_{x_2}(t) \hat \rho^{(\ell)}_{x_3}(t) \Bigg) \cr }
   \Eq(6.33)
   $$
Denote
   $$
I_{\g}^{(2)}(x_1, x_2| \hat \rho^{(\ell)}(t) ) = \sum_{x_3,x_4}
I_{\g}^{(4)}(x_1, x_2, x_3, x_4) \hat \rho^{(\ell)}_{x_3}(t) \hat
\rho^{(\ell)}_{x_4}(t)
   \Eq(6.34)
   $$
It is clear that 
   $$
I_{\g}^{(2)}(x_1, x_2) (\rho_{\b,\l,\si} - \z)^2 \le I_{\g}^{(2)}(x_1, x_2|
\hat \rho^{(\ell)}(t) ) \le I_{\g}^{(2)}(x_1, x_2) (\rho_{\b,\l,\si} +\z)^2
   \Eq(6.35)
   $$
as soon as $\displaystyle \max_{x} |\hat \rho^{(\ell)}_x(t)-
\rho_{\b,\l,\si}| \le \z$.

Introduce a symmetric matrix 
   $$
A(x_1, x_2| \hat \rho^{(\ell)}(t))=-I_{\g}^{(2)}(x_1, x_2) +{1 \over 2}
I_{\g}^{(2)}(x_1, x_2| \hat \rho^{(\ell)}(t) )
   \Eq(6.36)
   $$
and a diagonal matrix 
   $$
D(x_1, x_2|\hat \rho^{(\ell)}(t) ) =(\b \hat
\rho^{(\ell)}_{x_1}(t))^{-1} \ind{x_1=x_2}\; ,
   \Eq(6.36.1)
   $$
where $x_1, x_2 \in [\L]^{(\ell)}$. It is not hard to see that the
inverse matrix $B=(D - A)^{-1}$ exists if $\displaystyle \max_{x \in
[\L]^{(\ell)}} |\hat \rho^{(\ell)}_x(t)- \rho_{\b,\l,\si}| \le \z$.
Indeed, for such $\hat \rho^{(\ell)}_x(t)$ this matrix is given by a
convergent series
   $$
B=\left(\sum_{n=0}^{\infty} (D^{-1}A)^k \right) D^{-1}
   \Eq(6.37)
   $$
because 
   $$
\eqalign{
||D^{-1}A||&=\max_{x_1}\left( \b \hat \rho^{(\ell)}_{x_1}(t) 
\sum_{x_2} | A(x_1, x_2| \hat \rho^{(\ell)}(t))| \right) \cr
&\le a(\b) \cr
&<1, \cr}
   \Eq(6.38)
   $$
where $a(\b)$ is defined in \equ(2.21).  Moreover, the representation
\equ(6.37) and the fact that $A(x_1, x_2| \hat \rho^{(\ell)}(t))=0$ if
${\rm dist}\; (x_1, x_2) > \g^{-1}$ imply that
   $$
|B(x_1, x_2| \hat \rho^{(\ell)}(t))| \le {\b (\rho_{\b,\l,\si} +\z)
\over 1- a(\b)} a(\b)^{[\g {\rm dist}\; (x_1, x_2)]}
   \Eq(6.39)
   $$

Iterating \equ(6.33) and observing that ${d \over dt} \bar
\rho^{(\ell)}_x(t)=-\rho_{\b,\l,\si}+ \bar \rho^{(\ell)}_x$ for $x \in
[\L^c]^{(\ell)}$ we rewrite \equ(6.33) as
   $$
{d \over dt} \hat \rho^{(\ell)}_x(t)=\sum_{x_1 \in [\L]^{(\ell)}} 
\sum_{x_2 \in [\L^c]^{(\ell)}} B(x, x_1| \hat \rho^{(\ell)}(t))
A(x_1, x_2| \hat \rho^{(\ell)}(t)) (\bar \rho^{(\ell)}_{x_2}-
\rho_{\b,\l,\si})
   \Eq(6.40)
   $$
The right hand side of \equ(6.40) is an absolutely convergent series in
terms of $\hat \rho^{(\ell)}_x(t)$, $x \in [\L]^{(\ell)}$ and solving
this differential equation one finds $\hat \rho^{(\ell)}_x(t)$. The
solution exists at least up to $t=t_1$ at which the condition
   $$
\max_{x \in [\L]^{(\ell)}} \left|{d \over dt} \hat \rho^{(\ell)}_x(t) - 
\rho_{\b,\l,\si} \right| < \z
   \Eq(6.40.1)
   $$
is violated. Suppose that $t_1<1$, i.e.
   $$
\max_{x \in [\L]^{(\ell)}} \left|{d \over dt} \hat \rho^{(\ell)}_x(t) - 
\rho_{\b,\l,\si} \right| < \z
   \Eq(6.42)
   $$
for $t < t_1$ and
   $$
\left|{d \over dt} \hat \rho^{(\ell)}_{x_1} (t_1) - \rho_{\b,\l,\si}
\right|=\z
   \Eq(6.43)
   $$
for some $x_1 \in [\L]^{(\ell)}$. Then the representation \equ(6.40) is
valid for $t \in [0, t_1]$ and
   $$
\max_{x \in [\L]^{(\ell)}} \left|\hat \rho^{(\ell)}_x(t) -
\rho_{\b,\l,\si} \right| < \z
   \Eq(6.44)
   $$
for $t \le t_1$ as follows from the obvious identity
   $$
\hat \rho^{(\ell)}_x(t)=\rho_{\b,\l,\si} + \int_0^t {d \over ds} \hat
\rho^{(\ell)}_x(s) ds
   \Eq(6.41)
   $$
Plugging \equ(6.42)-\equ(6.44) into \equ(6.33) and using \equ(2.21.1) one
concludes that
   $$
\max_{x \in [\L]^{(\ell)}} \left|{d \over dt} \hat \rho^{(\ell)}_x(t) - 
\rho_{\b,\l,\si} \right| < \z
   \Eq(6.45)
   $$
for $t \le t_1$ which contradicts \equ(6.43). Hence 
   $$
\max_{x \in [\L]^{(\ell)}} \left|{d \over dt} \hat \rho^{(\ell)}_x(t) - 
\rho_{\b,\l,\si} \right| \le \z
   \Eq(6.46)
   $$
and
   $$
\max_{x \in [\L]^{(\ell)}} \left|\hat \rho^{(\ell)}_x(t) - 
\rho_{\b,\l,\si} \right| \le \z
   \Eq(6.47)
   $$
for all $t \in [0,1]$. Thus representation \equ(6.40) and estimate
\equ(6.39) are valid and being joined with \equ(6.41) they give
\equ(6.29).

The density configuration $\hat \rho (\L)$ was defined as the solution
of \equ(6.30) and we just checked that such a solution exists and
satisfies \equ(6.29). On the other hand, it follows from \equ(6.35) and
\equ(2.21.1) that the Hessian matrix ${\p^2 \over \p \rho^{(\ell)}_x \p
\rho^{(\ell)}_y} {\cal F}_{\g,\b,\l} (\rho^{(\ell)}(\L)| \bar
\rho^{(\ell)}(\L^c))$, $x,y \in [\L]^{(\ell)}$ is positive for any
$\rho^{(\ell)}(\L) \in (\rho_{\b,\l,\si} - \z, \rho_{\b,\l,\si} +
\z)^{\L}$ with the mass bounded from $0$ independently on
$\rho^{(\ell)}(\L)$ and $\bar \rho^{(\ell)}(\L^c)$. Thus the function
${\cal F}_{\g,\b,\l} (\cdot | \bar \rho^{(\ell)}(\L^c))$ with convex
domain $(\rho_{\b,\l,\si} - \z, \rho_{\b,\l,\si} + \z)^{\L}$ is convex
and therefore $\hat \rho (\L)$ is its unique minima. \qed 

\bigskip
\goodbreak
\centerline{{\it Variational Problem (Comparison of Two Minima)}}
\medskip
\nobreak

In this subsection we return back to the case $\l=\l(\b)$ and $\ell = \ell_1$.
First we need a stronger version of Lemma~\equ(s6.7).

\medskip\noindent
{\bf \Lemma (s6.7.1)} 

\nobreak
{\sl Consider a ${\cal D}^{(\ell_1)}$ measurable region $\L$ and take a
boundary condition $\bar \rho^{(\ell_1)}(\L^c)$ with $\displaystyle
\max_{x \in [\p \L]^{(\ell_2)}} |\bar \rho^{(\ell_2)}_x- \rho_{\b,\l,\si}|
\le \z$, where $\si$ is one of the phases, $+$ or $-$. Then the unique
minimum of ${\cal F}_{\g,\b,\l} (\rho^{(\ell_1)}(\L)| \bar
\rho^{(\ell_1)}(\L^c))$ among density configurations $\rho^{(\ell_1)}(\L)$
with $\displaystyle \max_{x \in [\L]^{(\ell_2)}} |\rho^{(\ell_2)}_x-
\rho_{\b,\l,\si}| \le \z$ is achieved on a
density configuration $\hat
\rho^{(\ell_1)}=(\hat \rho^{(\ell_1)}_x)$, $x \in [\L]^{(\ell_1)}$ such that
        $$
|\hat \rho^{(\ell_1)}_x - \rho_{\b,\l,\si}| \le {\z a(\b)^{[\g {\rm
dist}\;(x, \L^c)]} \over 1-a(\b)},
        \Eq(6.29c)
        $$
where $[ \cdot ]$ denotes the integer part.
}

\medskip\noindent
{\bf Proof.} The difference from  Lemma~\equ(s6.7)
is that we require here  closeness to $\rho_{\b,\l,\si}$ 
 only on the scale $\ell_2$,
 which is much larger than $\ell_1$.
As a consequence the variables $\rho^{(\ell_1)}_x$ may have
larger fluctuations and the free energy functional
${\cal F}_{\g,\b,\l}$ is no longer convex. The
  last argument in
 the proof of Lemma~\equ(s6.7) is then not valid anymore.
We will prove  that any minimizer
$\hat \rho^{(\ell_1)}_x$ (we do not have yet uniqueness)
satisfies the bound
       $$
|\hat \rho^{(\ell_1)}_x - \rho_{\b,\l,\si}| \le  \z
        \Eq(6e.29c)
        $$
After \equ(6e.29c) the proof becomes
 essentially  the same as for Lemma~\equ(s6.7)
and it will be omitted, we 
will just prove \equ(6e.29c).  
To this end, it is enough to show that if $ \rho^{(\ell_1)}_x 
\ge \rho_{\beta,\l,\si}+ \z$, $x \in [\L]^{(\ell_1)}$, then
   $$ 
{\p \over \partial \rho^{(\ell_1)}_x} {\cal F}_{\g,\b,\l}
(\rho^{(\ell_1)}(\L)| \bar 
\rho^{(\ell_1)}(\L^c))  >0
   \Eq(6.47.1)
   $$
and  that the reverse inequality holds if $ \rho^{(\ell_1)}_x \le
\rho_{\beta,\l,\si}- 
\z$. Since the two proofs are similar we only  consider the former.
 As in \equ(6.30) we have
   $$
{\p \over \partial \rho^{(\ell_1)}_x} {\cal F}_{\g,\b,\l}
(\rho^{(\ell_1)}(\L)| \bar 
\rho^{(\ell_1)}(\L^c))
={1 \over \b} \log   \rho^{(\ell_1 )}_x - \l -G_\g(x,\rho^{(\ell_1)})
   \Eq(6e.30)
   $$
where
     $$
G_\g(x,\rho^{(\ell_1)})= \sum_{x_1}
I_{\g,\ell_1}^{(2)}(x,x_1)   \rho^{(\ell_1)}_{x_1}   -
{1\over 3!}
\sum_{x_1,x_2,x_3} I_{\g,\ell_1}^{(4)}(x, x_1,x_2,x_3)  \rho^{(\ell_1)}_{x_1}
 \rho^{(\ell_1)}_{x_2}  \rho^{(\ell_1)}_{x_3} 
   \Eq(6e.31)
    $$
with the $ I_{\g,\ell_1}^{(n)}$ as in \equ(6.21.1) and \equ(6.21.2),
having added the subscript $\ell_1$ to specify  the scale used in the
definition. 
Since we are supposing that $ \rho^{(\ell_1)}_x 
\ge \rho_{\beta,\l,\si}+ \z$, there is $c>0$ so that
     $$
   {\log  \rho^{(\ell_1)}_x\over \beta} \ge { \log  (
\rho_{\beta,\l,\si}+ \z)\over \beta} 
\ge {\log\rho_{\beta,\l,\si}\over \beta} + {\z \over
\beta\rho_{\beta,\l,\si}} -c\z^2 
     $$
We next prove that there is $c'>0$ so that
     $$
\Big| G_\g(x,\rho^{(\ell_1)})  - [ \rho_{\beta,\l,\si} -
{\rho_{\beta,\l,\si}^3\over 3!}] \Big| \le \zeta
|1-{\rho_{\beta,\l,\si}^2\over 2}| + c'(\g^{\alpha_2}+\zeta^2) 
   \Eq(6e.32)
    $$
This is obtained by adding and subtracting 
$I_{\g,\ell_2}^{(n)}$ to $I_{\g,\ell_1}^{(n)}$ and then expanding.
In the terms containing $I_{\g,\ell_2}^{(n)}$  
 we can partial-sum the densities $\rho^{(\ell_1)}_x$ over $x$ in the
same cube  $C^{(\ell_2)}$, as $I_{\g,\ell_2}^{(n)}$ are constant 
in such cubes. We thus reconstruct  a density $\rho^{(\ell_2)}$
which, by   hypothesis,  differs
from $\rho_{\beta,\l,\si}$ at most by $\z$.
  In this way we obtain
the square bracket term in \equ(6e.32) and the terms with $\zeta$
on the r.h.s.\ of the same equation. The last term on
the r.h.s.\ of  \equ(6e.32) comes from the  the difference between the
$I_{\g}^{(n)}$'s.  We have thus proved \equ(6e.32). 

By recalling that
  $\rho_{\beta,\l,\si}$ satisfies the mean field equation, we get
   $$
{\p \over \partial \rho^{(\ell_1)}_x} {\cal F}_{\g,\b,\l}
(\rho^{(\ell_1)}(\L)| \bar 
\rho^{(\ell_1)}(\L^c))
\ge {\z \over \b \rho_{\beta,\l,\si}} - 
\z |1-{\rho_{\beta,\l,\si}^2\over 2}| - c\z^2- c'\g^{\alpha_2}
   \Eq(6e.33)
   $$
Since the coefficient that multiplies $\z$ on the r.h.s.\
of \equ(6e.33) is strictly positive (by the choice of $\b$
and $\l$), the  r.h.s.\
of \equ(6e.33) is positive
for $\g$ and $\z$ small enough. \equ(6.47.1) is thus proved and
since the proof of the reversed inequality for 
$ \rho^{(\ell_1)}_x \le \rho_{\beta,\l,\si}-
\z$ is completely similar, we can then conclude
that $ |\rho^{(\ell_1)}_x - \rho_{\beta,\l,\si}|\le
\z$.  The  proof of the lemma is now analogous to that of
   Lemma~\equ(s6.7) and it is omitted.
 \qed

\medskip
For any contour $\G$ the set 
   $$
{\rm S}=\p^{(\ell_3)} {\rm Ext}(\G)\bigcup \left( \bigcup_{m:\;
\si_m(\G)=\si(\G)} \p^{(\ell_3)} {\rm Int}_m(\G) \right)
   \Eq(6.48)
   $$
is occupied by the ground state of the phase $\si=\si(\G)$. Consider a
strip $\tilde \d^=(\G)$ of the width $\g^{-1}$ situated in the middle of
${\rm S}$. According to Lemma~\equ(s6.7) inside $\tilde \d^=(\G)$ the
density configurations minimizing ${\cal F}_{\g,\b,\l}(\rho^{(\ell_1)}|
\bar q^{(\d^= (\G))}, q_{\si}^{(\d^{\neq} (\G))}, \star)$ and ${\cal
F}_{\g,\b,\l}(\rho^{(\ell_1)}| \bar q^{(\d^= (\G))}, q_{-\si}^{(\d^{\neq}
(\G))}, \star)$ differ from $\rho_{\b,\si}$ by at most $(1-a(\b))^{-1}
\z a(\b)^{[\g {1 \over 3} \g^{-1-\a_3}]}$. Therefore the density
configuration which minimizes ${\cal F}_{\g,\b,\l}(\rho^{(\ell_1)}| \bar
q^{(\d^= (\G))}, q_{\si}^{(\d^{\neq} (\G))}, \star)$ under the additional
condition that this configuration coincides with $\rho_{\b,\si}$ in
$\tilde \d^=(\G)$ gives the value of the true minima of ${\cal
F}_{\g,\b,\l}(\rho^{(\ell_1)}| \bar q^{(\d^= (\G))}$, $q_{\si}^{(\d^{\neq}
(\G))}, \star)$ up to nonessential error. The same is true for the density
configuration minimizing ${\cal F}_{\g,\b,\l}(\rho^{(\ell_1)}| \bar
q^{(\d^= (\G))}, q_{-\si}^{(\d^{\neq} (\G))}, \star)$ with the same
additional condition. Obviously these modified minimizing configurations
coincide with each other not only in $\tilde \d^=(\G)$, where they both
are equal to $\rho_{\b,\si}$, but in the whole part of ${\rm S}$
stretching from $\d^=(\G)$ to $\tilde \d^=(\G)$, i.e. in
   $$
{\rm S}_1=\p^{(\ell_3/2)} {\rm Ext}(\G)\bigcup \left( \bigcup_{m:\;
\si_m(\G)=\si(\G)} \p^{(\ell_3/2)} {\rm Int}_m(\G) \right)\bigcup \tilde
\d^=(\G)
   \Eq(6.48.1)
   $$
Estimating from below the difference between the minimal value of ${\cal
F}_{\g,\b,\l}(\rho^{(\ell_1)}| \bar q^{(\d^= (\G))},$
$q_{-\si}^{(\d^{\neq} (\G))}, \star)$ and the minimal value of ${\cal
F}_{\g,\b,\l}(\rho^{(\ell_1)}| \bar q^{(\d^= (\G))}, q_{\si}^{(\d^{\neq}
(\G))}, \star)$ the contributions corresponding to ${\rm S}_1$ cancel each
other. Hence the initial variational problem in ${\rm Supp}(\G) \setminus
\d (\G)$ with a general boundary condition imposed on $\d(\G)$ is reduced
to a similar problem in a smaller volume ${\rm Supp}(\G) \setminus {\rm
S}_1$. 

For the sake of notational simplicity from now on we suppose that the
boundary condition is a standard one already for the initial problem,
i.e. it is equal to $\rho_{\b,\si}$ in $\d^=(\G)$. Thus all we need to
finish the Peierls estimate is the lower bound for
        $$
\eqalign{
\D{\cal F}_{\g,\b,\l}(\rho^{(\ell_1)},\G)= \ell_1^d \hskip-2em
 \sum_{x \in [{\rm Supp}(\G) \setminus \d (\G)]^{(\ell_1)}} 
&\left( {\cal I}_{\g,\ell_1} (x,\rho^{(\ell_1)})
- \l {\cal R}_{\g,\ell_1}(x,\rho^{(\ell_1)}) 
- {1\over 2!}{\cal R}_{\g,\ell_1}(x,\rho^{(\ell_1)})^2 \right. \cr
&+\left. {1\over 4!}{\cal R}_{\g,\ell_1}(x,\rho^{(\ell_1)})^4 -
F_{\b,\l}(\rho_{\b,\si}) \right) \cr}
        \Eq(6.49)
        $$
We note that for $x$ such that ${\rm dist}\;(x, \d(\G)) \le \g^{-1}$ the
values of ${\cal I}_{\g,\ell} (x,\rho^{(\ell)})$ and ${\cal R}_{\g,\ell}
(x,\rho^{(\ell)})$ depend on the boundary condition which is equal to
$\rho_{\b,\si}$ in $\d^=(\G)$ and $\rho_{\b,-\si}$ in
$\d^{\neq}(\G)$.

\medskip\noindent{\bf \Lemma (s6.8)} 

{\sl For any contour $\G=({\rm Supp}(\G), \eta^{\G})$
   $$
\min_{\rho^{(\ell_1)}:\; \eta(\rho^{(\ell_1)})=\eta^{\G}} \D{\cal
F}_{\g,\b,\l}(\rho^{(\ell_1)},\G) \ge c \ell_2^d \ell_3^{-d} \z^2 |{\rm
Supp}(\G)|
        \Eq(6.50)
        $$
}

\medskip\noindent
{\bf Proof.} Rewrite $\D{\cal F}_{\g,\b,\l}(\rho^{(\ell_1)},\G)$ in the
form 
        $$
\eqalign{
\D{\cal F}_{\g,\b,\l}(\rho^{(\ell_1)},\G)= \ell_1^d 
\sum_{x \in [{\rm Supp}(\G) \setminus \d (\G)]^{(\ell_1)}}
&\left( F_{\b,\l}({\cal R}_{\g,\ell_1}(x,\rho^{(\ell_1)}) ) -
F_{\b,\l}(\rho_{\b,\si}) 
\phantom{\rho^{(\ell_1)}_{x_1}\over \b}\right. \cr
&+ b_{\g,\ell_1}^{-1} \sum_{x_1 \in [B_{\g}(x)]^{(\ell_1)}}
{\rho^{(\ell_1)}_{x_1}\over \b} \log\rho^{(\ell_1)}_{x_1} \cr
& - \left. {{\cal R}_{\g,\ell_1}(x,\rho^{(\ell_1)}_{\phantom{x}}) \over
\b} \log {\cal R}_{\g,\ell_1}(x,\rho^{(\ell_1)}) \right) \cr}
        \Eq(6.51)
        $$
and observe that
        $$
 F_{\b,\l}({\cal R}_{\g,\ell_1}(x,\rho^{(\ell_1)}) ) -
F_{\b,\l}(\rho_{\b,\si}) \ge 0 
        \Eq(6.51.1)
        $$
by definition of $\rho_{\b,\si}$ and
   $$
b_{\g,\ell_1}^{-1} \sum_{x_1 \in [B_{\g}(x)]^{(\ell_1)}}
{\rho^{(\ell_1)}_{x_1}\over \b} \log\rho^{(\ell_1)}_{x_1} \ge {{\cal
R}_{\g,\ell_1}(x,\rho^{(\ell_1)}) \over \b} \log {\cal
R}_{\g,\ell_1}(x,\rho^{(\ell_1)})
        \Eq(6.52)
        $$
by convexity.

We call $C^{(\ell_2)}_x \in {\rm Supp}(\G)$ a {\it wrong box} if either
$\eta^{\G}_x=0$ or $\eta^{\G}_x \eta^{\G}_{x_1}=-1$ for at least one
cube $C^{(\ell_2)}_{x_1}$ adjacent to $C^{(\ell_2)}_{x}$. According to
the definition of the contour there exist at least $(3\ell_3)^{-d}|{\rm
Supp}(\G)|$ wrong boxes $C^{(\ell_2)}_x \in {\rm Supp}(\G)$ such that
they are at the distance longer than $5\g^{-1}$ from each other. In
particular $C^{(2\g^{-1})}_{x_1} \cap C^{(2\g^{-1})}_{x_2}= \emptyset$
for any two such boxes $C^{(\ell_2)}_{x_1}$ and
$C^{(\ell_2)}_{x_2}$.

Consider a wrong box $C^{(\ell_2)}_x$ for which $\eta^{\G}_x=0$. If
inside $C^{(2\g^{-1})}_x$ there exist at least $\ell_1^{-d}\ell_2^d$
points $x_1$ with
        $$
{\cal R}_{\g,\ell_1}(x_1,\rho^{(\ell_1)}) \not \in 
\left(\rho_{\b,-}-{\z \over 2},\;  \rho_{\b,-}+{\z \over 2}\right)
\bigcup
\left(\rho_{\b,+}-{\z \over 2},\;  \rho_{\b,+}+{\z \over 2}\right)
        \Eq(6.53)
        $$
then $C^{(2\g^{-1})}_x$ contributes to $\D{\cal
F}_{\g,\b,\l}(\rho^{(\ell_1)},\G)$ by at least 
        $$
c \ell_2^d \z^2 \min \big(F''_{\b,\l}(\rho_{\b,-}),
F''_{\b,\l}(\rho_{\b,+}) \big)
        \Eq(6.54)
        $$
This contribution comes from the terms 
        $$
\ell_1^d F_{\b,\l}({\cal R}_{\g,\ell_1}(x_1,\rho^{(\ell_1)}) ) -
\ell_1^d F_{\b,\l}(\rho_{\b,\si})
        \Eq(6.54.1)
        $$
in \equ(6.51).

In the opposite situation when
        $$
{\cal R}_{\g,\ell_1}(x_1,\rho^{(\ell_1)}) \not \in 
\left(\rho_{\b,-}-{\z \over 2},\;  \rho_{\b,-}+{\z \over 2}\right)
\bigcup
\left(\rho_{\b,+}-{\z \over 2},\;  \rho_{\b,+}+{\z \over 2}\right)
        \Eq(6.55)
        $$
for not more than $\ell_1^{-d}\ell_2^d$ points $x_1 \in
C^{(2\g^{-1})}_x$ we extract a contribution similar to \equ(6.54) from
the terms
   $$
\ell_1^d b_{\g,\ell_1}^{-1} \sum_{x_2 \in [B_{\g}(x_1)]^{(\ell_1)}}
{\rho^{(\ell_1)}_{x_2}\over \b} \log\rho^{(\ell_1)}_{x_2} - \ell_1^d {{\cal
R}_{\g,\ell_1}(x_1,\rho^{(\ell_1)}) \over \b} \log {\cal
R}_{\g,\ell_1}(x_1,\rho^{(\ell_1)})
        \Eq(6.56)
        $$
in \equ(6.51). 

Consider $x_1 \in C^{(2\g^{-1})}_x$ such that $C^{(\ell_2)}_x \in
[B_{\g}(x_1)]^{(\ell_1)}$ and observe that 
        $$
\rho^{(\ell_2)}_x=||[C^{(\ell_2)}_x]^{(\ell_1)}||^{-1} \sum_{x_2 \in
[C^{(\ell_2)}_x]^{(\ell_1)}} \rho^{(\ell_1)}_{x_2} 
        \Eq(6.57)
        $$
does not belong to $\left(\rho_{\b,-}-{\z},\;
\rho_{\b,-}+{\z}\right) \cup \left(\rho_{\b,+}-{\z},\;
\rho_{\b,+}+{\z}\right)$ as $\eta^{\G}_x=0$. Denote
   $$
{\cal R}_{x_1}= ||[B_{\g}(x_1)]^{(\ell_1)} \setminus
[C^{(\ell_2)}_x]^{(\ell_1)}||^{-1} \sum_{x_2 \in [B_{\g}(x_1)]^{(\ell_1)}
\setminus [C^{(\ell_2)}_x]^{(\ell_1)}} \rho^{(\ell_1)}_{x_2}
        \Eq(6.58)
        $$
Then by convexity
        $$
\eqalign{
\sum_{x_2 \in [B_{\g}(x_1)]^{(\ell_1)}} \rho^{(\ell_1)}_{x_2}
\log\rho^{(\ell_1)}_{x_2} &\ge ||[B_{\g}(x_1)]^{(\ell_1)} \setminus
[C^{(\ell_2)}_x]^{(\ell_1)}|| {\cal R}_{x_1} \log {\cal R}_{x_1} \cr
&+ ||[C^{(\ell_2)}_x]^{(\ell_1)}|| \rho^{(\ell_2)}_x \log
\rho^{(\ell_2)}_x \cr}
        \Eq(6.59)
        $$
implying the lower bound
        $$
\eqalign{
\ell_1^d {||[B_{\g}(x_1)]^{(\ell_1)} \setminus [C^{(\ell_2)}_x]^{(\ell_1)}||
\over \b b_{\g,\ell_1} } {\cal R}_{x_1} \log {\cal R}_{x_1} 
&+ \ell_1^d {||[C^{(\ell_2)}_x]^{(\ell_1)}|| \over \b b_{\g,\ell_1} }
\rho^{(\ell_2)}_x \log \rho^{(\ell_2)}_x \cr
&- \ell_1^d {{\cal R}_{\g,\ell_1}(x_1,\rho^{(\ell_1)}) \over \b} \log {\cal
R}_{\g,\ell_1}(x_1,\rho^{(\ell_1)}) \cr}
        \Eq(6.60)
        $$
for \equ(6.56). Now we apply the inequality
        $$
(1-\a)a\log a- \a b \log b -\big( (1-\a)a +\a b \big) \log \big( (1-\a)a
+\a b \big) \ge {\a \over 2} {(a-b)^2 \over \max (a, b)} -\a^2 {(a-b)^2
\over a} 
        \Eq(6.61)
        $$
which is true for any $a,b >0$ and $0 < \a <1$. This leads to the lover
bound
        $$
\ell_1^d {||[C^{(\ell_2)}_x]^{(\ell_1)}|| \over 2 \b b_{\g,\ell_1} }
{\left({\cal R}_{x_1} - \rho^{(\ell_2)}_x \right)^2 \over \max
\left({\cal R}_{x_1}, \rho^{(\ell_2)}_x \right)} - \ell_1^d
\left({||[C^{(\ell_2)}_x]^{(\ell_1)}|| \over  \b b_{\g,\ell_1}} \right)^2
{\left({\cal R}_{x_1} - \rho^{(\ell_2)}_x \right)^2 \over {\cal R}_{x_1}}
        \Eq(6.62)
        $$
for \equ(6.60). Observing that $( b_{\g,\ell_1})^{-1}
||[C^{(\ell_2)}_x]^{(\ell_1)}||$ has the order $\g^{\a_2 d}$ we conclude
that \equ(6.62) exceeds
        $$
c \ell_1^d {||[C^{(\ell_2)}_x]^{(\ell_1)}|| \over \b b_{\g,\ell_1}\;
\rho_{\b,-}} 
\left( {\cal R}_{\g,\ell_1}(x_1,\rho^{(\ell_1)}) - \rho^{(\ell_2)}_x
\right)^2 
\ge c \ell_1^d \g^{\a_2 d} {\z^2 \over \b \rho_{\b,-}}
        \Eq(6.63)
        $$
for $\g$ sufficiently small. The number of points $x_1 \in
C^{(2\g^{-1})}_x$ for which \equ(6.63) is true is not less than 
$b_{\g,\ell_1}/2$. Remind that for these points 
        $$
{\cal R}_{\g,\ell_1}(x_1,\rho^{(\ell_1)}) \in 
\left(\rho_{\b,-}-{\z \over 2},\;  \rho_{\b,-}+{\z \over 2}\right)
\bigcup
\left(\rho_{\b,+}-{\z \over 2},\;  \rho_{\b,+}+{\z \over 2}\right)
        \Eq(6.64)
        $$
and $[B_{\g}(x_1)]^{(\ell_1)} \ni C^{(\ell_2)}_x $. Thus \equ(6.63)
again gives us a lower bound
        $$
c \ell_2^d \z^2 (\b \rho_{\b,-})^{-1}
        \Eq(6.65)
        $$
similar to \equ(6.54).

Finally for the wrong box $C^{(\ell_2)}_x$ of the second type, i.e. when
$\eta^{\G}_x \eta^{\G}_{x_1}=-1$ for an adjacent box
$C^{(\ell_2)}_{x_1}$, we can consider $C^{(\ell_2)}_x \cup
C^{(\ell_2)}_{x_1}$ instead of $C^{(\ell_2)}_x$ and repeat all the
arguments above. They will work perfectly because
        $$
||[C^{(\ell_2)}_x]^{(\ell_1)} \cup [C^{(\ell_2)}_{x_1}]^{(\ell_1)}||^{-1} 
\sum_{x_2 \in [C^{(\ell_2)}_x]^{(\ell_1)} \cup
[C^{(\ell_2)}_{x_1}]^{(\ell_1)}} \rho^{(\ell_1)}_{x_2} 
        \Eq(6.66)
        $$
is again outside $\left(\rho_{\b,-}-{\z},\;
\rho_{\b,-}+{\z}\right) \cup \left(\rho_{\b,+}-{\z},\;
\rho_{\b,+}+{\z}\right)$. \qed

\medskip
To finish the proof of the Peierls estimate we need to compare all the
errors which are at most of the order $c \ell_1 \g |{\rm Supp}(\G)|= c
\g^{\a_1} |{\rm Supp}(\G)|$ (see \equ(6.20) and \equ(6.21)) with the
contribution coming from Lemma~\equ(s6.8). The last is of order $c
\ell_2^d \ell_3^{-d}|{\rm Supp}(\G)|= c \g^{d\a_2+d\a_3} |{\rm
Supp}(\G)|$ and dominates $c \g^{\a_1} |{\rm Supp}(\G)|$ since
$d\a_2+d\a_3 <\a_1$.

\bigskip
\bigskip
\goodbreak 
\centerline{{\bf 5. Auxiliary Model }}
\bigskip
\numsec= 5
\numfor= 1
\numtheo=1
In this section we study metastable models and we prove that the
corresponding measures satisfy the Dobrushin uniqueness condition and
hence exhibit an exponential decay of correlations. We note again that the
configuration $q^{(\L)}$ in a ${\cal D}^{(\ell_2)}$ measurable region $\L$
belongs to the ground state ensemble of the phase $\si$ iff in every cube
$C_x^{(\ell_2)} \in \L$ the density of particles
   $$
\rho^{(\ell_2)}_x(q^{(\L)})=\ell_2^{-d} \left| q^{(\L)} \cap
C^{(\ell_2)}_x \right|
   \Eq (4.1)
   $$
belongs to the interval 
   $$
(\rho_{\b,\si} - \z, \rho_{\b,\si} + \z)
   \Eq (4.2)
   $$
The partition function of the auxiliary model is given by \equ(3.8),
i.e. it is the integral over ground state configurations of the
corresponding contour partition function.

In the previous section we have shown that some partition functions
initially defined in terms of particle configurations can be approximated
by partition functions written in terms of density configurations related
to some scale $\ell$. Now we go further and show that the auxiliary model
for each of two phases can be {\it equivalently} rewritten in terms of
density configurations. Such an equivalent model is defined on the lattice
$\Z^d_{\ell_2}$ with the density variables $\rho^{(\ell_2)}_x$ taking
discrete values $n \ell_2^{-d}$,
$n=1,2, \ldots$ from the bounded interval \equ(4.2). The
corresponding Hamiltonian is of infinite range but with sufficiently fast
decaying interactions.  Quite naturally this Hamiltonian is close to
\equ(6.23) (with $\ell=\ell_2$) and it can be understood as a small
perturbation of a positive definite quadratic form. The treatment of
equivalent model is based on a specific approach [COPP] to the Dobrushin
uniqueness theorem developed initially for unbounded lattice spin systems.

{From} now on we fix $\ell_2$ as the scale at which we define density
configurations. Thus all regions are assumed to be ${\cal D}^{(\ell_2)}$
measurable and we often drop the superscript $(\ell_2)$ from notations.

In the next subsection we construct the effective Hamiltonian the exact
form of which is stated in Lemma~\equ(s4.2) at the very end of
subsection. Then in the last two subsections we prove Lemma~\equ(s4.3)
saying that the effective Hamiltonian satisfies the Dobrushin uniqueness
condition.

\bigskip
\goodbreak
\centerline{{\it Reduction to Density Model.}}
\medskip
\nobreak
The technical part of the reduction is based on the cluster or polymer
expansion technique. For the convenience of the reader Section~7 quotes
a version of the general cluster expansion theorem which is suitable for
our purposes. 

Consider a region $\L$ with the boundary condition $\bar q^{(\L^c)}$
belonging to the ground state ensemble of the phase $\si$. For the
partition function $Z^A_{\g,\b,\l}(\L|\bar q^{(\L^c)})$ we decompose the
integral in \equ(3.8) into a sum of integrals. In this decomposition the
external sum is taken over all density configurations $\rho_x,\; x \in
[\L]^{(\ell_2)}$ satisfying \equ(4.2). Given such a density configuration
$\rho(\L)$ an internal integral is taken over all particle configurations
$q^{(\L)}$ such that $\rho^{(\ell_2)}_x(q^{(\L)})= \rho^{(\ell_2)}_x(\L)$
for all $x \in [\L]^{(\ell_2)}$. The reduction we perform is nothing but
the calculation of
   $$
\log \int_{{\cal Q}^{(\L)}} dq^{(\L)} \; 
\ind{\rho^{(\ell_2)}(q)= \rho(\L)} \;\;
e^{-\b H_{\g,\l}(q^{(\L)}|\bar q^{(\L^c)})}
\sum_{ \{\G_i\}^{\si} \in \L} 
\prod_i W^T(\eta^{\G_i}| q^{(\d^=(\G_i))})
   \Eq(4.3)
   $$
as a function of $\rho_x(\L)$. Denote by $q_{x,i}$ the particles of
$q^{(\L)}$ situated inside $C_x^{(\ell_2)}$ and set $n_x= \ell_2^d \rho_x=
\left| q^{(\L)} \cap C^{(\ell_2)}_x \right|$. Then $\int_{{\cal Q}^{(\L)}}
dq$ is
   $$
\eqalign{
\left( \prod_{x \in [\L]^{(\ell_2)}} 
{\ell_2^{d n_x} \over n_x !}\; \ell_2^{-d n_x} \right)
\int  \ldots \int &\left( \prod_{x \in [\L]^{(\ell_2)}} 
\prod_{i=1}^{n_x} \ind{q_{x,i} \in C^{(\ell_2)}_x}\; dq_{x,i} \right) \;
e^{-\b H_{\g,\l}(q^{(\L)}|\bar q^{(\L^c)})} \cr
&\times \sum_{ \{\G_i\}^{\si} \in \L} 
\prod_i W^T(\eta^{\G_i}| q^{(\d^=(\G_i))}),}
   \Eq(4.4)
   $$
where the integral can be understood as an expectation with respect to
the system of independent particles $q_{x,i}$ uniformly distributed
in the corresponding boxes $C^{(\ell_2)}_x$. 

The Hamiltonian $H_{\g,\l}(q^{(\L)}|\bar q^{(\L^c)})$ can be decomposed into
the sum of a $\rho$-dependent Hamiltonian
   $$
H_{\g,\l}(\rho^{(\ell_2)}(q^{(\Lambda)})| \bar q^{(\Lambda^c)})
=H_{\g,\l}([q^{(\Lambda)}]^{(\ell_2)}| \bar q^{(\Lambda^c)})
=H_{\g,\l}(\rho(\L)| \bar q^{(\Lambda^c)})
   \Eq(4.5)
   $$
and error terms
   $$
\D H_{\g,\l}(q^{(\L)}|\bar q^{(\L^c)})=
H_{\g,\l}(q^{(\L)}|\bar q^{(\L^c)}) -
H_{\g,\l}(\rho^{(\ell_2)}(q^{(\Lambda)})| \bar q^{(\Lambda^c)})
   \Eq(4.6)
   $$
Clearly $\exp \left( -\b H_{\g,\l}(\rho| \bar q^{(\Lambda^c)} )
\right)$ can be taken outside the integral in \equ(4.4) leaving us with
the calculation of the $\log$ of the partition function
   $$
\eqalign{
\left( \prod_{x \in [\L]^{(\ell_2)}} \ell_2^{-d n_x} \right)
\int \ldots \int &\left( \prod_{x \in [\L]^{(\ell_2)}} 
\prod_{i=1}^{n_x} \ind{q_{x,i} \in C^{(\ell_2)}_x}\; dq_{x,i} \right) \;
e^{-\b \D H_{\g,\l}(q^{(\L)}|\bar q^{(\L^c)})} \cr
&\times \sum_{ \{\G_i\}^{\si} \in \L} 
\prod_i W^T(\eta^{\G_i}| q^{(\d^=(\G_i))})}
   \Eq(4.7)
   $$
where the first product is included for the convenience of treating this
partition function as an expectation over a system of independent
particles.

Observe that the error part of the Hamiltonian is given by
   $$
\D H_{\g,\l}(q^{(\L)}|\bar q^{(\L^c)})=
- \sum_{q_{i_1},q_{i_2} \in q}
\D J_{\g}^{(2)}(q_{i_1},q_{i_2})
+\sum_{q_{i_1},q_{i_2},q_{i_3},q_{i_4} \in q}
\D J_{\g}^{(4)}(q_{i_1},q_{i_2},q_{i_3},q_{i_4})
   \Eq(4.8)
   $$
where $\D J_{\g}^{(2)}(q_{i_1},q_{i_2})$ and $\D
J_{\g}^{(4)}(q_{i_1},q_{i_2},q_{i_3},q_{i_4})$ are much smaller than $1$.
The estimates
   $$
0 < \D J_{\g}^{(2)}(q_{i_1},q_{i_2}) <   c \g^{\a_2}
J_{\g}^{(2)}(q_{i_1},q_{i_2}) 
   \Eq(4.9)
   $$
and
   $$
0 < \D J_{\g}^{(4)}(q_{i_1},q_{i_2},q_{i_3},q_{i_4})  <   c \g^{\a_2} 
J_{\g}^{(4)}(q_{i_1},q_{i_2},q_{i_3},q_{i_4})
   \Eq(4.10)
   $$
are true unless some of the interacting particles are at the distance
larger than $\g^{-1}-\g^{-1+\a_2}$ from each other. In the last case $\D
J_{\g}^{(2)}(q_{i_1},q_{i_2})$ and $\D
J_{\g}^{(4)}(q_{i_1},q_{i_2},q_{i_3},q_{i_4})$ are extremely small
   $$
0 < \g^{-d}\D J_{\g}^{(2)}(q_{i_1},q_{i_2}),\quad
\g^{-3d}\D J_{\g}^{(4)}(q_{i_1},q_{i_2},q_{i_3},q_{i_4}) 
< c \g^{3d/2 -\a_2 /2}
   \Eq(4.11)
   $$
Denoting 
   $$
w_{\g}^{(2)}(q_{i_1},q_{i_2})=e^{\b \D J_{\g}^{(2)}(q_{i_1},q_{i_2})}
-1
   \Eq(4.12)
   $$
and 
   $$
w_{\g}^{(4)}(q_{i_1},q_{i_2},q_{i_3},q_{i_4})
=e^{-\b \D J_{\g}^{(4)}(q_{i_1},q_{i_2},q_{i_3},q_{i_4})}-1
   \Eq(4.13)
   $$
we have
   $$
e^{-\b \D H_{\g,\l}(q^{(\L)}|\bar q^{(\L^c)})}=
\prod_{q_{i_1},q_{i_2} \in q}
\left(1 + w_{\g}^{(2)}(q_{i_1},q_{i_2}) \right)
\prod_{q_{i_1},q_{i_2},q_{i_3},q_{i_4} \in q}
\left( 1 + w_{\g}^{(4)}(q_{i_1},q_{i_2},q_{i_3},q_{i_4})\right)
   \Eq(4.14)
   $$

First we obtain the polymer expansion for the $\log$ of the partition
function
   $$
\left( \prod_{x \in [\L]^{(\ell_2)}} \ell_2^{-d n_x} \right)
\int \ldots \int \left( \prod_{x \in [\L]^{(\ell_2)}} 
\prod_{i=1}^{n_x} \ind{q_{x,i} \in C^{(\ell_2)}_x}\; dq_{x,i} \right) \;
e^{-\b \D H_{\g,\l}(q^{(\L)}|\bar q^{(\L^c)})}
   \Eq(4.14.1)
   $$
containing no contours. Opening all brackets in \equ(4.14) we rearrange the
expression under the integral in \equ(4.14.1) in the following way.

Let {\it 2-link}, $L^{(2)}=(q_1, q_2)$, be a couple of particles $q_1,
q_2$ such that $W(L^{(2)})=w_{\g}^{(2)}(q_1,q_2) \not = 0$. Similarly a
{\it 4-link}, $L^{(4)}=(q_1,q_2,q_3,q_4)$, is a quadruple of particles
$q_1, q_2, q_3, q_4$ such that $W(L^{(4)})=w_{\g}^{(4)}(q_1,q_2,q_3,q_4)
\not = 0$. The quantities $W(L^{(2)})$ and $W(L^{(4)})$ are called the
statistical weights of the 2-link and 4-link respectively. Two links are
{\it connected} if they have a common particle. We stress that if two
links have no common particles but have a common space point occupied by
one particle from the first link and another particle from the second
link then these links are not connected. A {\it pre-diagram}, $\t$, is a
connected set of links. Denote by $q(\t)=(q_i(\t))$ the particles of
$q^{(\L)}$ which influence $\t$, i.e. the endpoints of links of $\t$.
The {\it statistical weight}, $w(\t)$, of the pre-diagram is the product
of the statistical weights of the contributing links. Two pre-diagrams
are {\it compatible} if they are not connected. Finally, a {\it
compatible collection of pre-diagrams} consists of mutually compatible
pre-diagrams.

The definitions above justify the representation for \equ(4.14.1) of the
form 
   $$
\left( \prod_{x \in [\L]^{(\ell_2)}} \ell_2^{-d n_x} \right)
\int \ldots \int \left( \prod_{x \in [\L]^{(\ell_2)}} 
\prod_{i=1}^{n_x} \ind{q_{x,i} \in C^{(\ell_2)}_x}\; dq_{x,i} \right) \;
\sum_{\{\t_j\} \not \in \L^c} \prod_j w(\t_j),
   \Eq(4.16)
   $$
where the sum goes over compatible collections, $\{\t_j\}$, of
pre-diagrams and $\{\t_j\} \not \in \L^c$ means that every $\t_j$ has at
least one particle inside $\L$.

We call two pre-diagrams {\it equivalent} if they can be transformed one
into another by shifting some particles such that every shifted particle
$q_{x,i}$ remains in its initial box $C_x^{(\ell_2)}$. The corresponding
equivalence classes, $\T$, are called {\it diagrams}. To have a
geometrical interpretation of the diagram $\T$ we identify it with the
pre-diagram $\t \in \T$ having all particles at the centers of the
corresponding cubes $C_x^{(\ell_2)}$. We say that two diagrams $\T_1$ and
$\T_2$ are compatible if any $\t_1 \in \T_1$ and $\t_2 \in \T_2$ are
compatible. Setting
   $$
W(\T)=\ell_2^{-d |q(\T)|} \int_{\t \in \T} w(\t)
   \Eq(4.17)
   $$
we rewrite the partition function \equ(4.14.1) in so called cluster form
   $$
\sum_{\{\T_j\} \not \in \L^c} \prod_j W(\T_j),
   \Eq(4.18)
   $$
where the sum is extended to all compatible collection of diagrams. The
transition from \equ(4.16) to \equ(4.18) relies on the fact that for
compatible $\t_{j_1}$ and $\t_{j_2}$ the corresponding sets of particles
$q(\t_{j_1})$ and $q(\t_{j_2})$ do not intersect each other.

\medskip\noindent{\bf \Lemma (s4.1)} 

{\sl Let 
   $$
a(\T)=|q(\T)|
   \Eq(4.19)
   $$
Then 
  $$
\sum_{\T' \not \sim \T} |W(\T')|e^{a(\T')} \le a(\T),
   \Eq(4.20)
   $$
where $\T' \not \sim \T$ denotes a diagram $\T'$ not compatible with a
given diagram $\T$.
}

\medskip\noindent
{\bf Proof.} We say that a diagram $\T$ is not compatible with a
particle $q$ and we denote it $\T \not \sim q$ if $q \in q(\T)$. {}From our
definition of compatibility of diagrams it is clear that \equ(4.20)
follows from
   $$
\sum_{\T \not \sim q} |W(\T)|e^{a(\T)} \le 1
   \Eq(4.21)
   $$
Remind that for each box $C_x^{(\ell_2)}$, $x \in [\L^{(\ell_2)}]$ the
number of particles inside $C_x^{(\ell_2)}$ is fixed as the density
configuration $\rho$ is fixed. Since $\rho$ belongs to interval
\equ(4.2) there are at most $c (\rho_{\b, \l, \si}+ \z) \g^{-d}$
$2$-links and at most $c (\rho_{\b, \l, \si}+ \z)^3 \g^{-3d}$ $4$-links
passing through any given particle. The statistical weights of links
satisfy
   $$
|W(L^{(2)})| \le c\g^{\a_2}\g^d
   \Eq(4.23)
   $$
and
   $$
|W(L^{(4)})| \le c\g^{\a_2}\g^{3d}
   \Eq(4.24)
   $$
as follows from \equ(4.9)-\equ(4.11) and \equ(2.4)-\equ(2.5). 

We provide a diagram with an abstract tree structure according to the
following algorithm. The root of the tree is the particle $q$. Links
which start from $q$ are called links of the first level. Links which
start at endpoints of the links of the first level and are different
from them are the links of the second level. Generally, links which
starts at the endpoints of $n$-th level links and are different from all
links of levels $1,2,\ldots,n$ are called links of level $n+1$.

Denote by $n(\T)$ the maximal level of links in $\T$. It is clear that
for $\g$ small enough
   $$
\sum_{\T \not \sim q:\;n(\T)=1 } |W(\T)|e^{a(\T)} \le 
\prod_{L^{(2)} \ni q} \Big(1+ e^2 |W(L^{(2)})|\Big)
\prod_{L^{(4)} \ni q} \Big(1+ e^4 |W(L^{(4)})|\Big) -1
\le 1
   \Eq(4.24.1)
   $$
By induction suppose that
   $$
\sum_{\T \not \sim q:\;n(\T)\le N } |W(\T)|e^{a(\T)} \le 1
   \Eq(4.24.2)
   $$
and consider $\T$ with $n(\T) \le N+1$. Take a link of the first level
in such $\T$. {}From every non root endpoint of this link ``grows'' a
subdiagram $\T_1$ with $n(\T_1) \le N$. Hence
   $$
\eqalign{
\sum_{\T \not \sim q:\;n(\T)\le N+1 } |W(\T)|e^{a(\T)} &\le 
\prod_{L^{(2)} \ni q} \Big(1+ |W(L^{(2)})| (e+1)^2 \Big) \cr
&\times \prod_{L^{(4)} \ni q} \Big(1+ |W(L^{(4)})| (e+ 1)^4 \Big) -1 \cr
&\le 1 \cr}
   \Eq(4.24.3)
   $$
Here $e$ correspond to the case when nothing is ``growing'' from a given
endpoint of the first level link while $1$ is the inductive estimate 
for the case when nonempty subdiagram $\T_1$ with $n(\T_1) \le N$ is
``growing'' from this endpoint. \qed 

\medskip
{From} Lemma~\equ(s4.1) applying Theorem~\equ(sA.1) one obtains the polymer
expansion
   $$
\sum_{\pi \not \in \L^c} W(\pi)
   \Eq(4.26)
   $$
for the $\log$ of \equ(4.18). The precise definition of the polymer
$\pi=[\T_j^{\e_j}]$ and its statistical weight $W(\pi)$ can be found in
Section~7. Geometrically a polymer is again a diagram-like object probably
with some links entering it more than once. We underline that constructing
pre-diagrams, diagrams and polymers one considers all particles entering
these object as distinct, say having unique indices or labels. For that
reason we call polymers from \equ(4.26) {\it labeled polymers}. To clarify
the dependence of \equ(4.26) on $\rho_x$ we perform another
factorization and define unlabeled polymers.

Suppose that from the total $n_{x}=\rho_{x} \ell_2^d$ particles situated
inside a box $C^{(\ell_2)}_x$ exactly $k(\pi)$ particles contribute to the
labeled polymer $\pi$. Replacing these $k(\pi)$ particles with another
$k(\pi)$ particles from the same box $C^{(\ell_2)}_x$ one obtains
different labeled polymer with the same statistical weight. Two labeled
polymers which can be transformed one into another after several
replacements, possibly taking place in different boxes, are called {\it
equivalent}. The corresponding equivalence classes are called {\it
unlabeled polymers} and are denoted $\tau$. In other words, unlabeled
polymer is obtained from a labeled one by dropping out labels of
particles.

Denote by $X(\tau) \subseteq [\L]^{(\ell_2)}$ the set of the centers of
all boxes $C^{(\ell_2)}_x$ containing particles from $\tau$. For $x \in
X(\tau)$ let $k_x(\tau)$ be the number of particles from $C^{(\ell_2)}_x$
contributing to $\tau$. Then the total number of different labeled
polymers $\pi \in \tau$ is a polynomial function of $\rho_x
\ell_2^d$, $x \in X(\tau)$
   $$
0< P(\tau) \le \prod_{x \in X(\tau)}(\rho_x \ell_2^d)^{k_x(\tau)}
   \Eq(4.27)
   $$
Setting $W(\tau)=W(\pi)$, where $\pi$ is an arbitrary labeled polymer from
$\tau$, we obtain the expression for the log of the partition function
\equ(4.14.1) 
   $$
\sum_{\tau \in \L} W(\tau) P(\tau)
   \Eq(4.28)
   $$
written in terms of $\rho_x$. Despite its involved structure we
need only few simple estimates on this sum. 

As follows from Corollary~\equ(sA.2) in Section~7 the sum of statistical
weights of all labeled (and hence unlabeled) polymers passing through
given particle and containing not less than $k$ links does not exceed
$\g^{-k \a_2 /2}$. Hence the sum of the statistical weights of all
labeled (or unlabeled) polymers containing two given particles $q_1$ and
$q_2$ on the distance $r> \g^{-1}$ from each other 
   $$
\sum_{\pi \ni q_1, q_2} |W(\pi)| = 
\sum_{\tau \ni q_1, q_2}|W(\tau)| P(\tau) 
\le \g^{-[\g r] \a_2 /2}
   \Eq(4.29)
   $$
Similarly for any given particle $q$ and sufficiently large absolute
constant $c$
   $$
\sum_{\pi \ni q:\; L(\tau) \ge c}|W(\pi)| =
\sum_{\tau \ni q:\; L(\tau) \ge c} |W(\tau)| P(\tau) 
\le \g^{4d}\;,
   \Eq(4.30)
   $$
where $L(\tau)$ denotes the number of links contributing to $\pi$ or
$\tau$. 

Polymer sum \equ(4.28) can be viewed as a Hamiltonian which we separate
in two parts. The first one is
   $$
\D H_{\g,\l}^{(1)}(\rho|\bar q^{(\L^c)})=
\sum_{\tau \not \in \L^c:\; L(\tau) < c} W(\tau) P(\tau)
   \Eq(4.31)
   $$
and
   $$
\D H_{\g,\l}^{(2)}(\rho|\bar q^{(\L^c)})=
\sum_{\tau \not \in \L^c:\; L(\tau) \ge c} W(\tau) P(\tau)
   \Eq(4.32)
   $$
It is not hard to see that $\D H_{\g,\l}^{(1)}(\rho|\bar q^{(\L^c)})$ is
simply a finite radius Hamiltonian of the polynomial type
   $$
\D H_{\g,\l}^{(1)}(\rho|\bar q^{(\L^c)})=\ell_2^d \sum_{D \not \in \L^c} 
W(D) \prod_{q \in D} \rho_q
   \Eq(4.33)
   $$
Here the sum is taken over connected sets of links (with no restriction
for a given link to enter this set more than once) containing less than
$c_{\equ(4.30)}$ links and the product is over all endpoints of the
links. The notation $\rho_q$ instead of $\rho_x$ is not
ambiguous as all endpoints of the links are assumed to be at the centers
of the corresponding boxes $C^{(\ell_2)}_x$. The statistical weights
$W(D)$ are obtained by resummation from \equ(4.31). 

{From} \equ(4.14.1) we pass to partition function \equ(4.7) containing
contours. For the $\log$ of the ratio between \equ(4.7) and \equ(4.14.1)
we also obtain a polymer expansion exploiting the theory of contour models
with interaction [DS], [BKL]. This expansion has the form
   $$
\sum_{\xi \not\in \L^c} W(\xi)
   \Eq(4.34)
   $$
where $\xi$ are another polymers constructed from labeled polymers $\pi$
and contours $\G$ in the same way as polymers $\pi$ are constructed from
diagrams $\T$. The statistical weights $W(\xi)$ are local functions of
$\rho_x$ and we interpret the whole sum as the Hamiltonian
   $$
\D H_{\g,\l}^{(3)}(\rho|\bar q^{(\L^c)})=\sum_{\xi \not\in \L^c} W(\xi)
   \Eq(4.35)
   $$
The only property of this Hamiltonian used later is the estimate
   $$
\sum_{\xi \ni q} |W(\xi)| \le \g^{4d} 
   \Eq(4.36)
   $$
Since all details can be found in [DS] and [BKL] we give only a sketch
of the proofs pointing out few technically important moments. 

Denote by $\nu(\cdot|q^{(\L^c)} )$ the Gibbs distribution of $\sum_{x
\in [\L]^{(\ell_2)}} \ell_2^d \rho_x$ particles given
by the Hamiltonian $\D H_{\g,\l}(q^{(\L)}|\bar q^{(\L^c)})$ with the
corresponding partition function \equ(4.14.1). For every contour $\G$
consider region $R(\G)=\p^{(\ell_3 /3)}{\rm Supp}(\G)$ with empty
boundary condition $\emptyset^{(R(\G)^c)}$ and define a modified
statistical weight
   $$
\widetilde W^T(\eta^{\G}|
q^{(\d^=(\G))})=\nu(q^{(\d^=(\G))}|\emptyset^{(R(\G)^c)})\;
W^T(\eta^{\G}| q^{(\d^=(\G))}) 
   \Eq(4.37)
   $$
Then the ratio of partition functions \equ(4.7) and \equ(4.14.1) can be
rewritten as
   $$
\eqalign{
\sum_{ \{\G_i\}^{\si} \in \L} \int \prod_i dq^{(\d^=(\G_i))}
&\exp \left( 
\sum_{\pi \not\in \L^c:\; \exists i, \; \pi \cap \d^=(\G_i)
\not=\emptyset,\atop \phantom{\pi \in \L:\; \exists i, \;} 
\pi \cap R(\G_i)^c \not=\emptyset} 
\Big( W(\pi | q^{(\cup _i \d^=(\G_i))})-W(\pi) \Big) \right) \cr
&\times \prod_i \widetilde W^T(\eta^{\G_i}| q^{(\d^=(\G_i))})}
   \Eq(4.37.1)
   $$
Here the statistical weight $W(\pi | q^{(\cup _i \d^=(\G_i))})$ is
defined respecting an additional boundary condition $q^{(\cup _i
\d^=(\G_i))}$ which is imposed in $\cup _i \d^=(\G_i)$. We remind that
only $\bar q^{(\L^c)}$ affects $W(\pi)$ if $\pi \cap \L^c \not= \emptyset$. 

The polymer sum in \equ(4.37.1) describes the interaction between
contours $\G_i$. It is important that every contributing to \equ(4.37.1)
polymer is sufficiently long and contains at least $[\g \ell_3 /3]$ links.
In view of \equ(4.29) such a polymer has very small statistical weight.
Therefore in complete similarity with \equ(4.12)-\equ(4.14) expanding
   $$
e^{ W(\pi | q^{(\cup _i \d^=(\G_i))})- W(\pi)}=1+\left( e^{ W(\pi |
q^{(\cup _i \d^=(\G_i))}) - W(\pi)}-1 \right)
   \Eq(4.37.2)
   $$
and integrating over $\prod_i dq^{(\d^=(\G_i))}$ one can derive for
\equ(4.37.1) a representation analogous to \equ(4.18). In this
representation newly defined diagrams are constructed from contours $\G$
connected via polymers $\pi$. Taking logarithm and applying
Theorem~\equ(sA.1) one obtains \equ(4.34) with correspondingly defined
polymers $\xi$. The key fact ensuring the condition \equ(A.3) of
Theorem~\equ(sA.1) is that for $\g$ small enough the sum of absolute
values of statistical weights of all polymers $\pi$ containing given
particle and being longer than $\ell_3 /3$ is much smaller than the
quantity $c \ell_2^d \ell_3^{-d} \z$ entering the Peierls estimate
\equ(3.7.3). 

The route to \equ(4.34) looks rather involved and tedious but it is a
standard one in the cluster expansion technique. On the other hand we need
only minor knowledge about $\xi$, namely \equ(4.36). This estimate is
obtained by the methods of [DS] and [BKL] along the following way.

The sum over all polymers $\xi$ containing given particle $q$ is equal
to 
   $$
\sum_{\G:\; \d^=(\G) \ni q}\;\; \sum_{\xi:\; \xi \ni \G} |W(\xi)|
+\sum_{\pi:\; \pi \ni q, L(\pi) \ge c_{\equ(4.30)}} 
\sum_{\xi:\; \xi \ni \pi} |W(\xi)|
   \Eq(4.38)
   $$  
By \equ(A.6) the first internal sum does not exceed $c W^T(\eta^{\G}|
q^{(\d^=(\G))})$ and the second internal sum does not exceed $c|W(\pi)|$.
In turn
   $$
\sum_{\G:\; \d^=(\G) \ni q} c W^T(\eta^{\G}| q^{(\d^=(\G))}) 
+\sum_{\pi:\; \pi \ni q, L(\pi) \ge c_{\equ(4.30)}} c|W(\pi)|
\le \g^{4d}
   \Eq(4.39)
   $$
because of \equ(3.9) and \equ(4.30).

The final result of this subsection can be states now as

\medskip\noindent{\bf \Lemma (s4.2)} 

\nobreak
{\sl The expression \equ(4.3) is equal to
   $$
\sum_{x \in [\L]^{(\ell_2)}} 
\log \left({\ell_2^{d n_x} \over n_x !} \right)
- \b H_{\g,\l}(\rho|\bar q^{(\L^c)}) + \sum_{i=1}^3 \D H_{\g,\l}^{(i)}(
\rho|\bar q^{(\L^c)}) 
   \Eq(4.40)
   $$
}

\medskip\noindent
Expression \equ(4.40) without the last sum is nothing but $\b
\widetilde {\cal F}_{\g,\b,\l}(\rho^{(\ell_2)} | \bar q^{(\L^c)})$
defined in \equ(6.18) with its main part $\b {\cal F}_{\g,\b,\l}(
\rho^{(\ell_2)} | \bar q^{(\L^c)})$ given by \equ(6.22)-\equ(6.23).

\bigskip
\goodbreak
\centerline{{\it Dobrushin Uniqueness (Basic Calculation).}}
\medskip
\nobreak

\noindent{\bf \Lemma (s4.3)} 

\nobreak
{\sl The effective Hamiltonian  \equ(4.40) satisfies the Dobrushin
uniqueness condition.}

\medskip\noindent 
{\bf Proof.} For the sake of simplicity we first check that the
Dobrushin uniqueness condition is true for the Hamiltonian ${\cal
F}_{\g,\b,\l} (\rho^{(\ell_2)})$. Then in the next subsection we show
that only minor modifications are necessary to treat \equ(4.40).

Take any site $x \in \Z^d_{\ell_2}$ and consider two boundary conditions
$\bar \rho^{(0)}$ and $\bar \rho^{(1)}$ on $\Z^d_{\ell_2} \setminus x$
such that they both belong to interval \equ(4.2) and differ only at a site
$y \in \Z^d_{\ell_2}$. For the definiteness we assume that $\bar
\rho_y^{(1)} > \bar \rho_y^{(0)}$. Denote by $\nu^{(0)}(d\rho_x)$ and
$\nu^{(1)}(d\rho_x)$ conditional Gibbs distributions defined by the
Hamiltonians ${\cal F}_{\g,\b,\l} (\rho_x|\bar \rho^{(0)} )$ and ${\cal
F}_{\g,\b,\l} (\rho_x|\bar \rho^{(1)} )$ respectively. The {\it
Vasserstein distance} between $\nu^{(0)}(d\rho_x)$ and
$\nu^{(1)}(d\rho_x)$ is
   $$
\eqalign{
R(\nu^{(0)},\nu^{(1)})&=
\int_{- \infty}^{\infty} dz \left|
\int_{-\infty}^z \left( \nu^{(0)}(d\rho_x)
- \nu^{(1)}(d\rho_x) \right) \right| \cr
&=\int_0^{\infty} dz \left| 
\int_0^z \left( \nu^{(0)}(d\rho_x)-
\nu^{(1)}(d\rho_x) \right) \right|\;, \cr}
   \Eq(4.41)
   $$
where the last equality utilizes the positivity of $\rho_x$.

The {\it Dobrushin uniqueness condition} is satisfied if one is able to find
a function $r_{xy}$ such that
   $$
R(\nu^{(0)},\nu^{(1)}) \le r_{xy} 
|\bar \rho^{(0)}_y - \bar \rho^{(1)}_y|
   \Eq(4.42)
   $$
and 
   $$
\sum_y r_{xy} < 1
   \Eq(4.43)
   $$

To check this condition we follow the strategy of [COPP] and define
$\nu_t(d\rho_x)$, $t \in [0,1]$ as a Gibbs measure corresponding to the
interpolated Hamiltonian 
   $$
{\cal F}_{\g,\b,\l} (\rho_x| t )=
{\cal F}_{\g,\b,\l} (\rho_x|\bar \rho^{(0)} )+
t \Big({\cal F}_{\g,\b,\l} (\rho_x|\bar \rho^{(1)} )
-{\cal F}_{\g,\b,\l} (\rho_x|\bar \rho^{(0)} ) \Big)
   \Eq(4.44)
   $$
This measure has the density
   $$
p(\rho_x |t)= {\exp \Big( -\b {\cal F}_{\g,\b,\l} (\rho_x| t ) \Big)
\over \int_0^{\infty} \exp \Big( -\b {\cal F}_{\g,\b,\l} (\rho_x| t )
\Big) d\rho_x }\; ,
   \Eq(4.45)
   $$
which is differentiable in $t$ and therefore from \equ(4.41) we have
   $$
R(\nu^{(0)},\nu^{(1)}) \le \int_0^1 dt \int_0^{\infty} dz
\left| \int_0^z {\p \over \p t} p(\rho_x |t) d \rho_x \right|
   \Eq(4.46)
   $$
Denoting 
   $$
\D {\cal F}_{\g,\b,\l} (\rho_x|\bar \rho^{(0)}, \bar \rho^{(1)} )=
\Big({\cal F}_{\g,\b,\l} (\rho_x|\bar \rho^{(1)} )
-{\cal F}_{\g,\b,\l} (\rho_x|\bar \rho^{(0)} ) \Big)
   \Eq(4.47)
   $$
one has
   $$
{\p \over \p t} p(\rho_x |t) 
= \b p(\rho_x |t)\; 
\left( \D {\cal F}_{\g,\b,\l} (\rho_x|\bar \rho^{(0)}, \bar \rho^{(1)} ) 
- \langle \D {\cal F}_{\g,\b,\l} (\rho_x|\bar \rho^{(0)}, \bar \rho^{(1)} )
\rangle_t \right)
   \Eq(4.48)
   $$
where $\langle \cdot \rangle_t$ denotes the expectation with respect to
$p(\rho_x |t)\; d \rho_x$. Observe that
   $$
\D {\cal F}_{\g,\b,\l} (\rho_x|\bar \rho^{(0)}, \bar \rho^{(1)} )=
\ell_2^d \Big( I_{\g}^{(2)}(x,y)- {1 \over 2!} I_{\g}^{(2)}(x,y| \bar
\rho) \Big)  
(\bar \rho^{(1)}_y - \bar \rho^{(0)}_y) \rho_x \; ,
   \Eq(4.49)
   $$
where $I_{\g}^{(2)}(x,y| \bar \rho)$ is defined by \equ(6.34) (with $\bar
\rho(\L \setminus x \setminus y)
=\bar \rho^{(0)}(\L \setminus x \setminus y)
=\bar \rho^{(1)}(\L \setminus x \setminus y)$ 
instead of $\hat \rho$) and satisfies
\equ(6.35). By direct calculation
   $$
\int_0^{\infty} dz \left| \int_0^z \Big( \rho_x -\langle \rho_x \rangle_t
\Big)   p(\rho_x |t)\; d \rho_x \right|
=\langle \rho_x^2 \rangle_t - \langle \rho_x \rangle_t^2
   \Eq(4.50)
   $$

For sufficiently small $\g$ and therefore sufficiently large $\ell_2^d$
the last expression can be estimated by Laplace method and it is equal to
   $$
\ell_2^{-d} \rho_*(t) + O(\ell_2^{-2d}),
    \Eq(4.51)
   $$
where $\ell_2^{d} \rho_*(t)^{-1}$ is the value of $\b {\p^2 \over \p
\rho_x^2}  {\cal F}_{\g,\b,\l} (\rho_x| t ) $ at the point,
$\rho_*(t)$, of minima of \goodbreak\noindent ${\cal F}_{\g,\b,\l}
(\rho_x| t ) $. As we know from Lemma~\equ(s6.7) such a minima exists,
is unique and lies strictly inside interval \equ(4.2). Hence
$O(\ell_2^{-2d})$ in \equ(4.51) is uniform in $t$ (see [F]). 

Combining \equ(4.46) - \equ(4.51) we obtain that
   $$
R(\nu^{(0)},\nu^{(1)}) \le 
\int_0^1 dt \big( \b \rho_* + O(\ell_2^{-d}) \big)
\Big( I_{\g}^{(2)}(x,y)- {1 \over 2!} I_{\g}^{(2)}(x,y| \bar \rho) \Big) 
(\bar \rho^{(1)}_y - \bar \rho^{(0)}_y)
    \Eq(4.52)
   $$
Setting 
   $$
r_{xy}=\max_{\bar \rho}
\Big| I_{\g}^{(2)}(x,y)- {1 \over 2!} I_{\g}^{(2)}(x,y| \bar \rho) \Big|
\int_0^1 dt \big( \b \rho_*(t) + O(\ell_2^{-d}) \big)
    \Eq(4.53)
   $$
one concludes that \equ(4.43) is true for sufficiently small $\g$
because of \equ(2.21.1), \equ(6.35) and the fact that $\sum_y
I_{\g}^{(2)}(x,y)=1$.

\bigskip
\goodbreak
\centerline{{\it Dobrushin Uniqueness (General Case).}}
\medskip
\nobreak

We turn now to a complete effective Hamiltonian \equ(4.40) and show that
only nonessential modification of the calculation of the previous
subsection are required to cover the complete \equ(4.40).

The first correction is due to the difference between ${\cal
F}_{\g,\b,\l}(\rho^{(\ell_2)} | \bar q^{(\L^c)})$ and $\widetilde {\cal
F}_{\g,\b,\l}(\rho^{(\ell_2)} |$ $\bar q^{(\L^c)})$. This difference is
discussed in detail in Lemma~\equ(s6.5). The corresponding modifications
of to our previous arguments are the following.

The difference between volumes of $B_{\g}(\cdot)$ and
$[B_{\g}(\cdot)]^{(\ell_2)}$ produces a factor $(1+ O(\g^{\a}))$, 
$\a >0$, in front of $I_{\g}^{(2)}(\cdot,\cdot)$ and
$I_{\g}^{(4)}(\cdot,\cdot,\cdot,\cdot)$ which clearly is not essential. 

The self interaction like $I_{\g}^{(2)}(\rho_x,\rho_x)$ or
$I_{\g}^{(4)}(\rho_x,\rho_x, \rho_y,\rho_z)$ produces nonlinear terms in
$\widetilde {\cal F}_{\g,\b,\l} (\rho_x| t )$ which is defined similarly
to \equ(4.44). Their contribution to $\widetilde {\cal F}_{\g,\b,\l}
(\rho_x| t )$ has the form
   $$
c_2 \rho_x^2 + c_3 \rho_x^3 +c_4 \rho_x^4 +t( c_5 \rho_x^2  +
c_6  \rho_x^3)(\bar \rho^{(1)}_y - \bar \rho^{(0)}_y) \; ,
    \Eq(4.54)
   $$
where $c_i$ does not depend on $t$, $\rho_x$ and $\rho^{(i)}_y$. Only
last, $t$-dependent, part from \equ(4.54) contributes to $\D \widetilde
{\cal F}_{\g,\b,\l} (\rho_x|\bar \rho^{(0)}, \bar \rho^{(1)} )$ which is
defined similarly to \equ(4.47). The corresponding effect on
$\int_0^{\infty} dz \left| \int_0^z {\p \over \p t} p(\rho_x |t) d
\rho_x \right|$ results in terms of the form
   $$
\b c_5 (\bar \rho^{(1)}_y - \bar \rho^{(0)}_y)\int_0^{\infty} dz \left|
\int_0^z \Big( \rho_x^2 -\langle \rho_x^2 \rangle_t \Big) p(\rho_x |t)\;
d \rho_x \right| 
=\b c_5 (\bar \rho^{(1)}_y - \bar \rho^{(0)}_y)(\langle \rho_x^3 \rangle_t
- \langle \rho_x^2 \rangle_t \langle \rho_x \rangle_t)
   \Eq(4.55)
   $$
and
   $$
\b c_6 (\bar \rho^{(1)}_y - \bar \rho^{(0)}_y)\int_0^{\infty} dz \left|
\int_0^z \Big( \rho_x^3 -\langle \rho_x^3 \rangle_t \Big) p(\rho_x |t)\;
d \rho_x \right|
=\b c_6 (\bar \rho^{(1)}_y - \bar \rho^{(0)}_y)(\langle \rho_x^4 \rangle_t
- \langle \rho_x^3 \rangle_t \langle \rho_x \rangle_t)
   \Eq(4.56)
   $$
in addition to $\ell_2^d \b \Big( I_{\g}^{(2)}(x,y)- {1 \over 2!}
I_{\g}^{(2)}(x,y| \bar \rho) \Big) (\bar \rho^{(1)}_y - \bar \rho^{(0)}_y)
(\langle \rho_x^2 \rangle_t - \langle \rho_x \rangle_t^2 )$. Clearly
   $$
|c_i| \le \ell_2^d \g^{\a}
   \Eq(4.57)
   $$
with some $\a >0$.

Applying Laplace method one has 
   $$
\langle \rho_x^{k+1} \rangle_t - \langle \rho_x^k \rangle_t \langle
\rho_x \rangle_t
={k \rho_*(t)^k \over 
\left. \b {\p^2 \over \p \rho_x^2} \widetilde {\cal F}_{\g,\b,\l} (\rho_x| t )
\right |_{\rho_*(t)}}+O(\ell_2^{-2d})\; ,
   \Eq(4.58)
   $$
where again $\rho_*(t)$ is the minima of $\widetilde {\cal F}_{\g,\b,\l}
(\rho_x| t )$. Since
   $$
\b {\p^2 \over \p \rho_x^2} \widetilde {\cal F}_{\g,\b,\l} (\rho_x| t )
=\ell_2^d \rho_x^{-1} + \b2c_2 + \b6c_3  \rho_x + \b12 c_4 \rho_x^2
+ \b t (2 c_5 + 6 c_6 \rho_x) (\bar \rho^{(1)}_y - \bar \rho^{(0)}_y) 
   \Eq(4.59)
   $$
and because of \equ(4.57) all corrections above does not destroy the
arguments of the previous subsection.

As the next step we incorporate in our calculation $\D H_{\g,\l}^{(1)}(
\rho|\bar q^{(\L^c)})$. According to \equ(4.33) this is the same type of
a polynomial correction which was just discussed. Thus the same
arguments work. The necessary smallness, like in \equ(4.57), of 
coefficients $W(D)$ is a consequence of the smallness of the statistical
weights of links \equ(4.23)-\equ(4.24) and the definition of $W(D)$ 
(see also \equ(4.29)).

Finally, to treat $\sum_{i=2,3} \D H_{\g,\l}^{(i)}(\rho|\bar
q^{(\L^c)})$ we simply observe that for any boundary condition $\bar
\rho$ given on $\Z^d_{\ell_2} \setminus x$
   $$
R(\nu(\rho_x| \bar \rho),\nu_{2,3}(\rho_x| \bar \rho)) < \g^{d}\; ,
   \Eq(4.60)
   $$
where $\nu(\rho_x| \bar \rho)$ is the conditional Gibbs distribution
given by the whole Hamiltonian \equ(4.40) while $\nu_{2,3}(\rho_x| \bar
\rho)$ is a similar distribution given by the Hamiltonian \equ(4.40)
without \goodbreak\noindent $\sum_{i=2,3} \D H_{\g,\l}^{(i)}(\rho|\bar
q^{(\L^c)})$. This estimate is obvious in view of definition \equ(4.41)
and bounds \equ(4.30) and \equ(4.36). One can comment that the
contribution of contours or long polymers to the free energy of the
auxiliary model is too small to affect anything at all. \qed

\bigskip
\bigskip
\goodbreak 
\centerline{{\bf 6. Properties of Auxiliary Model }}
\bigskip
\numsec=6
\numfor=1
\numtheo=1

In this section we use the Dobrushin uniqueness result established for
the auxiliary model in the previous section to prove
Statements~\equ(s3.3) and~\equ(s3.5). We begin with a construction which
is necessary for the proof of both statements. Namely, given a phase
$\si$ and a particle configuration $\bar q$ from the ground state
ensemble of the phase $\si$ we derive an appropriate representation for
the logarithm of partition function \equ(3.8). Lemma~\equ(s4.2) gives
   $$
Z^A_{\g,\b,\l}(\L|\bar q^{(\L^c)}) =
\sum_{\rho^{(\ell_2)}(\L)} 
\exp \left(
-\b \widetilde {\cal F}_{\g,\b,\l}(\rho^{(\ell_2)}(\L) | \bar q^{(\L^c)})
-\sum_{i=1}^3 \D H_{\g,\l}^{(i)}(\rho^{(\ell_2)} (\L)|\bar q^{(\L^c)})
\right)
   \Eq(5.1)
   $$
Below we consider density configurations, lattice volumes, etc related
to the scale $\ell_2$ and in most cases we omit the superscript
$(\ell_2)$ from the notations. On few occasions when we need other
scales we specify them explicitly.

In Section~4 we studied in detail the minimizers of ${\cal
F}_{\g,\b,\l}(\rho (\L) | \bar q^{(\L^c)})$ for bounded $\L$ and
$\L=\Z^d_{\ell_2}$. Now we need to study minimizers for
   $$
\widetilde H_{\g,\l}(\rho(\L)|\bar q^{(\L^c)}) 
=\b \widetilde {\cal F}_{\g,\b,\l}(\rho (\L)| \bar q^{(\L^c)})
+\D H_{\g,\l}^{(1)}(\rho(\L)|\bar q^{(\L^c)}),
   \Eq(5.2)
   $$
where both $\bar q^{(\L^c)}$ and $\rho(\L)$ belong to the ground state
ensemble of the phase $\si$. This is a finite range
translation-invariant Hamiltonian similar to ${\cal F}_{\g,\b,\l}(\rho
(\L) | \bar q^{(\L^c)})$ but not allowing a simple representation of the
type \equ(6.28). For a bounded $\L$ the existence of at least one
minimizer of $\widetilde H_{\g,\l}(\rho(\L)|\bar q^{(\L^c)})$ follows from
the compactness of its domain, $(\rho_{\b,\si} - \z, \rho_{\b,\si} +
\z)^{\L}$. This domain is convex and the minimizer is unique because the
function $\widetilde H_{\g,\l}(\rho(\L)|\bar q^{(\L^c)})$ (of the finite
number of variables $\rho_x$, $x \in \L$) is also convex. This is a
consequence of estimate \equ(2.21.1) from which it is not hard to see
that for any $\bar q^{(\L^c)}$ and $\rho(\L)$ the Hessian matrix of
$\widetilde H_{\g,\l}(\cdot|\bar q^{(\L^c)})$ calculated at $\rho(\L)$ is
positive definite with the mass bouded from $0$ independently of $\bar
q^{(\L^c)}$ and $\rho(\L)$. Depending on the context we denote the
unique minimizer of $\widetilde H_{\g,\l}(\rho(\L)|\bar q^{(\L^c)})$ by $\hat
\rho$, $\hat \rho(\L)$ or $\hat \rho^{\bar q^{(\L^c)}}$.

To clarify the structure and convexity of $\widetilde H_{\g,\l}(\rho(\L)|\bar
q^{(\L^c)})$ we introduce
   $$
\D \widetilde H_{\g,\l}(\rho(\L)|\bar q^{(\L^c)})= \widetilde
H_{\g,\l}(\rho(\L)|\bar q^{(\L^c)}) 
- \widetilde H_{\g,\l}(\hat \rho(\L)|\bar q^{(\L^c)})
   \Eq(5.2.1)
   $$
It is not hard to check that
   $$
\eqalign{
\D \widetilde H_{\g,\l}(\rho(\L)|\bar q^{(\L^c)})&=
\ell_2^d\; \sum_{x \in [\L]^{(\ell_2)}} \sum_{n=2}^{\infty} 
{1 \over n(n-1) \hat \rho^{n-1}_x } \D \rho_x^n \cr
&+\ell_2^d \sum_{D \not \subset [\L^c]^{(\ell_2)}} W(D)
\sum_{X_1(D), X_2(D)}\;\;  \prod_{x \in X_1(D)} \hat \rho_x(\L)
\prod_{x \in X_2(D)} \D \rho_x}
   \Eq(5.3)
   $$
Here $\D \rho_x=\rho_x -\hat \rho_x(\L)$ and $D$ is a connected family of
links with endpoints forming a set $X(D) \subset \Z^d_{\ell_2}$. The
number of links in $D$ is less than $c_{\equ(4.30)}$. The subsets
$X_1(D)$ and $X_2(D)$ form a partition of $X(D)$ with at least two
elements in $X_2(D)$. The second internal sum is taken over all
partitions of that type. For $W(D)$ one has the estimate
   $$
\sum_{D:\; X(D) \ni x \atop D \not= L^{(2)}, L^{(4)} } 
|W(D)| \rho_{\max}^{|X(D)|} \le \g^{\a_2 \over 3}
   \Eq(5.3.1)
   $$
as follows from \equ(4.29). Also
   $$
|\D \rho_x| \le 2\z
   \Eq(5.3.2)
   $$

The quadratic part of \equ(5.3) is
   $$
{1 \over 2!} \ell_2^d\; \sum_{x \in [\L]^{(\ell_2)}}{1 \over \hat
\rho } \D \rho_x^2 
-{1 \over 2!} \ell_2^d\; \sum_{x_1, x_2:\; x_1 \cup x_2 \not \in
[\L^c]^{(\ell_2)}}
\widetilde I^{(2)}(x_1, x_2 | \hat \rho(\L))\D \rho_{x_1} \D\rho_{x_2}
   \Eq(5.4)
   $$
with
   $$
{1 \over \hat \rho_{x_1}(\L)} -\sum_{x_2} \widetilde I^{(2)}(x_1, x_2
| \hat \rho(\L)) 
\ge m(\b) >0 
   \Eq(5.5)
   $$
for any $x_1 \in [\L]^{(\ell_2)}$. Here $\widetilde I^{(2)}(x_1, x_2 |
\hat \rho(\L))$ is an analogue of \equ(6.34) and satisfies the same
estimate \equ(6.35) for $\g$ small enough. ( By construction
$|I^{(2)}(x_1, x_2 | \hat \rho(\L))- \widetilde I^{(2)}(x_1, x_2 | \hat
\rho(\L))| \le c\g^{\a_2}I^{(2)}(x_1, x_2)$ ). Moreover, in view of
\equ(5.3.1) and assuming that $\z$ is chosen to be small enough with
respect to $m(\b)$
  $$ 
\D \widetilde H_{\g,\l}(\rho(\L)|\bar q^{(\L^c)})=
{1 \over 2!} \ell_2^d\; \sum_{x \in [\L]^{(\ell_2)}}
{m(\b) \over 2 } \D \rho_x^2 +U(\hat \rho(\L), \D\rho(\L))
   \Eq(5.6)
   $$
with positive convex $U(\hat \rho(\L), \D\rho(\L))$ having minima at
$\D\rho(\L) \equiv 0$.

Along with the minimizers of $\widetilde H_{\g,\l}(\rho(\L)|\bar q^{(\L^c)})$
for bounded regions $\L$ we also need a global minimizer, i.e. an
analogue of $\rho_{\b,\si}$. To find such a minimizer and to show its
uniqueness ( for given $\si$ ) we consider a suffuciently large (with
respect to the range of $\widetilde H_{\g,\l}(\cdot)$) periodic box $\L$ and
the corresponding Hamilltonian $\widetilde H_{\g,\l}(\rho(\L))$. All
convexity considerations remain true for $\widetilde H_{\g,\l}(\rho(\L))$
such that $\widetilde H_{\g,\l}(\rho(\L))$ has a unique minimizer $\hat
\rho(\L)$. Because of the translation invariance of $\widetilde
H_{\g,\l}(\rho(\L))$ this minimizer is a constant configuration $\hat
\rho_x(\L) \equiv \hat \rho,\; x \in \L$. (Otherwise one is able to
construt another minimizers by space translations). Since $\widetilde
H_{\g,\l}(\rho(\L))$ is of finite range we observe that $\hat \rho$ is
independent on $\L$ for all sufficiently large periodic $\L$ and hence
$\hat \rho$ is the global minimizer we are looking for. Note that the
specific energy $\tilde h_{\g,\b,\l}(s) = \lim_{ \L \to \Z^d_{\ell_2}}
|\L|^{-1} \widetilde H_{\g,\l}(\rho(\L))$ of any constant density
configuration $\rho (\L) \equiv s$ is just a finite sum and $\hat \rho$
is nothing but the value of $s$ at which $\tilde h_{\g,\b,\l}(s)$
achieves its minima. Thus given $\si$ such a minima is unique for $s \in
(\rho_{\b,\si} - \z, \rho_{\b,\si} + \z)$ and we have different values
$\hat \rho_{\b,\l,-}$ and $\hat \rho_{\b,\l,+}$ corresponding to
$\si=-1$ and $\si=+1$.

Denote by $\hat \l(\b,\g)$ the value of $\l$ at which $\tilde
h_{\g,\b,\l}(\hat \rho_{\b,\l,-}) = \tilde h_{\g,\b,\l}(\hat
\rho_{\b,\l,+})$ and set $\hat \rho_{\b,\si} = \hat \rho_{\b,\hat
\l(\b), \si}$. The important consequences of the representation of $\hat
\rho_{\b,\l,\si}$ via $\tilde h_{\g,\b,\l}(s)$ are the existence of
$\hat \l(\b,\g)$ and the estimates
   $$
|\rho_{\b,\si} - \hat \rho_{\b,\l,\si}| \le c\g^{\a_1 /2}
   \Eq(5.7.1)
   $$
and
   $$
|\l(\b) - \hat \l(\b,\g)| \le c\g^{\a_1}
   \Eq(5.7.2)
   $$

Indeed, let $h_{\g,\b,\l}(s)$ be the specific energy of $\rho \equiv s$
calculated via the Hamiltonian $\widetilde H_{\g,\l}(\rho(\L)) +\D
H_{\g,\l}^{(2)}(\rho(\L))$ with periodic $\L$. Then
   $$
|\tilde h_{\g,\b,\l}(s) - h_{\g,\b,\l}(s)| \le c \g^{4d}
   \Eq(5.7.3)
   $$
as follows from \equ(4.30).

Another way to calculate $h_{\g,\b,\l}(s)$ is the approach of Section~4.
Here the starting point is the partition function \equ(4.14.1) with
fixed density configuration $\rho^{(\ell_2)} \equiv s$. Then one
aproximate this partition function by using the density configurations
defined with respect to the scale $\ell_1 \ll \ell_2$. (Note that the
density configuration which is constant in the scale $\ell_2$ is not
necessarily a constant one in the finer scale $\ell_1$). As the first
step of the approximation one shifts all particles to the centers of the
corresponding boxes $C^{(\ell_1)}$ for the price of the error (in the
value of $\widetilde H_{\g,\l}(\rho^{(\ell_1)}(\L)) +\D
H_{\g,\l}^{(2)}(\rho^{(\ell_1)}(\L))$) which in absolute value does not
exceed $c \g^{\a_1}|\L|$ (see Lemma~\equ(s6.4)). Lemmas~\equ(s6.5)
and~\equ(s6.6) then show that up to the same error the Hamiltonian
$\widetilde H_{\g,\l}(\rho^{(\ell_1)}(\L)) +\D
H_{\g,\l}^{(2)}(\rho^{(\ell_1)}(\L))$ can be approximated by \equ(6.22).

Now the convexity considerations (see \equ(2.16)) imply that among all
density configurations $\rho^{(\ell_1)}(\L)$ such that
$\rho^{(\ell_2)}(\rho^{(\ell_1)}(\L)) \equiv s$ the configuration
$\rho^{(\ell_1)}(\L) \equiv s$ has the minimal value of the mean field
functional \equ(6.22). Hence (see \equ(6.19) and above) counting only
the contribution of $\rho^{(\ell_1)}(\L) \equiv s$ produces the error
which again in absolute value does not exceed $c \g^{\a_1}|\L|$ (in fact
it is much smaller). The specific energy of the constant density
configuration $\rho^{(\ell_1)} \equiv s$ calculated via \equ(6.22) is
nothing but the mean field specific energy $F_{\b,\l}(s)$ and therefore
   $$
|F_{\b,\l}(s) - h_{\g,\b,\l}(s)| \le c\g^{\a_1}
   \Eq(5.7.4)
   $$
Since $\tilde h_{\g,\b,\l}(s) = F_{\b,\l}(s) + \sum_{k=2}^K a_k
s^k$ with $|a_k| < c\g^{\a_2}$, i.e.  $\tilde h_{\g,\b,\l}(s)$ as
a small polynomial perturbation of $F_{\b,\l}(s)$, and
   $$
|F_{\b,\l}(s) - \tilde h_{\g,\b,\l}(s)| \le c\g^{\a_1},
   \Eq(5.7.5)
   $$
as follows from \equ(5.7.3) and \equ(5.7.4), one immediately obtains
\equ(5.7.1) and \equ(5.7.2). ( The existence of $\hat \l(\b,\g)$ follows
from an elementary linear analysis and continuity of $\tilde
h_{\g,\b,\l}(s)$~).

Finally observe that a straightforward modification of the arguments of
Lemma~\equ(s6.7) shows that $|\hat \rho_{\b,\l,\si} - \hat \rho^{\bar
q^{(\L^c)}}|$ satisfy \equ(6.29).

\medskip
With $\hat \rho_{\b,\l,\si}$ and $\hat \rho^{\bar q^{(\L^c)}}$ being
properly defined we construct now a convenient representation for the
$\log$ of the partition function \equ(5.1) using the following
interpolation trick. For $t \in [0,1]$ and $s \in [0,1]$ introduce an
interpolated Hamiltonian
   $$
\eqalign{
\D H_{\g,\l}(\D \rho(\L)|\bar q^{(\L^c)}; t,s)&=
\ell_2^d \sum_{x \in [\L]^{(\ell_2)}}\;\; \sum_{n=2}^{\infty} 
{1 \over n(n-1) \hat \rho^{n-1}_x } \D \rho_x^n \cr
&+ t \ell_2^d \sum_{D \not \subset [\L^c]^{(\ell_2)}} W(D)
\sum_{X_1(D),\atop X_2(D)}\;\;  \prod_{x \in X_1(D)} \hat \rho_x (\L)
\prod_{x \in X_2(D)} \D \rho_x \cr
&+ s \sum_{i=2,3} \D H_{\g,\l}^{(i)}(\rho (\L) |\bar q^{(\L^c)}) \cr}
   \Eq(5.8)
   $$
The Hamiltonian $\D H_{\g,\l}(\D \rho(\L)|\bar q^{(\L^c)}; t,s)$ satisfies
the Dobrushin uniqueness condition similarly to $\D H_{\g,\l}(\D
\rho(\L)|\bar q^{(\L^c)}; 1,1)$. We denote by $\langle \cdot
\rangle_{t,s,\L, \bar q^{(\L^c)}}$ the expectation with respect to the
corresponding Gibbs measure in the finite domain $\L$ with the boundary
condition $\bar q^{(\L^c)}$ and we denote by $\langle \cdot
\rangle_{t,s}$ the expectation with respect to the corresponding unique
limit Gibbs distribution.

{From} \equ(5.1) and \equ(5.8) by direct calculation one obtains
   $$
\eqalign{
&\log Z^A_{\g,\b,\l}(\L|\bar q^{(\L^c)}) =  
-\widetilde H_{\g,\l}(\hat \rho(\L)|\bar q^{(\L^c)}) \cr
&+ \ell_2^d \sum_{D \not \subset [\L^c]^{(\ell_2)}} W(D)
\sum_{X_1(D), \atop X_2(D)}\;\; 
\prod_{x \in X_1(D)} \hat \rho_x \int_0^1 dt\;
 \langle \prod_{x \in X_2(D)} \D \rho_x \rangle_{t,0,\L, \bar q^{(\L^c)}}\cr
&+\sum_{\tau \in \L:\; L(\tau) \ge c_{\equ(4.30)}} W(\tau) \int_0^1 ds\; 
\langle P(\tau, \D \rho) \rangle_{1,s,\L, \bar q^{(\L^c)}} \cr
&+\sum_{\xi \in \L}  \int_0^1 ds\; \langle W(\xi, \D \rho)
\rangle_{1,s,\L, \bar q^{(\L^c)}} \cr }
   \Eq(5.9)
   $$
Here $\hat \rho(\L) = \hat \rho^{\bar q^{(\L^c)}}$ is the minimizer of
$\widetilde H_{\g,\l}(\cdot |\bar q^{(\L^c)})$ and notations $P(\tau, \D
\rho)$ and $W(\xi, \D \rho)$ are used instead of $P(\tau)$ and $W(\xi)$
to underline the density configuration $\hat \rho(\L) + \D \rho(\L)$
with respect to which these quantities are calculated. We use two
parameters $s$ and $t$ instead of a single one for the technical
transparency. In models $\langle \cdot \rangle_{t,0,\L, \bar
q^{(\L^c)}}$, i.e. with $s=0$, the range of interaction is finite so it
is simpler to control decay of correlations.

Namely, a standard consequence of the Dobrushin uniqueness theorem [D2]
is the estimate
   $$
|\langle \D \rho_x \rangle_{t,0,\L, \bar q^{(\L^c)}}
- \langle \D \rho_x \rangle_{t,0}| \le 2\z
\sum_{n=2}^{\infty}
\sum_{{y_1=x \atop y_2 \ldots, y_{n-1} \in [\L]^{(\ell_2)}} \atop 
y_n \in [\L^c]^{(\ell_2)}} \prod_{i=1}^{n-1} r_{y_i y_{i+1}},
   \Eq(5.10)
   $$
where $r_{xy}$ are defined by \equ(4.53) with nonessential modification
discussed below \equ(4.59). Similarly
   $$
\left|\langle \prod_{x \in X} \D \rho_x \rangle_{t,0,\L, \bar q^{(\L^c)}}
- \langle \D \rho_x \rangle_{t,0} \right| \le (2\z)^{|X|}
\sum_{n=2}^{\infty}
\sum_{{y_1 \in X \atop y_2 \ldots, y_{n-1} \in [\L]^{(\ell_2)}} \atop 
y_n \in [\L^c]^{(\ell_2)}} \prod_{i=1}^{n-1} r_{y_i y_{i+1}}
   \Eq(5.11)
   $$
Now it follows from definition \equ(4.53) of $r_{xy}$ and estimate
\equ(2.21.1) that
   $$
|\langle \D \rho_x \rangle_{t,0,\L, \bar q^{(\L^c)}}
- \langle \D \rho_x \rangle_{t,0}| \le
{c\z \over 1-a(\b)} \left({1+a(\b) \over 2} \right)^{[\g {\rm
dist}\;(x, \L^c)]}
   \Eq(5.12)
   $$
and
   $$
\left|\langle \prod_{x \in X} \D \rho_x \rangle_{t,0,\L, \bar q^{(\L^c)}}
- \langle \D \rho_x \rangle_{t,0} \right| \le |X|(2\z)^{|X|}
{c\z \over 1-a(\b)}\left( {1+a(\b) \over 2} \right)^{[\g {\rm dist}\;(X,
\L^c)]}
   \Eq(5.13)
   $$
which are similar to \equ(6.29). Also given $x$
   $$
\sum_{n=2}^{\infty}
\sum_{y_1=x \atop y_2 \ldots, y_{n} \in \Z^d_{\ell_2}}
\prod_{i=1}^{n-1} r_{y_i y_{i+1}} \le {2 \over 1- a(\b)}
   \Eq(5.13.1)
   $$

In fact it easily follows from the structure and smallness of $|W(\tau)|$
and $|W(\xi)|$ expressed by \equ(4.30) and \equ(4.36) that the same
exponential decay of correlations takes place for models with $s \not=0$
which include the infinite range part $s \sum_{i=2,3} \D
H_{\g,\l}^{(i)}(\rho(\L) |\bar q^{(\L^c)})$ of the Hamiltonian \equ(5.8)

\goodbreak\medskip
The final representation of $\log Z^A_{\g,\b,\l}(\L|\bar q^{(\L^c)})$ is
   $$
\eqalign{
&\log Z^A_{\g,\b,\l}(\L|\bar q^{(\L^c)}) =  
-\widetilde H_{\g,\l}(\hat \rho(\L)|\bar q^{(\L^c)}) \cr
& -\ell_2^d \sum_{D \not \subset [\L^c]^{(\ell_2)}} W(D)
\sum_{X_1(D), \atop X_2(D)}\;\; \prod_{x \in X_1(D)} \hat \rho_x \int_0^1
dt\; \langle \prod_{x \in X_2(D)} \D \rho_x \rangle_{t,0}\cr
&-\sum_{\tau \not \in \L^c:\; L(\tau) \ge c} W(\tau) \int_0^1 ds\; 
\langle P(\tau,\D \rho) \rangle_{1,s} \cr
&-\sum_{\xi \not \in \L^c}  \int_0^1 ds\; \langle W(\xi, \D \rho)
\rangle_{1,s} \cr 
&+ \ell_2^d \hskip-1em\sum_{D \not \subset [\L^c]^{(\ell_2)}}
\hskip-1emW(D) \sum_{X_1(D),\atop X_2(D)}\;\; \prod_{x \in X_1(D)} \hat
\rho_x \int_0^1 dt\; \left(
\langle \prod_{x \in X_2(D)} \D \rho_x \rangle_{t,0,\L, \bar q^{(\L^c)}}
-\langle \prod_{x \in X_2(D)} \D \rho_x \rangle_{t,0} \right) \cr
&+\sum_{\tau \not \in \L^c:\; L(\tau) \ge c} W(\tau) \int_0^1 ds\; 
\left( \langle P(\tau, \D \rho) \rangle_{1,s,\L, \bar q^{(\L^c)}} 
- \langle P(\tau, \D \rho) \rangle_{1,s} \right)\cr
&+\sum_{\xi \not \in \L^c}  \int_0^1 ds\; 
\left( \langle W(\xi, \D \rho) \rangle_{1,s,\L, \bar q^{(\L^c)}} 
- \langle W(\xi, \D \rho) \rangle_{1,s} \right) \cr }
   \Eq(5.14)
   $$
Here speaking about $\langle f(\D \rho_x) \rangle_{t,s,\L, \bar
q^{(\L^c)}}$ we assume that $\D \rho_x=\rho_x - \hat \rho_x(\L)$ while
for $\langle f(\D \rho_x) \rangle_{t,s}$ we mean $\D \rho_x=\rho_x -
\hat \rho_{\b,\l,\si}$ such that $\langle f(\D \rho_x) \rangle_{t,s}$ is
truly independent on any boundary conditions. 

\goodbreak
\medskip\noindent
{\bf Proof of Statement \equ(s3.3).} {}From the interpolation trick we have
a representation for the metastable free energy
   $$
\eqalign{
&f^A_{\si,\l,\g}= 
\ell_2^d \hat \rho_{\b,\l,\si}(\log \hat \rho_{\b,\l,\si} -1) -
\ell_2^d \l \hat \rho_{\b,\l,\si}  \cr
&-\ell_2^d {1 \over 2!} \sum_{x \in \Z^d_{\ell_2}} 
J_{\g}^{(2)}(0,x) \hat \rho_{\b,\l,\si}^2 \cr
&+\ell_2^d {1\over 4!} \sum_{x_1,x_2,x_3 \in \Z^d_{\ell_2}}
J_{\g}^{(4)}(0, x_1, \dots, x_3)
\hat \rho_{\b,\l,\si}^4 \cr 
& -\ell_2^d \sum_{D:\; X(D) \ni 0 \atop D \not= L^{(2)},  L^{(4)}} 
{W(D) \over |X(D)|}
\hat \rho_{\b,\l,\si}^{|X(D)|} \cr
&-\sum_{\tau:\; X(\tau) \ni 0} 
{W(\tau) \over |X(\tau)|} P(\tau, \hat \rho_{\b,\l,\si}) \cr
&-\sum_{\xi:\; X(\xi) \ni 0}  
{W(\xi,\hat \rho_{\b,\l,\si}) \over |X(\xi)|} \cr 
& -\ell_2^d \sum_{D:\; X(D) \ni 0 \atop |X_2(D)| \ge 2} {W(D) \over |X(D)|}
\sum_{X_1(D),\atop X_2(D)}\;\; \hat \rho_{\b,\l,\si}^{|X_1(D)|} \int_0^1 dt\; 
\langle \prod_{x \in X_2(D)} \D\rho_x \rangle_{t,0}\cr
&-\sum_{\tau:\; X(\tau) \ni 0} 
{W(\tau) \over |X(\tau)|}\int_0^1 ds\; 
\langle P(\tau, \D \rho) \rangle_{1,s} \cr
&-\sum_{\xi:\; X(\xi) \ni 0}  \int_0^1 ds\; 
{\langle W(\xi,\D \rho) \rangle_{1,s} \over |X(\xi)|}  \cr 
}
   \Eq(5.15)
   $$
Here the definition of $X(\tau)$ and $X(\xi)$ is similar to that of
$X(D)$ and $\D \rho$ is defined with respect to $\hat \rho_{\b,\l,\si}$.

The difference $g_{\si}(\L | \bar q^{(\L^c)})$ between RHS of
\equ(5.14) and $f^A_{\si,\l,\g} |\L|$ can be estimated as
   $$
|g_{\si}(\L | \bar q^{(\L^c)})| \le c \ell_2^d 
{|\L \cap \L^c| \over \ell_2^{d-1}} \g^{-\a_2} \le c \g^{-1} |\L \cap \L^c| 
   \Eq(5.16)
   $$
Here $|\L \cap \L^c|$ is the $(d-1)$-dimensional volume (area) of
hypersurface separating $\L$ and $\L^c$ and $|\L \cap \L^c|\ell_2^{-d+1}
\g^{-\a_2}$ is the number of lattice points in $[\d \L]^{(\ell_2)}$. The
first factor $c \ell_2^d$ is the upper estimate for the sum of the
absolute values of the error terms associated with given site $x \in [\p
\L]^{(\ell_2)}$. To be more precise, the contribution to $g_{\si}(\L |
\bar q^{(\L^c)})$ is given by terms in \equ(5.14) and \equ(5.15)
crossing the boundary of $\L$ directly or indirectly. Directly crossing
terms are $D$-s with $X(D)$ intersecting both $\L$ and $\L^c$, $\tau$-s
with $X(\tau)$ intersecting both $\L$ and $\L^c$, $\xi$-s with $X(\xi)$
intersecting both $\L$ and $\L^c$ and links with endpoints in both $\L$
and $\L^c$. Indirectly intersection terms come from estimate \equ(5.11)
for the difference terms in the last three lines of \equ(5.14).
Graphically those terms can be represented by chains of $r$-s joining
something inside $\L$, say some $D \in \L$, with the sites outside $\L$.
It is important for us that geometrically any of the terms crossing the
boundary passes through some site $x \in \p \L$. This site $x$ may be:
one of the endpoints of the link, contributing itself or as a part of
$D$, $\tau$ or $\xi$, or the endpoint of some $r_{xy}$ or it may belong
to the support of some contour which is a part of some polymer $\xi$.
For all these involved weighted objects we always keep the property that
the sum of the absolute values of the statistical weights of all objects
passing through given point is less than absolute constant. Multiplying
this constant by the factor $\ell_2^d$, which depending on the notations
is present explicitly or implicitly in front of the sums just discussed,
we reproduce \equ(5.16). We need now some additional work to improve on
\equ(5.16).

The simplest observation is that using \equ(4.40) and \equ(4.36) one
obtains that the sum over the objects containing $\tau$ with $L(\tau) >
c_{\equ(4.30)}$ or $\xi$ as a constituting element is less than $c
\g^{4d}$.

Another source of smallness are various expectations $\langle \prod_x \D
\rho_x \rangle$ in the corresponding terms. In particular we check that
given $x$ and for any $k\ge 1$
  $$
\langle |\D\rho_x|^k \rangle \le c(\ell_2^d)^{-k{5 \over 12}}
   \Eq(5.17)
   $$
where the expectation means any one of those contributing to \equ(5.14)
and \equ(5.15). The power $-{5 \over 12}$ is taken for the definiteness
only and it can be replaced by $-{1 \over 2}+\e$. We prove \equ(5.17)
adapting to our situation relatively abstract Lemma~\equ(s5.1) in the
spirit of [R2]. The correspondence between current notations and those
of Section~8 is given by the following list of analogous objects:
$\phi_x$ and $\D \rho_x$, $\kappa$ and $\ell_2^d$, $a$ and $2\z$, $M$
and $2\z \ell_2^d$, $R$ and $c_{\equ(4.30)} \g^{-1}$, $J_{xy}$ and $\hat
\rho \widetilde I^{(2)}(x,y| \hat \rho)$ and so on. The bound \equ(5.17)
is an easy consequence of Lemma~\equ(s5.1) (which deals the finite range
interactions) as the infinite radius part, $s \sum_{i=2,3} \D
H_{\g,\l}^{(i)}(\rho^{(\ell_2)} |\bar q^{(\L^c)})$, of \equ(5.8) is so small
that it can not increase the expectation of $|\D \rho_x|$ more than in
$1+c\g^{4d}$ times. Note also that in \equ(5.14)
and \equ(5.15) we have terms $\langle |\D\rho_x|^k \rangle$ only with $k
\ge 2$.

In a similar way applying estimate \equ(5.29) to $|\langle \prod_{x \in
X_2(D)} \D \rho_x \rangle_{t,0,\L, \bar q^{(\L^c)}} -$ \hfil \goodbreak
\noindent $\langle \prod_{x \in X_2(D)} \D \rho_x \rangle_{t,0}|$ we
obtain improved \equ(5.13). Indeed, the RHS of \equ(5.11) is nothing but
the upper bound for the Vasserstein distance and a square root of this
estimate produces an exponential bound just with twice smaller exponent
than in \equ(5.13).

These improvements reduce by the same factor, $(\ell_2^d)^{-2
{5\over12}}$, the estimates of the cross boundary terms coming from
lines 2 and 5 in \equ(5.14) and line 7 in \equ(5.15). 

As an immediate consequence of these improvements we can see that that
the contribution, $e(\l)$, to the difference $\ell_2^{-d}(f^A_{+,\l,\g}
- f^A_{-,\l,\g})$ coming from the non energy terms ( lines 7,8,9 in
\equ(5.15) ) is a continuous function of $\l$ and $|e(\l)| \le c
(\ell_2^d)^{-2 {5\over12}}$. The energy parts ( lines 1-6 in \equ(5.15)
) of $\ell_2^{-d}f^A_{+,\l,\g}$ and $\ell_2^{-d}f^A_{-,\l,\g}$ coincide
with each other at $\l=\hat \l(\b, \g)$ and they contain explicitly
terms which are linear in $\l$. Hence shiftig $\hat \l(\b, \g)$ by at
most $c (\ell_2^d)^{-2 {5\over12}}$ one can find the solution,
$\l=\l(\b,\g)$ of \equ(3.10e3) which proves Statement~\equ(s3.3). \qed

\medskip
A useful consequence of our improvements is the estimate
   $$
\eqalign{
\Big|g_{\si}(\L| \bar q^{(\L^c)})
+\big( \widetilde H_{\g,\l} (\hat  \rho_{\b,\l,\g}(\L)| \hat
\rho_{\b,\l,\g}(\L^c)) 
&- {1 \over 2} \widetilde U_{\g} (\hat  \rho_{\b,\l,\g}(\L)| \hat
\rho_{\b,\l,\g}(\L^c)) \cr
- \widetilde H_{\g,\l} (\hat \rho(\L)| \bar q^{(\L^c)}) \big) \Big|
&\le c \g^{-1} (\ell_2^d)^{-2 {5\over12}} |\L \cap \L^c| \cr
& \le \g^{1/2}|\L \cap \L^c|}
   \Eq(5.31.1)
   $$
where $\widetilde U_{\g} (\hat \rho_{\b,\l,\g}(\L)| \hat
\rho_{\b,\l,\g}(\L^c))$ is defined as in \equ(2.3.1) but in terms of the
Hamiltonian $\widetilde H(\cdot)$. Observe that $\widetilde H_{\g,\l} (\hat
\rho_{\b,\l,\g}(\L)| \hat \rho_{\b,\l,\g}(\L^c)) - \widetilde H_{\g,\l}
(\hat \rho(\L)| \bar q^{(\L^c)})$ is rather small while the compensating
term $- {1 \over 2} \widetilde U_{\g} (\hat \rho_{\b,\l,\g}(\L)| \hat
\rho_{\b,\l,\g}(\L^c))$ is of order $c \g^{-1} |\L \cap \L^c|$.

\medskip\noindent
{\bf Proof of Statement~\equ(s3.5).} To check estimate \equ(3.10e6)
consider ${\rm Int}^{\not=}(\G)=$ \hfil \goodbreak \noindent $\cup_{m:\;
\si_m(\G) \not= \si(\G)} {\rm Int}_m(\G)$ and introduce a strip
$S(\G)=\p^{(\ell_3/2)} {\rm Int}^{\not=}(\G) \setminus \tilde
\d^{\not=}(\G)$, where $\tilde \d^{\not=}(\G)$ is defined similarly to
$\tilde \d^{=}(\G)$ (see \equ(6.48) and below). First we rewrite the
numerator of \equ(3.10) as
   $$
\int_{{\cal Q}^{({\rm Supp}(\G) \setminus S(\G))}} dq\; 
\ind{\eta(q)=\eta^{\G}} \;
e^{-\b H_{\g,\l}(q|\bar q^{(\d^= (\G))})}
Z^A_{\g,\b,\l} \big({\rm Int}^{\not=}(\G)\cup S(\G)|
q^{(\tilde \d^{\not=}(\G))}\big)
   \Eq(5.32)
   $$
Observe that for any $q$ contributing to \equ(5.32) the restriction of
$q$ to $\tilde \d^{\not=}(\G)$ belongs to the ground state ensemble of
the phase $-\si(\G)$. Given such $q^{(\tilde \d^{\not=}(\G))}$ denote by
$\hat \rho ({\rm Int}^{\not=}(\G)\cup S(\G)) = \hat \rho^{q^{(\tilde
\d^{\not=}(\G))}}$ the minimizer of \equ(5.2). Then using representation
\equ(5.14) and definition \equ(5.15) one can see that

   $$
\eqalign{
Z^A_{\g,\b,\l} \big({\rm Int}^{\not=}(\G)\cup S(\G)| q^{(\tilde
\d^{\not=}(\G))}\big)
&=Z^A_{\g,\b,\l} \big(S(\G)| q^{(\tilde \d^{\not=}(\G))},
\hat \rho^{q^{(\tilde \d^{\not=}(\G))}}({\rm Int}^{\not=}(\G)) \big) \cr
&\times \exp \Big( \ell_2^{-d} f^A_{-\si(\G),\l,\g} | {\rm
Int}^{\not=}(\G)| \cr 
&-{1 \over 2}U(\hat \rho^{q^{(\tilde \d^{\not=}(\G))}}({\rm Int}^{\not=}(\G))|
\hat \rho^{q^{(\tilde \d^{\not=}(\G))}}(\p {\rm Int}^{\not=}(\G)))  \cr
&-\big(\widetilde H_{\g,\l}(\hat \rho^{q^{(\tilde \d^{\not=}(\G))}}
({\rm Int}^{\not=}(\G))|\hat \rho^{q^{(\tilde \d^{\not=}(\G))}}
(\p {\rm Int}^{\not=}(\G))) \cr
&- \widetilde H_{\g,\l} (\hat \rho_{\b,\l,\si}({\rm
Int}^{\not=}(\G))|  \hat \rho_{\b,\l,\si}(\p {\rm Int}^{\not=}(\G))) \big) \cr
&+\tilde g_{-\si(\G)}({\rm Int}^{\not=}(\G)| \hat \rho^{q^{(\tilde
\d^{\not=}(\G))}} 
(\p {\rm Int}^{\not=}(\G))) \Big) \cr}
   \Eq(5.33)
   $$
The term $\tilde g_{-\si(\G)}({\rm Int}^{\not=}(\G)| \hat \rho^{q^{(\tilde
\d^{\not=}(\G))}}(\p {\rm Int}^{\not=}(\G)))$ is the sum over ``polymer
type'' terms cross the boundary of ${\rm Int}^{\not=}(\G)$ and it can be
estimated (see \equ(5.31.1)) by
   $$
|\tilde g_{-\si(\G)}({\rm Int}^{\not=}(\G)| \hat \rho^{q^{(\tilde
\d^{\not=}(\G))}}(\p {\rm Int}^{\not=}(\G)))| \le c\g^{1/2}
|{\rm Int}^{\not=}(\G) \cap {\rm Int}^{\not=}(\G)^c|
   \Eq(5.34)
   $$
On the other hand the strip $S(\G)$ is so wide that in the $\g^{-1}$
neighborhood of ${\rm Int}^{\not=}(\G) \cap {\rm Int}^{\not=}(\G)^c$
the difference 
$|\hat \rho^{q^{(\tilde \d^{\not=}(\G))}}  - \hat \rho_{\b,\l,\si}|$
is exponentially 
small as follows from a straightforward analogue of \equ(6.29). Thus
   $$
\eqalign{
Z^A_{\g,\b,\l} \big({\rm Int}^{\not=}(\G)\cup S(\G)| q^{(\tilde
\d^{\not=}(\G))}\big)
&=Z^A_{\g,\b,\l} \big(S(\G)| q^{(\tilde \d^{\not=}(\G))},
\hat \rho_{\b,\l,\si}({\rm Int}^{\not=}(\G)) \big) \cr
&\times \exp \Big( \ell_2^{-d} f^A_{-\si(\G),\l,\g} | {\rm
Int}^{\not=}(\G)| \cr 
&-{1 \over 2}U(\hat \rho_{\b,\l,\si}({\rm Int}^{\not=}(\G))|
\hat \rho_{\b,\l,\si}(\p {\rm Int}^{\not=}(\G)))  \cr
&+\bar g_{-\si(\G)}({\rm Int}^{\not=}(\G)) \Big) \cr}
   \Eq(5.35)
   $$
where
   $$
|\bar g_{-\si(\G)}({\rm Int}^{\not=}(\G))|
\le c\g^{1/2}|{\rm Int}^{\not=}(\G) \cup {\rm
Int}^{\not=}(\G)|
   \Eq(5.36)
   $$
Here $\bar g_{-\si(\G)}({\rm Int}^{\not=}(\G))$ collects the
contribution of $\tilde g_{-\si(\G)}({\rm Int}^{\not=}(\G)| \hat
\rho^{q^{(\tilde \d^{\not=}(\G))}}(\p {\rm Int}^{\not=}(\G)))$ together
with corrections due to replacement of $U(\hat \rho^{q^{(\tilde
\d^{\not=}(\G))}}({\rm Int}^{\not=}(\G))| \hat \rho^{q^{(\tilde
\d^{\not=}(\G))}}(\p {\rm Int}^{\not=}(\G)))$ by $U(\hat
\rho_{\b,\l,\si}({\rm Int}^{\not=}(\G))| \hat \rho_{\b,\l,\si}(\p {\rm
Int}^{\not=}(\G)))$ and the estimate of $\widetilde H_{\g,\l}(\hat
\rho^{q^{(\tilde \d^{\not=}(\G))}} ({\rm Int}^{\not=}(\G))|$ \hfil
\goodbreak \noindent $\hat \rho^{q^{(\tilde \d^{\not=}(\G))}} (\p {\rm
Int}^{\not=}(\G))) - \widetilde H_{\g,\l} (\hat \rho_{\b,\l,\si}({\rm
Int}^{\not=}(\G))| \hat \rho_{\b,\l,\si}(\p {\rm Int}^{\not=}(\G)))$

Using \equ(5.7.1) we obtain from \equ(5.35) our last estimate
   $$
\eqalign{
Z^A_{\g,\b,\l} \big({\rm Int}^{\not=}(\G)\cup S(\G)| q^{(\tilde
\d^{\not=}(\G))}\big)
&=Z^A_{\g,\b,\l} \big(S(\G)| q^{(\tilde \d^{\not=}(\G))},
\rho_{\b,\l,\si}({\rm Int}^{\not=}(\G)) \big) \cr
&\times \exp \Big( \ell_2^{-d} f^A_{-\si(\G),\l,\g} | {\rm
Int}^{\not=}(\G)| \cr 
&-{1 \over 2}U(\rho_{\b,\l,\si}({\rm Int}^{\not=}(\G))|
\rho_{\b,\l,\si}(\p {\rm Int}^{\not=}(\G)))  \cr
&+\bar g_{-\si(\G)}({\rm Int}^{\not=}(\G)) + \D \cr}
   \Eq(5.37)
  $$
where $|\D| \le c \g^{1/4} |S(\G)|$.

A similar estimate is true for the denominator of \equ(3.10) which after
integration over $q^{(\tilde \d^{\not=}(\G)}$ implies \equ(3.10e6).
\qed

\bigskip
\bigskip
\goodbreak 
\centerline{{\bf 7. Appendix. Polymer Expansion Theorem }}
\bigskip
\numsec=7
\numfor=1
\numtheo=1

Consider a finite or countable set $\Th$ the elements of which are called
(abstract) {\it diagrams} and denoted $\th,\th'$, ets. Fix some reflexive
and symmetric relation on $\Th\times\Th$. A pair $\th,\th' \in
\Th\times\Th$ is called incompatible ($\th\not\sim\th'$) if it satisfies
given relation and compatible ($\th\sim\th'$) in the opposite case. A
collection $\{\th_j\}$ is called a {\it compatible collection of diagrams}
if any two its elements are compatible. Every diagram $\th$ is assigned a
complex-valued {\it statistical weight} denoted by $w(\th)$, and for any
finite $\L\subseteq\Th$ an (abstract) {\it partition function} is defined
as $$Z(\L)=\sum_{\{\th_j\}\subseteq\L} \prod_j w(\th_j), \Eq(A.1)$$ where
the sum is extended to all compatible collections of diagrams
$\th_i\in\L$. The empty collection is compatible by definition, and it is
included in $Z(\L)$ with statistical weight $1$.

A {\it polymer} $\pi=\pg$ is an (unordered) finite collection of different
diagrams $\th_i \in \Th$ taken with positive integer multiplicities
$\e_i$, such that for every pair $\th',\ \th'' \in \pi $ there exists a
sequence $\th'=\th_{i_1},\ \th_{i_2},\ldots, \th_{i_s}=\th'' \in \pi$ with
$\th_{i_j}\not\sim\th_{i_{j+1}},\ j=1,2,\ldots,s-1$. The notation
$\pi\subseteq \L$ means that $\th_i \in \L$ for every $\th_i \in \pi$.

With every polymer $\pi$ we associate an (abstract) graph $G(\pi)$
which consists of $\sum_i \e_i$ vertices labeled by the diagrams from
$\pi$ and edges joining every two vertices labeled by incompatible
diagrams. It follows from the definition of $G(\pi)$ that it is
connected and we denote by $r(\pi)$ the quantity
$$r(\pi)=\prod_i (\e_i!)^{-1} \sum_{G' \subset G(\pi)} (-1)^{|G'|},
\Eq(A.2)$$
where the sum is taken over all connected subgraphs $G'$ of $G(\pi)$
containing all of $\sum_i \e_i$ vertices and $|G'|$ denotes
the number of edges in $G'$. For any $\th \in \pi$ we denote by
$\e(\th,\pi)$ the multiplicity of $\th$ in the polymer $\pi$.

The polymer expansion theorem below is a modification of results of [Se]
and [KP] proven in [MSu]. See also [D3] for similar results.

\medskip

{\bf \Theorem (sA.1)} {\sl Suppose that there exists a function
$a(\th):\ \Th \mapsto {\bf R}^{+}$ such that for any diagram $\th$
$$\sum_{\th':\ \th'\not\sim\th} |w(\th')| e^{a(\th')} \le
a(\th).\Eq(A.3)$$ 
Then, for any finite $\L$, 
$$\log Z(\L)=\sum_{\pi\subseteq\L} w(\pi), \Eq(A.4)$$
where the statistical weight of a polymer $\pi=\pg$ equals
$$w(\pi)=r(\pi) \prod_i w(\th_i)^{\e_i}.  \Eq(A.5)$$
Moreover, the series \equ(A.4) for $\log Z(\L)$ is absolutely convergent
in view of the estimate
$$\sum_{\pi:\ \pi\ni\th} \e(\th,\pi) |w(\pi)| \le 
|w(\th)| e^{a(\th)}, \Eq(A.6)$$
which is true for any diagram $\th$.}
\medskip

{\bf \Corollary (sA.2)} {\sl For any function $b(\th):\ \Th
\mapsto {\Bbb R}^{+}$ consider modified statistical weights of diagrams
$$\tilde w(\th)=w(\th) e^{b(\th)}. \Eq(A.7)$$
and suppose that still
$$\sum_{\th':\ \th'\not\sim\th} |\tilde w(\th')| e^{a(\th')} \le
a(\th).\Eq(A.8)$$ 
Then for any family, $\Pi$, of polymers such that any $\pi \in \Pi$
contains given diagram $\t$
$$\sum_{\pi \in \Pi} \e(\th,\pi) |w(\pi)| \le
|\tilde w(\th)| e^{a(\th)} 
\min_{\pi \in \Pi} \left( \prod_{\t' \in \pi} e^{b(\t')} \right)
\Eq(A.9)$$}

\bigskip
\bigskip
\goodbreak 
\centerline{{\bf 8. Appendix. A Technical Lemma }}
\bigskip
\numsec=8
\numfor=1
\numtheo=1

Consider a spin model on the lattice $\Z^d$ given by the formal
Hamiltonian 
   $$
H(\phi)=\sum_x \phi_x^2 + {1 \over 2} \sum_{x \not= y} J_{xy} \phi_x
\phi_y + \sum_{A \in {\cal A}} K_A \phi_A
   \Eq(5.19)
   $$
Here the spin variable $\phi_x$ takes $M$ discrete values, including
$0$, from the bounded interval $[-a,a]$. The last sum runs over some
family $\cal A$ of sets $A=\{x_1, \ldots, x_{|A|} \in \Z^d\}$ containing
$|A|$ not necessarily different sites $x$ and $\phi_A= \prod_{x \in A}
\phi_x$. The interaction, $J_{xy}$ and $K_A$, is of finite range $R <
\infty$, i.e. $J_{xy}=0$ if dist$(x,y) \ge R$ and $K_A=0$ if diam$(A) \ge
R$. Suppose that
   $$
\sum_{y \not= x}| J_{xy}|=1-\a,\quad 0 < \a <1
   \Eq(5.20)
   $$
and
   $$
\sum_{A \ni x} a^{|A|} |K_A| \ll \a
   \Eq(5.21)
   $$
and denote by $\mu(\phi(\L))$ the corresponding Gibbs distribution in
the finite domain $\L \subset \Z^d$ with zero ($\equiv$ empty) 
boundary condition in $\L^c$. 

\medskip\noindent
{\bf \Lemma (s5.1)} 

\nobreak
{\sl For any $c>1$ and sufficiently large inverse temperature 
   $$
\kappa >{M+R \over c}
   \Eq(5.22)
   $$
one has
   $$
\mu(|\phi_x| \ge \kappa^{-5/12}) \le e^{-c \kappa^{1/6}},\quad \forall x \in \L
   \Eq(5.22.1)
   $$
}

\medskip\noindent 
{\bf Proof.} For an arbitrary configuration $\phi(\L)$ define {\it spots}
$S(\phi(\L))$ as $R$-connected components of sites $x \in \L$ with
$|\phi_x| \ge \kappa^{-5/12}$. Taking $x \in S$ set 
   $$
h(x,\phi(\L))=\sum_{y \not= x} J_{xy} \phi_x \phi_y \ind{|\phi_x| \ge |\phi_y|}
+ \sum_{A \in {\cal A}:\; A \ni x} K_A \phi_A 
\ind{|\phi_x| \ge \max_{y \in A}|\phi_y|}
   \Eq(5.23)
   $$
Then
   $$
H(\phi(S)| \phi(\L \setminus S))= \sum_{x \in S} h(x,\phi(\L))
   \Eq(5.24)
   $$
and
   $$
h(x,\phi(\L))\ge {\a \over 2} \phi_x^2
   \Eq(5.25)
   $$
Here in \equ(5.24)-\equ(5.25) we used \equ(5.20), \equ(5.21) and the
fact that $|\phi_x| \ge |\phi_y|$ for any $x \in S$ and $y \in
\p^{(R)}S$.

Now the probability that $\phi(\L)$ contains a spot $S$ with the fixed
value of $\phi(S)$ can be estimated as
   $$
\eqalign{
\mu(\phi(S)) &= {\sum_{\phi(\L \setminus S)} 
\exp \big(-\kappa H(\phi(S)|\phi(\L \setminus S)) 
-\kappa H(\phi(\L \setminus S)) \big)
\over \sum_{\phi(\L)} \exp \big( -\kappa H(\phi(\L)) \big) } \cr
&\le \exp \big( -\kappa {\a \over 2} \sum_{x \in S} \phi_x^2 \big)
{\sum_{\phi(\L \setminus S)} \exp \big(
-\kappa H(\phi(\L \setminus S)) \big)
\over \sum_{\phi(\L):\; \phi(S)\equiv 0} 
\exp \big( -\kappa H(\phi(\L)) \big) } \cr
&\le  \exp \big( -{\a \over 2} \kappa^{1/6} |S| \big)
{\sum_{\phi(\L \setminus S)} 
\exp \big( -\kappa H(\phi(\L \setminus S)) \big)
\over \sum_{\phi(\L):\; \phi(S)\equiv 0} 
\exp \big( -\kappa H(\phi(\L)) \big) } \cr
&=\exp \big( -{\a \over 2} \kappa^{1/6} |S| \big) \cr}
   \Eq(5.26)
   $$
Hence the probability of the spot $S$ 
   $$
\mu(S)=\sum_{\phi(S)} \mu(\phi(S)) \le (c_{\equ(5.22)} \kappa)^{|S|}
\exp \big( -{\a \over 2} \kappa^{1/6} |S| \big) \le 
\exp \big( -{\a \over 4} \kappa^{1/6} |S| \big), 
   \Eq(5.27)
   $$
where the last inequality is true for $\kappa$ large enough.

Finally for any $x \in \L$
   $$
\eqalign{
\mu(|\phi_x| \ge \kappa^{-5/12}) &=\sum_{S \ni x} \mu(S) \cr
&\le \sum_{|S|=1}^{\infty} (c_{\equ(5.22)} \kappa)^{|S|} 
\exp \big( -{\a \over 4} \kappa^{1/6} |S| \big) \cr
&\le \sum_{|S|=1}^{\infty}
\exp \big( -{\a \over 8} \kappa^{1/6} |S| \big) \cr
&\le e^{-c \kappa^{1/6}} \cr},
   \Eq(5.28)
   $$
where again $\kappa$ is large enough. \qed

\medskip
Let now $\mu^{(0)}(\phi_x, \phi_y)$ and $\mu^{(1)}(\phi_x, \phi_y)$ be a
distribution at sites $x, y \in \L$ of conditional Gibbs measures
in $\L$ with two arbitrary boundary conditions $\bar \phi^{(0)}(\L^c)$
and $\bar \phi^{(1)}(\L^c)$ respectively. The Vasserstein distance
$R(\mu^{(0)},\mu^{(1)})$ between $\mu^{(0)}$ and $\mu^{(1)}$ is given by
some coupling $\mu(\phi^{(0)}_x, \phi^{(0)}_y; \phi^{(1)}_x,
\phi^{(1)}_y)$. Denote by $\langle \cdot \rangle^{(0)}$, 
$\langle \cdot \rangle^{(1)}$ and $\langle \cdot \rangle$ the
expectation with respect to $\mu^{(0)}$, $\mu^{(1)}$ and $\mu$ and
suppose that $2a <1$. Then
   $$
\eqalign{
|\langle \phi_x, \phi_y \rangle^{(0)} -
\langle \phi_x, \phi_y \rangle^{(1)}| 
&=\langle 
(\phi^{(0)}_x- \phi^{(1)}_x)\phi^{(0)}_y 
+ (\phi^{(0)}_y- \phi^{(1)}_y) \phi^{(1)}_x \rangle \cr
&\le \langle (\phi^{(0)}_x- \phi^{(1)}_x)^2 \rangle^{1/2} 
\langle (\phi^{(0)}_y)^2 \rangle^{1/2}
+ \langle (\phi^{(0)}_y- \phi^{(1)}_y)^2 \rangle^{1/2} 
\langle (\phi^{(1)}_x)^2 \rangle^{1/2} \cr
&\le \langle |\phi^{(0)}_x- \phi^{(1)}_x| \rangle^{1/2} 
\langle (\phi^{(0)}_y)^2 \rangle^{1/2}
+ \langle |\phi^{(0)}_y- \phi^{(1)}_y| \rangle^{1/2} 
\langle (\phi^{(1)}_x)^2 \rangle^{1/2} \cr
&\le \max \left( \Big(\langle \phi_x^2 \rangle^{(0)}\Big)^{1/2},
\Big(\langle \phi_y^2 \rangle^{(1)}\Big)^{1/2} \right) \cr
&\times \max \left( R(\mu^{(0)}(\phi_x), \mu^{(1)}(\phi_x))^{1/2},
R(\mu^{(0)}(\phi_y), \mu^{(1)}(\phi_y))^{1/2} \right), \cr}
   \Eq(5.29)
   $$
Similarly one can estimate via the Vasserstein distance and the second
moment the difference $|\langle \prod_{x \in X} \phi_x \rangle^{(0)} -
\langle \prod_{x \in X} \phi_x \rangle^{(1)}|$ for any $X \subset \L$.

\bigskip
\bigskip
\goodbreak 
\centerline{{\bf References }}
\bigskip

\item{~[Bax]} R.J.Baxter, {\it Exactly solved models in statistical
mechanics}, London-New York: Academic Press (1982).

\item{~[Bak]} Baker

\item{~[BKL]} J.Bricmont, K.Kuroda and J.L.Lebowitz, ``First Order Phase
Transitions in Lattice and Continuous Systems: Extension of Pirogov-Sinai
Theory'', {\it Comm. Math. Phys.} {\bf 101}, 501-538 (1985).

\item{~~[BP]} T.Bodineau and E.Presutti, ``Phase Diagram of Ising Systems
with Additional Long Range Forces'', submitted to {\it Comm. Math. Phys.}
(1996). 

\item{~~[BZ]} A. Bovier and M. Zahradnik, ``The low temperature phase of
Kac-Ising models'', {\it J. Stat. Phys.} to appear.
   
\item{[COPP]} M.Cassandro, E.Olivieri, A.Pellegrinotti and E.Presutti,
``Existence and Uniqueness of DLR measures for Unbounded Spin Systems'',
{\it Z. Wahr. verv. Geb.} {\bf 41}, 313-334 (1978).

\item{~~[CP]} M.Cassandro and E. Presutti, ``Phase transitions in Ising
systems with long but finite range interactions'' {\it Markov Processes
and Related Fields} {\bf 2}, 241--262 (1996).

\item{~~[D1]} R.L.Dobrushin, ``Existence of Phase Transition in Two and
Three Dimensional Ising Models'', {\it Th. Prob. Appl.} {\bf 10}, 193-313
(1965).

\item{~~[D2]} R.L.Dobrushin, ``Prescribing a System of Random Variables by
Conditional Distributions'', {\it Th. Prob. Appl.} {\bf 15}, 458-456
(1970). 

\item{~~[D3]} R.L.Dobrushin, ``Estimates of Semi-invariants for the Ising
Model at Low Temperatures'', Topics in statistical and theoretical
physics, {\it Amer. Math. Soc. Transl. (2)}, {\bf V177}, 59--81 (1996).

\item{~~[DS]} E.I.Dinaburg and Ya.G.Sinai, ``Contour Models with Interaction
and their Applications'', {\it Sel. Math. Sov.} {\bf 7}, 291-315 (1988).

\item{~~[DZ]} R.L.Dobrushin and M.Zahradnik, ``Phase Diagrams for 
Continuous Spin Models. Extension of Pirogov-Sinai Theory'', in {\it
Mathematical Problems of Statistical Mechanics and Dynamics}, R.L.Dobrushin
ed., Dordrecht, Boston: Kluwer Academic Publishers, 1-123 (1986).

\item{~~~[F]} M.V.Fedoryuk, {\it Asymptotic: Integrals and Series}, Moscow:
Nauka (1987).

\item{~~[FF]} B.U.Felderhof and M.E.Fisher, ``Phase Transitions in
One-Dimensional Cluster-Inter\-action Fluids. IA. Thermodynamics,
IB. Critical Behavior, II Simple Logarithmic Model'', {\it Annals of Phys.}
{\bf 58} N1, 176-216, 217-267, 268-280 (1970).

\item{~~[Ge]} H-O.Georgii, 

\item{~~[Gr]} R.B.Griffiths, ``Peierls' Proof of Spontaneous Magnetization
in a Two Dimensional Ising Ferromagnet'', {\it Phys. Rev.} {\bf 136A},
437-439 (1964)

\item{~~~[J]} K.Johansson, ``On separation of phases in one-dimensional
gases'', {\it Comm. Math. Phys.}  {\bf 169}, 521--561 (1995).

\item{~~[vK]}

\item{~~[KP]} R.Kotecky and D.Preiss, ``Cluster Expansion for Abstract
Polymer Models'', {\it Comm. Math. Phys.}, {\bf 103}, 491-498 (1986)

\item{~[KUH]} M.Kac, G.Uhlenbeck and P.C.Hemmer, ``On the Van der Waals
Theory of Vapor-Liquid equilibrium'', {\it J. Mat. Phys} {\bf 4},
216-228, 229-247 (1963), {\it J. Mat. Phys} {\bf 5}, 60-74 (1964).

\item{~~[LL]} J.L.Lebowitz and E.H.Lieb, ``Phase Transition in a Continuum
Classical System with Finite Interactions'', {\it Phys. Let.} {\bf 39A},
N2, 98-100 (1972).

\item{~[LMP]} J.L.Lebowitz, A.E.Mazel and E.Presutti, in preparation

\item{~~[LP]} J.L.Lebowitz and O.Penrose, ``Rigorous Treatment of the Van
der Waals Maxwell Theory of the Liquid-Vapor Transition'', {\it
J. Mat. Phys} {\bf 7}, 98-113 (1966).

\item{~~[MS]} A.E.Mazel and Yu.M.Suhov, Ground States of Boson Quantum Lattice 
Model, {\it Amer. Math. Soc. Transl. (2)}, {\bf V171}, 185-226 (1996). 

\item{~~~[O]} L.Onsager, ``Crystal Statistics I. A Two-Dimensional Model
with an Order-Disorder Transition'', {\it Phys. Rev.} {\bf 65}, 117-149
(1944)

\item{~~~[P]} R.Peierls, ``On Ising's Model of Ferromagnetism'', {\it
Proc. Camb. Phil. Soc.} {\bf 32}, 477-481 (1936).

\item{~~[PS]} S.A.Pirogov and Ya.G.Sinai, ``Phase Diagrams of Classical
Lattice Systems'', {\it Theor. and Math. Phys.} {\bf 25}, 358-369,
1185-1192 (1975).

\item{~~[R1]} D.Ruelle, ``Existence of a Phase Transition in a Continuous
Classical system'', {\it Phys. Rev. Let.} {\bf 27}, 1040-1041 (1971).

\item{~~[R2]} D.Ruelle, ``Probability Estimates for Continuous Spin
Systems'',  {\it Comm. Math. Phys.} {\bf 50}, 189-194 (1976).

\item{~~~[S]} Ya.G.Sinai, {\it Theory of Phase Transitions}, Budapest:
Academia Kiado and London: Pergamon Press (1982)

\item{~~[Se]} E.Seiler, ``Gauge Theories as a Problem of Constructive
Quantum Field Theory and Statistical Mechanics'', {\it Lect. Notes in
Physics}, {\bf 159}, Berlin: Springer-Verlag (1982).

\item{~~[WR]} B.Widom and J.S.Rowlinson, ``New Model for the Study of
Liquid-Vapor Phase Transitions'', {\it J. Chem. Phys.} {\bf 52}, 1670-1684
(1970). 

\item{~~~[Z]} M.Zahradnik, ``An Alternate Version of Pirogov-Sinai
Theory'', {\it Comm. Math. Phys.} {\bf 93}, 559-581 (1984).

\end